\documentclass[twocolumn,aps,prc,floatfix,superscriptaddress]{revtex4-2}

\usepackage{graphicx}
\usepackage{dcolumn}
\usepackage{bm}
\usepackage{hyperref}
\usepackage{amsmath,bm}
\usepackage{amssymb}
\usepackage{longtable}
\usepackage{placeins}
\usepackage[mathlines]{lineno}
\usepackage{xcolor}
\usepackage{color}
\usepackage{ulem}
\allowdisplaybreaks[4]

\newcommand{\mspl}{m_{\mathrm{n}\text{-}\mathrm{p}}^{\ast}}

\renewcommand{\sout}{\bgroup \color{red} \ULdepth=-.5ex \ULset}

\newcommand{\hA}{\hat{A}}
\newcommand{\hB}{\hat{B}}

\newcommand{\Ahatp}{\left( \frac{\vec{k}^{\prime}{}^2+\vec{k}{}^2}{2} \right)}
\newcommand{\Bhat}{\vec{k}^{\prime}\cdot\vec{k}}
\newcommand{\Bhatp}{ \left( \vec{k}^{\prime}\cdot\vec{k} \right)}
\newcommand{\DiracrR}{\hat{\delta}(\vec{r}_1 - \vec{r}_2)}
\newcommand{\Psig}{\hat{P}_{\sigma}}
\newcommand{\Pm}{\hat{P}_{M}}
\newcommand{\Ptau}{\hat{P}_{\tau}}
\newcommand{\rhor}{ \rho(\vec{r})}
\newcommand{\rhotr}{ \rho_{\tau}(\vec{r})}

\newcommand{\Skypax}[2]{x_{#1}^{[#2]}}
\newcommand{\Skypat}[2]{t_{#1}^{[#2]}}

\begin{document}

\title{Extended Skyrme effective interactions with higher-order momentum-dependence for transport models and neutron stars}

\author{Si-Pei Wang}
\affiliation{
School of Physics and Astronomy, Shanghai Key Laboratory for
Particle Physics and Cosmology, and Key Laboratory for Particle Astrophysics and Cosmology (MOE),
Shanghai Jiao Tong University, Shanghai 200240, China
}
\author{Xin Li}
\affiliation{
School of Physics and Astronomy, Shanghai Key Laboratory for
Particle Physics and Cosmology, and Key Laboratory for Particle Astrophysics and Cosmology (MOE),
Shanghai Jiao Tong University, Shanghai 200240, China
}
\affiliation{College of Physics, Henan Normal University, Xinxiang 453007, China}

\author{Rui Wang}
\affiliation{%
Istitudo Nazionale di Fisica Nucleare (INFN), Sezione di Catania, I-95123 Catania, Italy
}%

\author{Jun-Ting Ye}
\author{Lie-Wen Chen}
 \email{Correspinding author: lwchen@sjtu.edu.cn}
\affiliation{
School of Physics and Astronomy, Shanghai Key Laboratory for
Particle Physics and Cosmology, and Key Laboratory for Particle Astrophysics and Cosmology (MOE),
Shanghai Jiao Tong University, Shanghai 200240, China
}

\date{\today}

\begin{abstract}
The recently developed extended Skyrme effective interaction based on the so-called N3LO Skyrme pseudopotential
is generalized to the general N$n$LO case
by incorporating the derivative terms up to 2$n$th-order into the central term of the pseudopotential.
The corresponding expressions of Hamiltonian density and single-nucleon potential are derived within the Hartree-Fock approximation under general nonequilibrium conditions.
The inclusion of the higher-order derivative terms provides additional higher-order momentum dependence for the single-nucleon potential, and in particular, we find that the N5LO single-nucleon potential with momentum dependent terms up to $p^{10}$ can give a nice description for the empirical nucleon optical potential up to energy of $2$ GeV.
At the same time,
the density-dependent terms in the extended Skyrme effective interaction are extended correspondingly in the spirit of the Fermi momentum expansion, which allows highly flexible variation of density behavior for both the symmetric nuclear matter equation of state and the symmetry energy.
Based on the Skyrme pseudopotential up to N3LO, N4LO and N5LO,
we construct a series of interactions with the nucleon optical potential having different high-momentum behaviors and with the symmetry potentials featuring different linear isospin-splitting coefficients for nucleon effective mass, by which we study the properties of nuclear matter and neutron stars.
Furthermore, within the lattice BUU transport model, some benchmark simulations with selected interactions are performed for the Au+Au collisions at a beam energy of $1.23$~GeV/nucleon, and the predicted collective flows for protons are found to nicely agree with the data measured by HADES collaboration.
\end{abstract}

\maketitle

\section{INTRODUCTION}
The equation of state~(EOS) of dense nuclear matter,
which is directly related to the in-medium effective nuclear interactions, is of fundamental importance for both nuclear physics and astrophysics~\cite{Li:1997px,Lattimer:2000kb,Danielewicz:2002pu,Lattimer:2004pg,Baran:2004ih,Steiner:2004fi,Li:2008gp,Oertel:2016bki}.
In the past four decades,
the EOS of symmetric nuclear matter~(SNM) from nuclear saturation density $\rho_0$ to suprasaturation density of about $5\rho_0$ has been relatively well constrained, based on analyses on the data from the giant monopole resonance in finite nuclei~\cite{Blaizot:1980tw,Youngblood:1999zza,Li:2007bp,Li:2022suc,Shlomo:2006ole,Garg:2018uam} as well as the measurements of collective flow~\cite{Danielewicz:2002pu,LeFevre:2015paj} and kaon production~\citep{Aichelin:1985rbt,Fuchs:2000kp,Hartnack:2005tr,Fuchs:2005zg} in high energy heavy-ion collisions~(HICs).
For the symmetry energy $E_{\rm sym}(\rho)$,
which characterizes the isospin dependent part of the EOS for asymmetric nuclear matter~(ANM),
significant progress has been made in constraining its density dependence around and below the nuclear saturation density $\rho_0$ by analyzing finite nuclei properties~\cite{Chen:2005ti,Centelles:2008vu,Chen:2010qx,Agrawal:2012pq,Zhang:2013wna,Brown:2013mga,Roca-Maza:2012uor,Zhang:2014yfa,Danielewicz:2013upa,Zhang:2015ava,Danielewicz:2016bgb,Xu:2020xib,Qiu:2023kfu}, isospin probes in HICs at energies less than about $100$~MeV/nucleon~\cite{Chen:2004si,Li:2005jy,Kowalski:2006ju,Shetty:2005qp,Tsang:2008fd,Wada:2011qm,Morfouace:2019jky,Zhang:2020azr} and microscopic many-body calculations (see, e.g., Refs.~\cite{Tews:2012fj,Wlazlowski:2014jna,Akmal:1998cf,Baldo:2014rda,Carbone:2014mja}), but its high density behavior remains elusive.
In recent years, many works have been devoted to constraining the high density behavior of the symmetry energy by using the multimessenger data, especially the simultaneous determination of the mass and radius of neutron stars by NICER as well as the gravitational wave observation in binary neutron star merger, but the uncertainty of the constraints is still very large~\citep{FiorellaBurgio:2018dga,Zhang:2018bwq,Zhou:2019omw,Zhou:2019sci,Li:2021thg,Krastev:2021reh,Yue:2021yfx,Koehn:2024set}.
Dense neutron-rich nuclear matter
can be present or produced in neutron-rich nuclei, heavy-ion collisions~(HICs) induced by neutron-rich nuclei, neutron stars and their mergers, as well as in core-collapse supernovae. The ongoing accumulation of experimental data from radioactive beam facilities around the world, together with multimessenger signals of neutron stars and their mergers from astrophysical observations will provide unique opportunities to pin down the EOS of dense neutron-rich matter~\cite{Lovato:2022vgq,Sorensen:2023zkk}.

The HICs provide a unique approach to produce dense nuclear matter in terrestrial laboratories.
In HICs, the state of matter produced during the collisions can be modified by adjusting the beam energy, collision geometry as well as the isospin asymmetry of the collision system.
In order to describe the dynamics of the collisions and extract the EOS of dense nuclear matter from the collision products which generally involves dynamics far from thermal equilibrium, microscopic transport models such as the Boltzmann-Uehling-Uhlenbeck~(BUU) equation~\cite{Bertsch:1988ik} and the quantum molecular dynamics~(QMD) model~\cite{Aichelin:1991xy} have been developed and widely applied.
In recent years, the Transport Model Evaluation Project~(TMEP) has been pursued to test the robustness of transport models and then try to narrow down the uncertainties of their predictions~\cite{TMEP:2016tup,TMEP:2017mex,TMEP:2019yci,TMEP:2021ljz,TMEP:2022xjg,TMEP:2023ifw}.
One direct and fundamental input for the transport model is the single-nucleon potential (nuclear mean-field potential), which is generally obtained from an in-medium effective nuclear interactions based on the Hartree-Fock (HF) approach.
It is directly connected to the nuclear matter EOS through the corresponding energy-density functional~(EDF)~\cite{Bertsch:1988ik,Aichelin:1991xy}.
Therefore, probing the single-nucleon potential is a key issue of transport model simulations for HICs at intermediate energies.

The single-nucleon potential in neutron-rich matter generally depends
on the density and isospin asymmetry of the medium as well as the momentum and isospin of the nucleon.
The momentum/energy dependence of the single-nucleon potential plays an important role for nuclear systems at finite temperatures or in nonequilibrium conditions, including not only HICs, but also proto-neutron stars, neutron star mergers and supernovae.
The effective nuclear interactions used in transport model simulations are therefore necessary to accurately describe the momentum/energy dependence of the single-nucleon potential.
In previous works~\cite{Carlsson:2008gm,Raimondi:2011pz},
a Skyrme-like quasilocal EDF up to N3LO has been constructed by including additional higher-order derivative terms (higher-power momentum dependence) in the conventional Skyrme interaction, which is notorious with the incorrect high energy behavior of the nucleon optical potential when nucleon kinetic energy is above about $300$ MeV/nucleon.
Actually, the Skyrme interaction can be recognized as a low-momentum expansion of the finite-range interaction, and a quasilocal functional can be obtained from a finite-range interaction through the density-matrix expansion \cite{Carlsson:2010da,Dobaczewski:2010qp}.
Indeed, the construction of the quasilocal interaction based on the density-matrix expansion is quite important, as it provides a general procedure for exploring the universal EDF of nuclear systems and examining the validity of each term in an order-by-order way.
Through the partial wave decomposition, it has been shown that the Skyrme pseudopotential up to N3LO provides a very good approximation of finite-range potential up to about saturation density, and thus it is flexible enough to substitute finite-range interaction in nuclear structure calculations
\cite{Davesne:2016fqg}.

Moreover, the additional higher-order derivatives in the effective interaction extend the momentum-dependent (MD) terms and density-gradient terms in the corresponding nuclear EDF, which could be important for investigating HICs using transport models.
We would like to emphasize that, one of important advantages for the polynomial form of the mean-field potential provided by the Skyrme pseudopotential is that it can significantly reduce the computational complexity of the transport model, and we will further discuss this point later.
Based on the Skyrme pseudopotential up to N3LO, the extended Skyrme interactions have been built within the mean-field approximation in Ref.~\cite{Wang:2018yce} to reproduce the empirical results on the nucleon optical potential up to energy of $1$ GeV (corresponding to nucleon momentum of $1.5$ GeV/$c$) obtained by Hama \textit{et al.} from analyzing the proton-nucleus elastic scattering data~\cite{Hama:1990vr,Cooper:1993nx}.
These extended Skyrme interactions have been applied in the lattice BUU transport model to successfully describe the nuclear collective dynamics (giant resonances)~\cite{Wang:2019ghr,Wang:2020xgk,Wang:2020ixf,Song:2021hyw,Song:2023fnc} and the light-nuclei production at energies of $0.25$-$1.0$~GeV/nucleon~\cite{Wang:2023gta} measured by the FOPI Collaboration~\cite{FOPI:2010xrt}.
Similar idea has also been employed in QMD-type transport models~\cite{Yang:2023umm}.

Besides the HICs experiments at intermediate and high energies in terrestrial laboratories, the neutron stars and their dynamics also provide natural laboratories to shed light on the EOS of dense nuclear matter.
Recent multimessenger astrophysical observations, such as the gravitational wave signal from the binary neutron-star merger~\cite{LIGOScientific:2018cki}, the relativistic Shapiro delay measurements~\cite{NANOGrav:2019jur,Fonseca:2021wxt}, the x-ray data obtained by Neutron Star Interior Composition Explorer~(NICER) and x-ray multimirror (XMM-Newton) telescope~\cite{Riley:2019yda,Miller:2019cac,Riley:2021pdl,Miller:2021qha,Choudhury:2024xbk} have been used to determine the maximum mass~($M_{\mathrm{TOV}}$), the mass-radius relation~(M-R), and the tidal deformability ($\Lambda$) of neutron stars, thereby further to constrain the EOS of neutron-rich matter at suprasaturation densities.
To make the Skyrme pseudopotential applicable in neutron stars, it is proposed recently in Ref.~\cite{Wang:2023zcj} to extend the density-dependent~(DD) terms based on the Fermi momentum expansion, which is considered to be a model-independent parametrization of the nuclear matter EOS \cite{Patra:2022yqc}.
By this means, the density behavior of nuclear matter bulk properties, including properties of SNM and the symmetry energy, can be highly flexible.
A series of extended Skyrme interactions up to N3LO with different slope parameter $L$ of the symmetry energy have been constructed in Ref.~\cite{Wang:2023zcj}.
These interactions are developed not only by fitting the empirical nucleon optical potential, but also by incorporating the empirical properties of SNM, the microscopic calculations of pure neutron matter~(PNM) as well as neutron stars properties from astrophysical observations through modifying the extended DD terms.

It should be pointed out that
for the Skyrme pseudopotential up to N3LO, it is failing to describe the saturated behavior of nuclear optical potential when the kinetic energy exceeds $1$ GeV, although it can give a nice description of the empirical nuclear optical potential below $1$ GeV~\cite{Wang:2023zcj}.
As a matter of fact,
the single-nucleon potential in the extended Skyrme interactions up to N3LO rises rapidly when nucleon kinetic energy exceeds $1$ GeV if the empirical nuclear optical potential below $1$ GeV is well reproduced~\cite{Wang:2018yce}, mainly due to its polynomial structure in the framework of Skyrme pseudopotential,
and this feature
may hinder the interaction from its application in transport model simulations for heavy-ion collisions when the incident energy exceeds $1$ GeV.
To address this problem, in the present work, we provide a general approach to extend the Skyrme pseudopotential by including derivative terms up to 2$n$th-order in the central term of the Skyrme pseudopotential, corresponding to the Skyrme pseudopotential up to N$n$LO.
We derive the corresponding expressions of Hamiltonian density as well as single-nucleon potential, whose MD part is now extended to $p^{2n}$.
This extension provides extra flexibility of the high-momentum behavior of the single-nucleon potential, and with an optimization method we find that the N5LO Skyrme pseudopotential can provide a good description of the empirical nucleon optical potential up to energy of $2$~GeV with a saturated behavior above $1$ GeV.
Therefore, the N5LO Skyrme pseudopotential can be applied in transport model simulations for HICs at incident energy up to $2$~GeV/nucleon, such as the experiments conducted and to be conducted by HIAF~\cite{Yang:2013yeb,Xiaohong:2018weu} in China, HADES~\cite{HADES:2009aat,HADES:2017def,HADES:2020lob,HADES:2022osk} and FAIR~\cite{Sturm:2010yit,CBM:2016kpk} at GSI in Germany, J-PARC-HI~\cite{Sako:2014fha,J-PARCHeavy-Ion:2016ikk,Sako:2019hzh} in Japan, and NICA~\cite{Kekelidze:2016wkp} and BM@N~\cite{Kapishin:2016ojm} at JINR in Russia.
Note the density of the nuclear matter produced in HICs at incident energies of about $2$~GeV/nucleon is expected to exceed $3\rho_{0}$ during the collision process~\cite{Li:2002yda}, and therefore the N5LO Skyrme pseudopotential can be principally applied to probe the EOS of nuclear matter with density up to $3\rho_{0}$.

Furthermore, in the present work, we follow the Fermi momentum expansion approach in Ref.~\cite{Wang:2023zcj} to construct the DD terms in the N$n$LO Skyrme pseudopotential.
Particularly, for the N5LO Skyrme pseudopotential, the adjustable macroscopic parameters to characterize the density behavior of the symmetry energy, which correspond to the expansion coefficients of the symmetry energy at $\rho_{0}$, shall include: the magnitude $E_{\mathrm{sym}}(\rho_0)$, the slope parameter $L$, the curvature parameter $K_{\mathrm{sym}}$, the skewness parameter $J_{\mathrm{sym}}$, the kurtosis parameter $I_{\mathrm{sym}}$ and the hyper-skewness parameter $H_{\mathrm{sym}}$.
Similarly, the adjustable characteristic parameters of the SNM are also increased to six quantities: the nuclear saturation density $\rho_{0}$, the binding energy $E_{0}(\rho_{0})$ as well as its expansion coefficients at $\rho_{0}$ from the second to the fifth order, i.e., $K_{0}$, $J_{0}$, $I_{0}$ and $H_{0}$.

Another important quantity in the construction of the extended Skyrme effective interaction is the (first-order) symmetry potential, which represents the isospin-dependent part of the single-nucleon potential in ANM.
The uncertainty of the symmetry energy may result from our poor understanding of the symmetry potential, and this can be evidenced from the single-nucleon potential decomposition of $E_{\mathrm{sym}}(\rho)$ and $L$~\cite{Brueckner:1964zz,Dabrowski:1972mbb,Dabrowski:1973zz,Xu:2010fh,Xu:2010kf,Chen:2011ag} based on the Hugenholtz-Van Hove (HVH) theorem \cite{Hugenholtz:1958zz,SATPATHY199985}.
Similar to the structure of the single-nucleon potential, the symmetry potential also includes MD terms up to $p^{2n}$, with a total of $n$ adjustable parameters to control its momentum behavior.
In order to ensure that the symmetry potential behaves well in the momentum range up to $2$ GeV/$c$ (avoiding sharp changes or fluctuations), we assume a scaling between the coefficients of each $p^{2n}$ term. In this case, we can use one physical quantity, i.e., the linear isospin splitting coefficient $\Delta m_{1}^{\ast}(\rho_{0})$, to feature the different symmetry potentials.

To demonstrate the validity and clearly illustrate the comparison of the extension at different orders, we construct in this work a series of interactions based on the Skyrme pseudopotential up to N3LO, N4LO and N5LO, respectively.
The single-nucleon potentials of these three models are consistent with the empirical nucleon optical potential up to energy of $1$ GeV, resulting in similar values of the isoscalar nucleon effective mass at saturation density $m^{\ast}_{s,0}/m \approx 0.77$, while their behaviors above $1$ GeV differ significantly due to the high-order momentum dependence.
At the same time, we use the parametrization to construct eight symmetry potentials with $\Delta m_{1}^{\ast}(\rho_{0})$ values of $\pm 0.7$, $\pm 0.5$, $\pm 0.3$, and $\pm 0.1$, respectively.
Based on the N3LO model, a series of characteristic quantities related to nuclear matter bulk properties have been obtained by fitting various nuclear theoretical/experimental constraints and the astrophysical observations \cite{Wang:2023zcj}.
Here for demonstration, we choose one parameter set, i.e., ``SP6L45'' in Table~\uppercase\expandafter{\romannumeral 2} in Ref.~\cite{Wang:2023zcj}, to construct the new interactions.
The values of $I_{0}$ and $I_{\mathrm{sym}}$ in N4LO and N5LO model as well as $H_{0}$ and $H_{\mathrm{sym}}$ in N5LO model are taken to be the corresponding calculated values from the N3LO model.
Thus, the properties of SNM and the symmetry energy among different models are constructed as consistent as possible, allowing for a clearer observation of the impacts caused by different models (different momentum behaviors of single-nucleon potential).
Based on these new constructed extended Skyrme interactions, we investigate the properties of nuclear matter and neutron stars, and we find that these interactions can give reasonable descriptions of neutron-star properties.
Furthermore, the interactions with $\Delta m_{1}^{\ast}(\rho_{0})=0.3$ are applied in the lattice BUU model \cite{Wang:2019ghr,Wang:2020ixf} to simulate the fixed-target Au+Au collision at $E_\mathrm{beam}=1.23$~GeV/nucleon~\cite{HADES:2020lob,HADES:2022osk} (corresponding to $\sqrt{s_{N N}}=2.4 \,\mathrm{GeV}$) conducted by HADES collaboration.
Additionally, to investigate the effect of the isoscalar nucleon effective mass $m_{s,0}^{\ast}$ on the HICs, we also introduce another single-nucleon potential constructed based on the N3LO model from Ref.~\cite{Wang:2023zcj}, for which the value of $m^{\ast}_{s,0}$ equals to $0.83m$.
We show that these four interactions can provide good predictions of the proton collective flows in the HADES experiments.
Moreover, we find that the proton elliptic flows $v_2$ is sensitive to the single-nucleon potential as well as the $m^{\ast}_{s,0}$.
Specifically, a larger $m^{\ast}_{s,0}$ (indicating weaker energy dependence at low energies) predicts a smaller magnitude of $v_2$, while the rapidly rising single-nucleon potential, compared to the saturated ones, results in a larger magnitude of $v_2$.

This paper is organized as follow:
In Sec.~\ref{sec:ImSkyForm}, we introduce the Skyrme pseudopotential up to N5LO by extending the central terms and the DD terms, and display the expressions of Hamiltonian density and single-nucleon potential.
Ignoring the high-order terms, these expressions in the N5LO model will reduce to the corresponding forms in the N4LO (N3LO) model.
In Sec.~\ref{sec:fitting}, we present the fitting strategy as well as the theoretical/experimental data and constraints and the parametrization used in our fitting, and we construct 24 new parameter sets of the extended Skyrme interactions.
In Sec.~\ref{sec:properties}, we present the bulk properties of cold nuclear matter, the single-nucleon potential behaviors and neutron-star properties of the 24 interactions.
The cumulative contributions from the MD term up to different orders for the nuclear matter bulk properties and for the single-nucleon potential are also discussed in this section.
In Sec.~\ref{sec:LBUU}, we briefly introduce the BUU equation and the lattice Hamiltonian method, by which these new interactions are used to simulate the Au+Au collisions at $\sqrt{s_{N N}}=2.4 \,\mathrm{GeV}$.
In this section, we also present the simulation result for proton collective flows and compare them with the HADES data.
Finally, we summarize our conclusions and make a brief outlook in Sec.~\ref{sec:summary}.

For completeness, we include several Appendices.
In Appendix~\ref{sec:App_NnLO}, we introduce our basic assumption to extend the central term of the Skyrme interaction and give the corresponding forms up to N$n$LO.
With HF approximation, the expressions of Hamiltonian density and single-nucleon potential are derived under general nonequilibrium conditions.
We also discuss the advantage of the polynomial form of the MD part of single particle potentials in Appendix~\ref{sec:App_NnLO}.
The expressions of the characteristic quantities of the SNM EOS and the symmetry energy are provided in Appendix~\ref{sec:App_quantities}, where we also present the expressions of the fourth-order symmetry energy, the isoscalar and isovector nucleon effective masses as well as the linear isospin splitting coefficient.
In Appendix~\ref{sec:Skyrme_paras}, we list the Skyrme parameters for these new interactions.

\section{Theoretical framework}
\label{sec:ImSkyForm}
\subsection{New extended Skyrme interaction basd on N5LO pseudopotential}
In the previous works \cite{Carlsson:2008gm,Raimondi:2011pz}, the generalization of Skyrme interaction has been constructed up to the N3LO by including the additional fourth and sixth-order derivative terms, which is also referred to as N3LO Skyrme pseudopotential.
This theoretical framework has been applied in studies of the EOS of nuclear matter \cite{Davesne:2013aja,Davesne:2014wya,Davesne:2014rva,Davesne:2015dba,Davesne:2015lca} and the properties of finite nuclei \cite{Carlsson:2009mq,Becker:2017kkt}.
The Hamiltonian density and the single-nucleon potential has been derived within the HF approximation \cite{Wang:2018yce}, and it can be seen that compared to the conventional Skyrme interaction, the higher-order derivative terms provide additional momentum dependence for the single-nucleon potential, allowing it to better reproduce the empirical nucleon optical potential \cite{Hama:1990vr,Cooper:1993nx}.
Consequently, the N3LO Skyrme pseudopotential can be employed in transport models, such as the lattice BUU method, to investigate both the collective dynamics of finite nuclei \cite{Wang:2020xgk,Song:2023fnc} and the HICs simulations at intermediate energies \cite{Wang:2023gta}.
In a very recent work \cite{Wang:2023zcj}, the DD term in the N3LO Skyrme pseudopotential is extended in the spirit of Fermi momentum expansion, resulting in a highly flexible density behavior of SNM EOS and the symmetry energy.
This allows for simultaneous descriptions of both the various microscopic calculations of the EOS of PNM (see Ref.~\cite{Zhang:2022bni} and references therein) and multimessenger astrophysical observations of neutron stars \cite{LIGOScientific:2018cki,NANOGrav:2019jur,Fonseca:2021wxt,Miller:2019cac,Riley:2019yda,Miller:2021qha,Riley:2021pdl}.
Finally, we can describe the properties of finite nuclei, neutron stars and HICs in a unified theoretical framework.

However, because the single-nucleon potential rapidly deviates from its saturated behavior when energy exceeds $1$ GeV, these interactions may not be suitable for studying the HICs of incident energies beyond $1$ GeV/nucleon, such as the fixed Au$+$Au collision at $1.23$~GeV/nucleon by HADES collaboration~\cite{HADES:2020lob,HADES:2022osk} and the central collisions at the energies of $1$-$2$ GeV/nucleon measured by FOPI collaboration \cite{FOPI:2010xrt}.
To solve this problem, we propose extending the central terms to provide additional momentum dependence for the single-nucleon potential.
We assume that the 2$n$th-order derivative terms in the central term come from $\left(\hat{A} + \hat{B}  \right)^{n}$, where: $\hat{A}=\frac{1}{2}\left[\hat{\vec{k}}^{\prime}{}^2 \delta \left( \vec{r}_1 -\vec{r}_2  \right) + \delta \left( \vec{r}_1 -\vec{r}_2  \right) \hat{\vec{k}}{}^2\right]$, $\hat{B}=\hat{\vec{k}}^{\prime} \cdot \delta \left( \vec{r}_1 -\vec{r}_2  \right) \hat{\vec{k}}$, and $ \hat{\vec{k}}=-i\left( \hat{\vec{\nabla}}_1-\hat{\vec{\nabla}}_2 \right)/2 $
is the relative momentum operator, while $ \hat{\vec{k}}^{\prime} $ is the conjugate operator of $ \hat{\vec{k}}$.
In Appendix~\ref{sec:App_NnLO}, we present the general form of the extended central term up to arbitrary order, along with the derivations of the corresponding Hamiltonian density and single-nucleon potential.
Compared to the N3LO model, the central term in N5LO model includes additional eighth- and tenth-order derivative terms, which ensure that single-nucleon potential remains saturated up to $2$ GeV.
To clearly illustrate the impact of the high-order derivative terms in an order-by-order way, we provide the forms of the central terms for the N3LO, N4LO, and N5LO models, respectively, and subsequently construct the corresponding interactions based on these.

\begin{widetext}
The central term of the N3LO Skyrme pseudopotential is written as
\begin{equation}
\label{eq:VN3LO}
\small
\begin{aligned}
V_{\mathrm{N} 3 \mathrm{LO}}^C  =
& t_0\left(1+x_0 \hat{P}_\sigma\right)+t_1^{[2]}\left(1+x_1^{[2]} \hat{P}_\sigma\right) \frac{1}{2}\left(\hat{\vec{k}}^{\prime 2}+\hat{\vec{k}}^2\right)+t_2^{[2]}\left(1+x_2^{[2]} \hat{P}_\sigma\right) \hat{\vec{k}}^{\prime} \cdot \hat{\vec{k}}+t_1^{[4]}\left(1+x_1^{[4]} \hat{P}_\sigma\right)\left[\frac{1}{4}\left(\hat{\vec{k}}^{\prime 2}+\hat{\vec{k}}^2\right)^2+\left(\hat{\vec{k}}^{\prime} \cdot \hat{\vec{k}}\right)^2\right] \\
& +t_2^{[4]}\left(1+x_2^{[4]} \hat{P}_\sigma\right)\left(\hat{\vec{k}}^{\prime} \cdot \hat{\vec{k}} \right)\left(\hat{\vec{k}}^{\prime 2}+\hat{\vec{k}}^2\right)+t_1^{[6]}\left(1+x_1^{[6]} \hat{P}_\sigma\right)\left(\hat{\vec{k}}^{\prime 2}+\hat{\vec{k}}^2\right)\left[\frac{1}{2}\left(\hat{\vec{k}}^{\prime 2}+\hat{\vec{k}}^2\right)^2+6\left(\hat{\vec{k}}^{\prime} \cdot \hat{\vec{k}}\right)^2\right] \\
& +t_2^{[6]}\left(1+x_2^{[6]} \hat{P}_\sigma\right)\left(\hat{\vec{k}}^{\prime} \cdot \hat{\vec{k}}\right)\left[3\left(\hat{\vec{k}}^{\prime 2}+\hat{\vec{k}}^2\right)^2+4\left(\hat{\vec{k}}^{\prime} \cdot \hat{\vec{k}}\right)^2\right] ,
\end{aligned}
\end{equation}
and the extensions of $V_{\mathrm{N} 3 \mathrm{LO}}^C$ include the eighth-order term $v^{[8]}$ and the tenth-order term $v^{[10]}$ based on Eq.~(\ref{eq:def_Vn}) and Eq.~(\ref{eq:VP_2n}), i.e.,
\begin{equation}
\label{eq:v8_new}
\small
\begin{aligned}
v^{[8]} = & \, t_{1}^{[8]}\left( 1 + x_{1}^{[8]} \Psig \right) \left[ \Ahatp^4 + 6 \Ahatp^2 \Bhatp^2 + \Bhatp^4 \right]  \\
& + t_{2}^{[8]}\left( 1 + x_{2}^{[8]} \Psig \right) \left[ 4 \Ahatp^3 \Bhatp  + 4 \Ahatp \Bhatp^3 \right] ,
\end{aligned}
\end{equation}
and
\begin{equation}
\label{eq:v10_new}
\small
\begin{aligned}
v^{[10]} = & \, t_{1}^{[10]}\left( 1 + x_{1}^{[10]} \Psig \right) \left[ \Ahatp^5 + 10 \Ahatp^3 \Bhatp^2 + 5 \Ahatp \Bhatp^4      \right] \\
& + t_{2}^{[10]}\left( 1 + x_{2}^{[10]} \Psig \right) \left[ 5\Ahatp^4 \Bhatp + 10 \Ahatp^2 \Bhatp^3 + \Bhatp^5  \right],
\end{aligned}
\end{equation}
where $\hat{P}_\sigma$ is the spin-exchange operator.
\end{widetext}
Thus the central term of N4LO and N5LO Skyrme pseudopotential can be expressed as
\begin{align}
V_{\mathrm{N4LO}}^{C}=&V_{\mathrm{N3LO}}^{C}+v^{[8]}, \\ V_{\mathrm{N5LO}}^{C}=&V_{\mathrm{N3LO}}^{C}+v^{[8]}+v^{[10]}.
\end{align}
Based on Fermi momentum expansion proposed in Ref.~\cite{Wang:2023zcj}, the DD term can be written as
\begin{equation}
\label{eq:V_DD}
V^{\mathrm{DD}}_{N} = \sum_{n=1}^{N} \frac{1}{6} t_{3}^{[2n-1]} \left( 1+ x_{3}^{[2n-1]} \Psig \right) \rho^{\frac{2n-1}{3}} ( \vec{R} ),
\end{equation}
where $\vec{R}=\left( \vec{r}_1 + \vec{r}_2 \right)/2$.
For brevity, the factor $\hat{\delta}\left(\vec{r_1}-\vec{r_1} \right)$ is omitted from Eqs.~(\ref{eq:VN3LO}), (\ref{eq:v8_new}), (\ref{eq:v10_new}) and (\ref{eq:V_DD}).
The general form of the Skyrme pseudopotential includes the spin-independent terms, the spin-orbit terms, as well as the tensor components (see, e.g., Refs.~\cite{Carlsson:2008gm,Raimondi:2011pz,Davesne:2014wya,Davesne:2014rva}).
In the present work, we focus on the spin-averaged quantities, where the spin-orbit and tensor terms make no contribution.
By extending the central term and DD term systematically, the Skyrme interactions used in this work are then written as
\begin{align}
\label{eq:vN3}
v^{\mathrm{N3LO}}_{sk}=& V_{\mathrm{N3LO}}^C + V_{3}^{\mathrm{DD}}, \\
\label{eq:vN4}
v^{\mathrm{N4LO}}_{sk}=& V_{\mathrm{N4LO}}^C + V_{4}^{\mathrm{DD}}, \\
\label{eq:vN5}
v^{\mathrm{N5LO}}_{sk}=& V_{\mathrm{N5LO}}^C + V_{5}^{\mathrm{DD}}.
\end{align}
The $t_0$, $x_0$; $t_{i}^{[n]}$, $x_{i}^{[n]}$ ($n=2,4,6,8,10$ and $i=1,2$); $t_{3}^{[n]}$, $x_{3}^{[n]}$ ($n=1,3,5,7,9$) are Skyrme parameters, and the total numbers of these parameters are 20, 26 and 32 for N3LO, N4LO and N5LO interactions, respectively.

\subsection{Hamiltonian density and single-nucleon potential in one-body transport model}
In the BUU microscopic transport model, since the system is in a nonequilibrium state, the Hamiltonian density $\mathcal{H}(\vec{r})$ and the single-nucleon potential $U_{\tau}(\vec{r},\vec{p})$ are expressed as functions of the phase space distribution function of nucleons (Wigner function) $f_{\tau}(\vec{r},\vec{p})$, with $\tau=1$ (or n) for neutrons and $-1$ (or p) for protons.
With the HF approximation, the Hamiltonian density with the Skyrme interactions defined in Eqs.~(\ref{eq:vN3})-(\ref{eq:vN5}) can be expressed as (detailed derivation can be found in Appdendix~\ref{sec:App_NnLO})
\begin{equation}
\label{eq:Hdensity}
\mathcal{H} \left( \vec{r} \right) =
\mathcal{H} ^{ \mathrm{kin}  } \left( \vec{r} \right)
+ \mathcal{H}^{ \mathrm{loc}  } \left( \vec{r} \right)
+ \mathcal{H}^{ \mathrm{MD}  } \left( \vec{r} \right)
+ \mathcal{H}^{ \mathrm{grad}  } \left( \vec{r} \right)
+ \mathcal{H}^{ \mathrm{DD}  } \left( \vec{r} \right),
\end{equation}
where $\mathcal{H} ^{ \mathrm{kin}  } \left( \vec{r} \right)$, $\mathcal{H}^{ \mathrm{loc}  } \left( \vec{r} \right)$, $\mathcal{H}^{ \mathrm{MD}  } \left( \vec{r} \right)$, $\mathcal{H}^{ \mathrm{grad}  } \left( \vec{r} \right)$ and $\mathcal{H}^{ \mathrm{DD}  } \left( \vec{r} \right)$ are the kinetic, local, MD, gradient (GD) and DD terms, respectively.
The kinetic term and the local term are the same for N3LO, N4LO and N5LO Skyrme pseudopotential, where they are expressed as
\begin{equation}
\label{eq:H_kin}
\mathcal{H} ^{ \mathrm{kin}  } \left( \vec{r} \right)
=\sum_{\tau=n,p} \int \frac{{\rm d}^3 p}{(2\pi\hbar)^3}\frac{p^2}{2 m_{\tau}} f_\tau \left( \vec{r}, \vec{p} \right)
\end{equation}
and
\begin{equation}
\label{eq:H_loc}
\mathcal{H}^{ \mathrm{loc}  } \left( \vec{r} \right)
= \frac{1}{4} t_0 \left[ \left( 2+x_0 \right) \rho^2 - \left( 2 x_0+1 \right) \sum_{\tau=n, p} \rho_\tau^2\right],
\end{equation}
respectively.
The $\rho_{\tau} \left( \vec{r} \right) = \int f_\tau \left( \vec{r}, \vec{p} \right) \frac{{\rm d}^3 p}{(2\pi\hbar)^3} $ is the nucleon density and the $\rho \left( \vec{r} \right) = \rho_n \left( \vec{r} \right) + \rho_p \left( \vec{r} \right) $ is the total nucleon density.

\begin{widetext}
For N5LO interaction, the MD term is expressed as
\begin{equation}
\label{eq:H_MD}
\begin{aligned}
\mathcal{H}^{\mathrm{MD}}(\vec{r})=
& \frac{C^{[2]}}{16 \hbar^2} \mathcal{H}^{\mathrm{md}[2]}(\vec{r})+\frac{D^{[2]}}{16 \hbar^2} \sum_{\tau=n, p} \mathcal{H}_\tau^{\mathrm{md}[2]}(\vec{r})
+\frac{C^{[4]}}{32 \hbar^4} \mathcal{H}^{\mathrm{md}[4]}(\vec{r})+\frac{D^{[4]}}{32 \hbar^4} \sum_{\tau=n, p} \mathcal{H}_\tau^{\mathrm{md}[4]}(\vec{r}) \\
& +\frac{C^{[6]}}{16 \hbar^6} \mathcal{H}^{\mathrm{md}[6]}(\vec{r})+\frac{D^{[6]}}{16 \hbar^6} \sum_{\tau=n, p} \mathcal{H}_\tau^{\mathrm{md}[6]}(\vec{r})
+ \frac{C^{[8]}}{128 \hbar^{8}} \mathcal{H}^{\mathrm{md}[8]} (\vec{r}) + \frac{D^{[8]}}{128 \hbar^{8}} \sum_{\tau=n,p} \mathcal{H}_\tau^{\mathrm{md}[8]}(\vec{r}) \\
& + \frac{C^{[10]}}{256 \hbar^{10}} \mathcal{H}^{\mathrm{md}[10]} (\vec{r}) + \frac{D^{[10]}}{256 \hbar^{10}} \sum_{\tau=n,p} \mathcal{H}_\tau^{\mathrm{md}[10]}(\vec{r}),
\end{aligned}
\end{equation}
where $\mathcal{H}^{\mathrm{md}\left[ n \right]}(\vec{r})$ and $\mathcal{H}_{\tau}^{\mathrm{md}\left[ n \right]}(\vec{r})$ are defined as
\begin{align}
\mathcal{H}^{\mathrm{md}[n]}(\vec{r})&= \int \frac{{\rm d}p^3}{(2\pi\hbar)^3}\frac{{\rm d}^3 p'}{(2\pi\hbar)^3}\left(\vec{p}-\vec{p}^{\, \prime}\right)^n f(\vec{r}, \vec{p}) f\left(\vec{r}, \vec{p}^{\, \prime}\right), \\
\mathcal{H}_{\tau}^{\mathrm{md}[n]}(\vec{r})&= \int \frac{{\rm d}p^3}{(2\pi\hbar)^3}\frac{{\rm d}^3 p'}{(2\pi\hbar)^3}\left(\vec{p}-\vec{p}^{\, \prime}\right)^n f_{\tau}(\vec{r}, \vec{p}) f_{\tau}\left(\vec{r}, \vec{p}^{\, \prime}\right),
\end{align}
with $f(\vec{r}, \vec{p})=f_{n}(\vec{r}, \vec{p})+f_{p}(\vec{r}, \vec{p})$.
The gradient term can be expressed as
\begin{equation}
\label{eq:Hgrad}
\small
\begin{aligned}
\mathcal{H}^{\mathrm{GD}}(\vec{r})=
& \frac{1}{16} E^{[2]}\left\{2 \rho(\vec{r}) \nabla^2 \rho(\vec{r})-2[\nabla \rho(\vec{r})]^2\right\}+\frac{1}{16} F^{[2]} \sum_{\tau=n, p}\left\{2 \rho_\tau(\vec{r}) \nabla^2 \rho_\tau(\vec{r})-2\left[\nabla \rho_\tau(\vec{r})\right]^2\right\} \\
& +\frac{1}{32} E^{[4]}\left\{2 \rho(\vec{r}) \nabla^4 \rho(\vec{r})-8 \nabla \rho(\vec{r}) \nabla^3 \rho(\vec{r})+6\left[\nabla^2 \rho(\vec{r})\right]^2\right\} \\
& +\frac{1}{32} F^{[4]} \sum_{\tau=n, p}\left\{2 \rho_\tau(\vec{r}) \nabla^4 \rho_\tau(\vec{r})-8 \nabla \rho_\tau(\vec{r}) \nabla^3 \rho_\tau(\vec{r})+6\left[\nabla^2 \rho_\tau(\vec{r})\right]^2\right\} \\
& +\frac{1}{16} E^{[6]}\left\{2 \rho(\vec{r}) \nabla^6 \rho(\vec{r})-12 \nabla \rho(\vec{r}) \nabla^5 \rho(\vec{r})+30 \nabla^2 \rho(\vec{r}) \nabla^4 \rho(\vec{r})-20\left[\nabla^3 \rho(\vec{r})\right]^2\right\} \\
& +\frac{1}{16} F^{[6]} \sum_{\tau=n, p}\left\{2 \rho_\tau(\vec{r}) \nabla^6 \rho_\tau(\vec{r})-12 \nabla \rho_\tau(\vec{r}) \nabla^5 \rho_\tau(\vec{r})+30 \nabla^2 \rho_\tau(\vec{r}) \nabla^4 \rho_\tau(\vec{r})-20\left[\nabla^3 \rho_\tau(\vec{r})\right]^2\right\} \\
& + \frac{E^{[8]}}{128} \left\{ 2 \rhor \nabla^{8} \rhor - 16 \nabla \rhor \nabla^{7} \rhor + 56 \nabla^2 \rhor \nabla^{6} \rhor - 112 \nabla^{3} \rhor \nabla^{5} \rhor + 70 \left[ \nabla^{4} \rhor \right]^2 \right\} \\
& + \frac{F^{[8]}}{128} \sum_{\tau=n, p} \left\{2 \rhotr \nabla^{8} \rhotr - 16 \nabla \rhotr \nabla^{7} \rhotr + 56 \nabla^2 \rhotr \nabla^{6} \rhotr - 112 \nabla^{3} \rhotr \nabla^{5} \rhotr + 70 \left[ \nabla^{4} \rhotr \right]^2
\right\} \\
& + \frac{E^{[10]}}{256} \Big\{ 2 \rhor \nabla^{10} \rhor - 20 \nabla \rhor \nabla^{9} \rhor + 90 \nabla^{2} \rhor \nabla^{8} \rhor - 240 \nabla^{3} \rhor \nabla^{7} \rhor  \\
& \qquad \qquad + 420 \nabla^{4} \rhor \nabla^{6} \rhor - 252 \left[ \nabla^{5} \rhor \right]^2 \Big\} \\
& + \frac{F^{[10]}}{256} \sum_{\tau=n, p} \Big\{ 2 \rhotr \nabla^{10} \rhotr - 20 \nabla \rhotr \nabla^{9} \rhotr + 90 \nabla^{2} \rhotr \nabla^{8} \rhotr - 240 \nabla^{3} \rhotr \nabla^{7} \rhotr \\
& \qquad \qquad \qquad + 420 \nabla^{4} \rhotr \nabla^{6} \rhotr - 252 \left[ \nabla^{5} \rhotr \right]^2 \Big\}.
\end{aligned}
\end{equation}
In Eqs.~(\ref{eq:H_MD}) and (\ref{eq:Hgrad}), we have recombined the Skyrme parameters $\Skypat{1}{2n}$, $\Skypax{1}{2n}$, $\Skypat{2}{2n}$ and $\Skypax{2}{2n}$ into the parameters $C^{[2n]}$, $D^{[2n]}$, $E^{[2n]}$ and $F^{[2n]}$, which are defined in Eqs.~(\ref{eq:C2n})-(\ref{eq:F2n}).
The DD term comes from $V_{5}^{\mathrm{DD}}$ in Eq.~(\ref{eq:V_DD}) of the N5LO interaction is expressed as
\begin{equation}
\label{eq:H_DD}
\mathcal{H}^{\mathrm{DD}}(\vec{r}) = \sum_{n=1}^{5} \frac{1}{24} t_3^{[2n-1]} \left[ \left(2+x_3^{[2n-1]}\right) \rho^2 -\left(2 x_3^{[2n-1]}+1\right) ( \rho_n^2 + \rho_p^2 ) \right] \rho^{\frac{2n-1}{3}}.
\end{equation}
By ignoring the $C^{[10]}$, $D^{[10]}$, $E^{[10]}$, $F^{[10]}$ and $t_3^{[9]}$ terms (or as well as the $C^{[8]}$, $D^{[8]}$, $E^{[8]}$, $F^{[8]}$ and $t_3^{[7]}$ terms), Eq.~(\ref{eq:H_MD}), Eq.~(\ref{eq:Hgrad}) and Eq.~(\ref{eq:H_DD}) will reduce to the expressions corresponding to N4LO (N3LO) interactions (same in the following expressions).

Based on the Landau Fermi liquid theory, the single-nucleon potential reflects the net nuclear medium effects, and it can be obtained by taking the variation of $\mathcal{H}\left( \vec{r} \right)$ with respect to $f_{\tau} \left( \vec{r},\vec{p} \right)$.
Since $\mathcal{H}\left( \vec{r} \right)$ includes terms with density gradients, the single-nucleon potential can be calculated as (details can be found in Ref.~\cite{Kolomietz:2017qkb})
\begin{equation}
\label{eq:U=dHdf}
U_{\tau} \left( \vec{r}, \vec{p} \right)
= \frac{\delta \mathrm{H}^{\mathrm{pot}}}{ \delta f_\tau\left( \vec{r}, \vec{p} \right)}
= \frac{ \partial \left[ \mathcal{H}^{\mathrm{loc}} \left( \vec{r} \right) + \mathcal{H}^{\mathrm{DD}} \left( \vec{r} \right) + \mathcal{H}^{\mathrm{grad}} \left( \vec{r} \right) \right] }
{\partial \rho_\tau(\vec{r})}
+ \sum_{n} (-1)^n \nabla^n \frac{\partial \mathcal{H}^{\mathrm{grad}} \left( \vec{r} \right) }{ \partial \left[ \nabla^n \rho_\tau(\vec{r})\right]}
+ \frac{\delta \mathrm{H}^{\mathrm{MD}}}{ \delta f_\tau\left( \vec{r}, \vec{p} \right)},
\end{equation}
where $\mathrm{H}^{\mathrm{pot}}= \int d \vec{r} \left[ \mathcal{H}^{\mathrm{loc}}(\vec{r})+\mathcal{H}^{\mathrm{DD}}(\vec{r})+\mathcal{H}^{\mathrm{MD}}(\vec{r})+\mathcal{H}^{\mathrm{grad}}(\vec{r}) \right]$ is the potential part of the Hamiltonian with $\mathrm{H}^{\mathrm{MD}}= \int d \vec{r} \mathcal{H}^{\mathrm{MD}}(\vec{r}) $ being the MD part.
Substitute Eq.~(\ref{eq:Hdensity}) into Eq.~(\ref{eq:U=dHdf}), and this yields
\begin{equation}
\label{eq:U_general}
\small
\begin{aligned}
U_\tau(\vec{r}, \vec{p})= &
\frac{1}{2} t_0\left[\left(2+x_0\right) \rho\left(\vec{r}\right)-\left(2 x_0+1\right) \rho_\tau \left(\vec{r}\right) \right]
+ \sum_{n=1}^{5} \left\{ \frac{1}{12} t_{3} ^{\left[2n-1 \right]}\left[\left(2+x_{3} ^{\left[2n-1 \right]}\right) \rho(\vec{r})-\left(2 x_{3} ^{\left[2n-1 \right]}+1\right) \rho_\tau(\vec{r})\right] \rho(\vec{r})^{\frac{2n-1}{3}}  \right\} \\
& +\sum_{n=1}^{5} \left\{ \frac{t_{3} ^{\left[2n-1 \right]}}{24} \frac{2n-1}{3} \left[\left(2+x_{3} ^{\left[2n-1 \right]}\right) \rho(\vec{r})^2-\left(2 x_{3} ^{\left[2n-1 \right]}+1\right) \sum_{\tau=n, p} \rho_\tau(\vec{r})^2\right] \rho(\vec{r})^{ \frac{2n-1}{3}-1 }  \right\}
+\frac{1}{8 \hbar^2} C^{[2]} U^{\mathrm{md}[2]}(\vec{r}, \vec{p})   \\
& +\frac{1}{8 \hbar^2} D^{[2]} U_\tau^{\mathrm{md}[2]}(\vec{r}, \vec{p}) +\frac{1}{16 \hbar^4} C^{[4]} U^{\mathrm{md}[4]}(\vec{r}, \vec{p})+\frac{1}{16 \hbar^4} D^{[4]} U_\tau^{\mathrm{md}[4]}(\vec{r}, \vec{p})+\frac{1}{8 \hbar^6} C^{[6]} U^{\mathrm{md}[6]}(\vec{r}, \vec{p}) \\
&+\frac{1}{8 \hbar^6} D^{[6]} U_\tau^{\mathrm{md}[6]}(\vec{r}, \vec{p}) + \frac{C^{[8]}}{64 \hbar^{8}} U^{\mathrm{md}[8]} (\vec{r},\vec{p}) + \frac{D^{[8]}}{64 \hbar^{8}} U_\tau^{\mathrm{md}[8]}(\vec{r},\vec{p}) +
\frac{C^{[10]}}{128 \hbar^{10}} U^{\mathrm{md}[10]} (\vec{r},\vec{p}) + \frac{D^{[10]}}{128 \hbar^{10}} U_\tau^{\mathrm{md}[10]}(\vec{r},\vec{p}) \\
& +\frac{1}{2} E^{[2]} \nabla^2 \rho(\vec{r})+\frac{1}{2} F^{[2]} \nabla^2 \rho_\tau(\vec{r})+  E^{[4]} \nabla^4 \rho(\vec{r})+ F^{[4]} \nabla^4 \rho_\tau(\vec{r})+ 8 E^{[6]} \nabla^6 \rho(\vec{r})+ 8 F^{[6]} \nabla^6 \rho_\tau(\vec{r}) \\
& +4 E^{[8]} \nabla^{8} \rhor + 4 F^{[8]} \nabla^{8} \rhotr
+ 8 E^{[10]} \nabla^{10} \rhor + 8 F^{[10]} \nabla^{10} \rhotr,
\end{aligned}
\end{equation}
where $U^{\mathrm{md}[n]}(\vec{r}, \vec{p})$ and $U_{\tau}^{\mathrm{md}[n]}(\vec{r}, \vec{p})$ are defined as
\begin{align}
U^{\mathrm{md}[n]} (\vec{r},\vec{p}) =& \int \frac{{\rm d}^3 p'}{(2\pi\hbar)^3}\left(\vec{p}-\vec{p}\,{}^{\prime}\right)^{n} f\left(\vec{r}, \vec{p}\,{}^{\prime}\right),\label{E:UMD1}\\
U^{\mathrm{md}[n]}_{\tau} (\vec{r},\vec{p}) =& \int \frac{{\rm d}^3 p'}{(2\pi\hbar)^3} \left(\vec{p}-\vec{p}\,{}^{\prime}\right)^{n} f_{\tau} \left(\vec{r}, \vec{p}\,{}^{\prime}\right).
\label{E:UMDt1}
\end{align}

\subsection{Equation of state of cold nuclear matter}
The EOS of isospin asymmetric nuclear matter with total nucleon density $\rho=\rho_{n} + \rho_{p}$ and isospin asymmetry $\delta = (\rho_{n}-\rho_{p})/\rho$ are defined as its binding energy per nucleon.
In uniform infinite system, all the gradient terms ($E^{[n]}$ and $F^{[n]}$ terms) in the Hamiltonian density [Eq.~(\ref{eq:Hdensity})] vanish.
At zero temperature, $f_{\tau} ( \vec{r},\vec{p} )$ becomes a step function, i.e., $f_{\tau} ( \vec{r},\vec{p} )=g_{\tau} \theta(p_{F_{\tau}}-|\vec{p}|)$, with $g_{\tau}$ being the degeneracy and $p_{F_{\tau}}=\hbar(3 \pi^2 \rho_\tau)^{1/3}$ being the Fermi momentum of nucleons with isospin $\tau$.
In this case, the EOS of isospin asymmetric nuclear can be analytically expressed as
\begin{equation}
\label{eq:EOS_N5LO}
\small
\begin{aligned}
E(\rho,\delta) =& \, \frac{3}{5} \frac{\hbar^2 a^2}{2m} F_{5/3} \rho^{2/3} + \frac{1}{8} t_{0} \bigg[ 2\left(x_{0} +2 \right) - \left( 2 x_{0} +1 \right) F_{2} \bigg] \rho^{3/3} + \frac{1}{48} t_{3}^{[1]} \bigg[ 2\left(x_{3}^{[1]} +2 \right) - \left( 2 x_{3}^{[1]} +1 \right) F_{2} \bigg] \rho^{4/3} \\
&+ \frac{9 a^2}{64}\left[\frac{8}{15} C^{[2]} F_{5 / 3}+\frac{4}{15} D^{[2]} F_{8 / 3}\right] \rho^{5 / 3}
+\frac{1}{48} t_3^{[3]}\left[2\left(x_3^{[3]}+2\right)-\left(2 x_3^{[3]}+1\right) F_2\right] \rho^{6/3} \\
&+ \frac{9 a^4}{128} \left[
C^{[4]}\left(\frac{68}{105} F_{7/3}+\frac{4}{15} \delta G_{7 / 3}+\frac{4}{15} H_{5 / 3}\right) + \frac{16}{35} D^{[4]} F_{10 / 3} \right] \rho^{7/3}
+\frac{1}{48} t_3^{[5]}\left[2\left(x_3^{[5]}+2\right)-\left(2 x_3^{[5]}+1\right) F_2\right] \rho^{8/3} \\
&+ \frac{9 a^6}{64}\left[
C^{[6]}\left(\frac{148}{135} F_3+\frac{4}{5} \delta G_3+\frac{4}{5} H_{5 / 3} F_{2 / 3}\right) +\frac{128}{135} D^{[6]} F_4 \right] \rho^{9/3}
+\frac{1}{48} t_3^{[7]}\left[2\left(x_3^{[7]}+2\right)-\left(2 x_3^{[7]}+1\right) F_2\right] \rho^{10/3} \\
&+ \frac{9 a^8}{512}\left[
C^{[8]} \left( \frac{180}{77} F_{11/3} + \frac{44}{21} \delta G_{11/3} + \frac{16}{15}H_{5/3}F_{4/3} + \frac{36}{35}H_{7/3} \right) + \frac{512}{231} D^{[8]} F_{14/3}
\right] \rho^{11/3} \\
&+ \frac{1}{48} t_3^{[9]}\left[2\left(x_3^{[9]}+2\right)-\left(2 x_3^{[9]}+1\right) F_2\right] \rho^{12/3} \\
&+ \frac{9 a^{10}}{1024}\left[
C^{[10]} \left(\frac{1564}{273} F_{13/3} + \frac{116}{21} \delta G_{13/3} + \frac{4}{3} H_{5/3} F_{2} + \frac{88}{21} H_{7/3} F_{2/3} \right) + \frac{512}{91} D^{[10]} F_{16/3}
\right] \rho^{13/3},
\end{aligned}
\end{equation}
\end{widetext}
where $a=(3\pi^2/2)^{1/3}$, and $m$ is nucleon rest mass in vacuum.
In Eq.~(\ref{eq:EOS_N5LO}), $F_x$, $G_x$ and $H_x$ are defined as
\begin{align*}
F_x & =\left[(1+\delta)^x+(1-\delta)^x\right] /2 ,  \\
G_x & =\left[(1+\delta)^x-(1-\delta)^x\right] / 2 ,\\
H_x & =\left[(1+\delta)(1-\delta)\right]^x.
\end{align*}

The EOS can be expanded as a power series in $\delta$, i.e.,
\begin{equation}
E\left(\rho, \delta\right) = E_{0}\left(\rho\right) + E_{\mathrm{sym}}\left(\rho\right)\delta^2 + E_{\mathrm{sym},4}\left(\rho\right)\delta^4 + \mathcal{O}\left(\delta^6\right),
\end{equation}
where $E_0(\rho)$ is the EOS of the SNM.
The symmetry energy $E_{\mathrm{sym}}(\rho)$ and the fourth-order symmetry energy $E_{\mathrm{sym},4}(\rho)$ are defined as
\begin{equation}
\label{eq:Esym_def}
E_{\mathrm{sym}}(\rho)  =\left.\frac{1}{2 !} \frac{\partial^2 E(\rho, \delta)}{\partial \delta^2}\right|_{\delta=0},
\end{equation}
and
\begin{equation}
\label{eq:Esym4_def}
E_{\mathrm{sym}, 4}(\rho)  =\left.\frac{1}{4 !} \frac{\partial^4 E(\rho, \delta)}{\partial \delta^4}\right|_{\delta=0},
\end{equation}
respectively.
The expressions of $E_{0}(\rho)$, $E_{\mathrm{sym}}(\rho)$ and $E_{\mathrm{sym}, 4}(\rho)$ are presented in Appendix~\ref{sec:App_quantities}.

The pressure of the isospin asymmetric nuclear matter can be expressed as
\begin{equation}
\label{eq:press}
P \left(\rho,\delta \right) = \rho^2 \frac{\partial E(\rho, \delta)}{\partial \rho}.
\end{equation}
The saturation density $\rho_{0}$ is defined where the pressure of the SNM is zero (except for $\rho=0$), i.e.,
\begin{equation}
\label{eq:rho_0}
\left. P \left(\rho_0, \delta=0 \right) = \rho_{0}^2 \frac{d E(\rho, 0)}{d \rho} \right| _{\rho=\rho_0}=0 .
\end{equation}
Around the saturation density $\rho_{0}$, both $E_0(\rho)$ and $E_{\mathrm{sym}}(\rho)$ can be expanded as power series in a dimensionless variable $\chi \equiv \frac{\rho-\rho_0}{3 \rho_0}$, i.e.,
\begin{equation}
\begin{aligned}
E_0(\rho)=&\,E_0\left(\rho_0\right)+L_0 \chi+\frac{K_0}{2 !} \chi^2\\
&+\frac{J_0}{3 !} \chi^3 +\frac{I_0}{4 !} \chi^4+ \frac{H_0}{5 !} \chi^5 +\mathcal{O}\left(\chi^6\right),
\end{aligned}
\end{equation}
and
\begin{equation}
\begin{aligned}
E_{\mathrm{sym}}(\rho)= & \, E_{\mathrm{sym}}\left(\rho_0\right)+L \chi+\frac{K_{\mathrm{sym}}}{2 !} \chi^2 \\
&+\frac{J_{\mathrm{sym}}}{3 !} \chi^3
+\frac{I_{\mathrm{sym}}}{4 !} \chi^4 +\frac{H_{\mathrm{sym}}}{5 !} \chi^5
+\mathcal{O}\left(\chi^6\right).
\end{aligned}
\end{equation}
The first five coefficients of $\chi^n$ in the two expansions are
\begin{align}
\label{eq:L_def}
L_0 &  =\left.3 \rho_0 \frac{d E_0(\rho)}{d \rho}\right|_{\rho=\rho_0},
L  =\left.3 \rho_0 \frac{d E_{\mathrm{sym}}(\rho)}{d \rho}\right|_{\rho=\rho_0},\\
\label{eq:K_def}
K_0 &  =\left. (3\rho_0)^2 \frac{d^2 E_0(\rho)}{d \rho^2}\right|_{\rho=\rho_0},
K_{\mathrm{sym}}  =\left.(3\rho_0)^2 \frac{d^2 E_{\mathrm{sym}}(\rho)}{d \rho^2}\right|_{\rho=\rho_0},\\
\label{eq:J_def}
J_0 &  =\left.(3\rho_0)^3 \frac{d^3 E_0(\rho)}{d \rho^3}\right|_{\rho=\rho_0},
J_{\mathrm{sym}}=\left.(3\rho_0)^3 \frac{d^3 E_{\mathrm{sym}}(\rho)}{d \rho^3}\right|_{\rho=\rho_0},\\
\label{eq:I_def}
I_0 & =\left.(3\rho_0)^4 \frac{d^4 E_0(\rho)}{d \rho^4}\right|_{\rho=\rho_0},
I_{\mathrm{sym}}=\left.(3\rho_0)^4 \frac{d^4 E_{\mathrm{sym}}(\rho)}{d \rho^4}\right|_{\rho=\rho_0},\\
\label{eq:H_def}
H_0 & =\left.(3\rho_0)^5 \frac{d^5 E_0(\rho)}{d \rho^5}\right|_{\rho=\rho_0},
H_{\mathrm{sym}}=\left.(3\rho_0)^5 \frac{d^5 E_{\mathrm{sym}}(\rho)}{d \rho^5}\right|_{\rho=\rho_0},
\end{align}
respectively.
Obviously, we have $L_0=0$ by the definition of $\rho_0$ in Eq.~(\ref{eq:rho_0}).
$K_0$ is the incompressibility coefficient of SNM which characterizes the curvature of $E_0(\rho)$ at $\rho_0$.
$J_0$, $I_0$ and $H_0$ represent higher-order contributions and are commonly referred to as the skewness, kurtosis and hyper-skewness coefficients of SNM.
$L$, $K_{\mathrm{sym}}$, $J_{\mathrm{sym}}$, $I_{\mathrm{sym}}$ and $H_{\mathrm{sym}}$ are the slope coefficient, curvature coefficient, skewness coefficient, kurtosis coefficient and hyper-skewness coefficient of the symmetry energy at $\rho_0$.
The expressions of these characteristic parameters are presented in Appendix~\ref{sec:App_quantities}.

\begin{widetext}

\subsection{Single-nucleon potential, symmetry potential, nucleon effective masses and linear isospin splitting coefficient in cold nuclear matter}
In the case of zero-temperature and uniform nuclear matter, the single-nucleon potential [Eq.~(\ref{eq:U_general})] reduces to an analytical function of $\rho$, $\delta$, and the magnitude of nucleon momentum $p=\left|\vec{p}\right|$, i.e.,
\begin{equation}
\label{eq:Utau_N5LO}
\small
\begin{aligned}
U_\tau(\rho, \delta,p)= & \frac{1}{4} t_0\left[2\left(x_0+2\right)-\left(2 x_0+1\right)(1+\tau \delta)\right] \rho \\
&+ \sum_{n=1}^{5} \frac{1}{24} t_3^{[2n-1]}\left[(\frac{2n-1}{3}+2)\left(x_3^{[2n-1]}+2\right)-\left(2 x_3^{[2n-1]}+1\right)\left(\frac{1}{2} \frac{2n-1}{3} F_2+1+\tau \delta\right)\right] \rho^{\frac{2n-1}{3}+1} \\
& +\frac{1}{4} C^{[2]}\left[\frac{1}{3} \frac{k_F^3}{\pi^2}\left(\frac{p}{\hbar}\right)^2+\frac{1}{5} \frac{k_F^5}{\pi^2} F_{5 / 3}\right]+\frac{1}{8} D^{[2]}\left[\frac{1}{3} \frac{k_F^3}{\pi^2}\left(\frac{p}{\hbar}\right)^2(1+\tau \delta)+\frac{1}{5} \frac{k_F^5}{\pi^2}(1+\tau \delta)^{5 / 3}\right] \\
& +\frac{1}{8} C^{[4]}\left[\frac{1}{3} \frac{k_F^3}{\pi^2}\left(\frac{p}{\hbar}\right)^4+\frac{2}{3} \frac{k_F^5}{\pi^2}\left(\frac{p}{\hbar}\right)^2 F_{5 / 3}+\frac{1}{7} \frac{k_F^7}{\pi^2} F_{7 / 3}\right] \\
& +\frac{1}{16} D^{[4]}\left[\frac{1}{3} \frac{k_F^3}{\pi^2}\left(\frac{p}{\hbar}\right)^4(1+\tau \delta)+\frac{2}{3} \frac{k_F^5}{\pi^2}\left(\frac{p}{\hbar}\right)^2(1+\tau \delta)^{5 / 3}+\frac{1}{7} \frac{k_F^7}{\pi^2}(1+\tau \delta)^{7 / 3}\right] \\
& +\frac{1}{4} C^{[6]}\left[\frac{1}{3} \frac{k_F^3}{\pi^2}\left(\frac{p}{\hbar}\right)^6+\frac{7}{5} \frac{k_F^5}{\pi^2}\left(\frac{p}{\hbar}\right)^4 F_{5 / 3}+\frac{k_F^7}{\pi^2}\left(\frac{p}{\hbar}\right)^2 F_{7 / 3}+\frac{1}{9} \frac{k_F^9}{\pi^2} F_3\right] \\
& +\frac{1}{8} D^{[6]}\left[\frac{1}{3} \frac{k_F^3}{\pi^2}\left(\frac{p}{\hbar}\right)^6(1+\tau \delta)+\frac{7}{5} \frac{k_F^5}{\pi^2}\left(\frac{p}{\hbar}\right)^4(1+\tau \delta)^{5 / 3}+\frac{k_F^7}{\pi^2}\left(\frac{p}{\hbar}\right)^2(1+\tau \delta)^{7 / 3}+\frac{1}{9} \frac{k_F^9}{\pi^2}(1+\tau \delta)^3\right] \\
& + \frac{1}{32} C^{[8]} \left[ \frac{1}{3} \frac{k_{F}^{3}}{\pi^2} \left(\frac{p}{\hbar}\right)^{8}
+ \frac{12}{5} \frac{k_{F}^{5}}{\pi^2} \left(\frac{p}{\hbar}\right)^{6} F_{5/3}
+ \frac{18}{5} \frac{k_{F}^{7}}{\pi^2} \left(\frac{p}{\hbar}\right)^{4} F_{7/3}
+ \frac{4}{3} \frac{k_{F}^{9}}{\pi^2} \left(\frac{p}{\hbar}\right)^{2} F_{3}
+ \frac{1}{11} \frac{k_{F}^{11}}{\pi^2}  F_{11/3} \right] \\
&+ \frac{1}{64} D^{[8]} \left[ \frac{1}{3} \frac{k_{F}^{3}}{\pi^2} \left(\frac{p}{\hbar}\right)^{8} (1+\tau \delta) + \frac{12}{5} \frac{k_{F}^{5}}{\pi^2} \left(\frac{p}{\hbar}\right)^{6} (1+\tau \delta)^{5/3} + \frac{18}{5} \frac{k_{F}^{7}}{\pi^2} \left(\frac{p}{\hbar}\right)^{4} (1+\tau \delta)^{7/3} \right. \\
& \qquad \qquad \left. + \frac{4}{3} \frac{k_{F}^{9}}{\pi^2} \left(\frac{p}{\hbar}\right)^{2} (1+\tau \delta)^{3}  + \frac{1}{11} \frac{k_{F}^{11}}{\pi^2}  F_{11/3} (1+\tau \delta)^{11/3} \right] \\
& + \frac{1}{64} C^{[10]} \left[ \frac{1}{3} \frac{k_{F}^{3}}{\pi^2} \left(\frac{p}{\hbar}\right)^{10}
+ \frac{11}{3} \frac{k_{F}^{5}}{\pi^2} \left(\frac{p}{\hbar}\right)^{8} F_{5/3}
+ \frac{66}{7} \frac{k_{F}^{7}}{\pi^2} \left(\frac{p}{\hbar}\right)^{6} F_{7/3}
+ \frac{22}{3} \frac{k_{F}^{9}}{\pi^2} \left(\frac{p}{\hbar}\right)^{4} F_{3} \right. \\
& \qquad \qquad \left. + \frac{5}{3} \frac{k_{F}^{11}}{\pi^2} \left(\frac{p}{\hbar}\right)^{2} F_{11/3}
+ \frac{1}{13} \frac{k_{F}^{13}}{\pi^2}  F_{13/3} \right] \\
& + \frac{1}{128} D^{[10]} \left[ \frac{1}{3} \frac{k_{F}^{3}}{\pi^2} \left(\frac{p}{\hbar}\right)^{10} (1+\tau \delta)
+ \frac{11}{3} \frac{k_{F}^{5}}{\pi^2} \left(\frac{p}{\hbar}\right)^{8} (1+\tau \delta)^{5/3}
+ \frac{66}{7} \frac{k_{F}^{7}}{\pi^2} \left(\frac{p}{\hbar}\right)^{6} (1+\tau \delta)^{7/3}  \right. \\
& \qquad \qquad \left. + \frac{22}{3} \frac{k_{F}^{9}}{\pi^2} \left(\frac{p}{\hbar}\right)^{4} (1+\tau \delta)^{3}
+ \frac{5}{3} \frac{k_{F}^{11}}{\pi^2} \left(\frac{p}{\hbar}\right)^{2} (1+\tau \delta)^{11/3}
+ \frac{1}{13} \frac{k_{F}^{13}}{\pi^2}  (1+\tau \delta)^{13/3} \right],
\end{aligned}
\end{equation}
where $\tau$ equals $1$ [$-1$] for neutrons [proton] and $k_F=\left(3 \pi^2 \rho / 2\right)^{1/3}$ is the Fermi wave number of nucleons in the SNM.

Expanding $U_\tau(\rho, \delta,p)$ as a power series in $\tau\delta$, we can obtain
\begin{equation}
\label{eq:Utau_series}
U_\tau(\rho, \delta, p) = U_0(\rho,p)+\sum_{i=1,2, \cdots} U_{\mathrm{sym}, i}(\rho,p)(\tau \delta)^i
= U_0(\rho,p)+U_{\mathrm{sym}, 1}(\rho,p)(\tau \delta)+U_{\mathrm{sym}, 2}(\rho, p)(\tau \delta)^2+\cdots,
\end{equation}
where
\begin{equation}
\label{eq:U0_N5LO}
\small
\begin{aligned}
U_{0}(\rho,p) \equiv & U_{\tau}(\rho,\delta=0,p)     \\
=& \frac{3}{4}t_{0}\rho + \sum_{n=1}^{5} \frac{t_3^{[2n-1]}}{16} \left( \frac{2n-1}{3}+2 \right) \rho^{\frac{2n-1}{3}+1} +\frac{1}{8} \left(2C^{[2]} +D^{[2]} \right) \left[\frac{1}{3} \frac{k_F^3}{\pi^2}\left(\frac{p}{\hbar}\right)^2+\frac{1}{5} \frac{k_F^5}{\pi^2}\right] \\
& +\frac{1}{16} \left(2C^{[4]}+D^{[4]}\right) \left[\frac{1}{3} \frac{k_F^3}{\pi^2}\left(\frac{p}{\hbar}\right)^4 + \frac{2}{3} \frac{k_F^5}{\pi^2}\left(\frac{p}{\hbar}\right)^2 +\frac{1}{7} \frac{k_F^7}{\pi^2} \right] \\
&+\frac{1}{8} \left(2C^{[6]}+D^{[6]}\right) \left[\frac{1}{3} \frac{k_F^3}{\pi^2}\left(\frac{p}{\hbar}\right)^6 + \frac{7}{5} \frac{k_F^5}{\pi^2}\left(\frac{p}{\hbar}\right)^4 +\frac{k_F^7}{\pi^2}\left(\frac{p}{\hbar}\right)^2 +\frac{1}{9} \frac{k_F^9}{\pi^2} \right] \\
&+ \frac{1}{64} \left(2C^{[8]}+D^{[8]}\right) \left[ \frac{1}{3} \frac{k_{F}^{3}}{\pi^2} \left(\frac{p}{\hbar}\right)^{8}
+ \frac{12}{5} \frac{k_{F}^{5}}{\pi^2} \left(\frac{p}{\hbar}\right)^{6}
+ \frac{18}{5} \frac{k_{F}^{7}}{\pi^2} \left(\frac{p}{\hbar}\right)^{4}
+ \frac{4}{3} \frac{k_{F}^{9}}{\pi^2} \left(\frac{p}{\hbar}\right)^{2}
+ \frac{1}{11} \frac{k_{F}^{11}}{\pi^2}  \right] \\
&+ \frac{1}{128} \left(2C^{[10]}+D^{[10]}\right) \left[ \frac{1}{3} \frac{k_{F}^{3}}{\pi^2} \left(\frac{p}{\hbar}\right)^{10}
+ \frac{11}{3} \frac{k_{F}^{5}}{\pi^2} \left(\frac{p}{\hbar}\right)^{8}
+ \frac{66}{7} \frac{k_{F}^{7}}{\pi^2} \left(\frac{p}{\hbar}\right)^{6}
+ \frac{22}{3} \frac{k_{F}^{9}}{\pi^2} \left(\frac{p}{\hbar}\right)^{4} + \frac{5}{3} \frac{k_{F}^{11}}{\pi^2} \left(\frac{p}{\hbar}\right)^{2}
+ \frac{1}{13} \frac{k_{F}^{13}}{\pi^2}  \right]
\end{aligned}
\end{equation}
is the single-nucleon potential in SNM and $U_{\mathrm{sym}, i}$ can be expressed as
\begin{equation}
U_{\mathrm{sym}, i}(\rho,p) \left.\equiv \frac{1}{i !} \frac{\partial^i U_n(\rho, \delta,p)}
{\partial \delta^i}\right|_{\delta=0} =\left.\frac{(-1)^i}{i !} \frac{\partial^i U_p(\rho, \delta,p)}{\partial \delta^i}\right|_{\delta=0}.
\end{equation}
Neglecting higher-order terms ($\delta^2,\delta^3,\cdots$) in Eq.~(\ref{eq:Utau_series}) leads to the well-known Lane potential \cite{Lane:1962zz}:
\begin{equation}
U_\tau(\rho, \delta, p) \approx U_0(\rho,p)+U_{\mathrm{sym }}(\rho,p)(\tau \delta).
\end{equation}
In the following, we abbreviate the first-order symmetry potential $U_{\mathrm{sym}, 1}$ as $U_{\mathrm{sym}}$, and it can be expressed as
\begin{equation}
\label{eq:Usym}
\begin{aligned}
U_{\mathrm{sym}}(\rho,p)=&
-\frac{1}{4} t_0\left(2 x_0+1\right) \rho -\sum_{n=1}^{5}  \frac{1}{24} t_{3}^{[2n-1]}\left(2 x_{3}^{[2n-1]}+1\right) \rho^{\frac{2n-1}{3}+1} +\frac{D^{[2]}}{8}\left[\frac{1}{3} \frac{k_F^3}{\pi^2}\left(\frac{p}{\hbar}\right)^2+\frac{1}{3} \frac{k_F^5}{\pi^2}\right] \\
& +\frac{D^{[4]}}{16}\left[\frac{1}{3} \frac{k_F^3}{\pi^2}\left(\frac{p}{\hbar}\right)^4+\frac{10}{9} \frac{k_F^5}{\pi^2}\left(\frac{p}{\hbar}\right)^2+\frac{1}{3} \frac{k_F^7}{\pi^2}\right]
+\frac{D^{[6]}}{8}\left[\frac{1}{3} \frac{k_F^3}{\pi^2}\left(\frac{p}{\hbar}\right)^6+\frac{7}{3} \frac{k_F^5}{\pi^2}\left(\frac{p}{\hbar}\right)^4 +\frac{7}{3} \frac{k_F^7}{\pi^2}\left(\frac{p}{\hbar}\right)^2+\frac{1}{3} \frac{k_F^9}{\pi^2}\right] \\
&+ \frac{D^{[8]}}{64} \left[
\frac{1}{3} \frac{k_F^3}{\pi^2} \left(\frac{p}{\hbar}\right)^8 +
4\frac{k_F^5}{\pi^2} \left(\frac{p}{\hbar}\right)^6 +
\frac{42}{5}\frac{k_F^7}{\pi^2} \left(\frac{p}{\hbar}\right)^4 +
4\frac{k_F^9}{\pi^2} \left(\frac{p}{\hbar}\right)^2 +
\frac{1}{3}\frac{k_F^{11}}{\pi^2}
\right] \\
&+ \frac{D^{[10]}}{128} \left[
\frac{1}{3} \frac{k_F^3}{\pi^2} \left(\frac{p}{\hbar}\right)^{10} +
\frac{55}{9} \frac{k_F^5}{\pi^2} \left(\frac{p}{\hbar}\right)^8 +
22 \frac{k_F^7}{\pi^2} \left(\frac{p}{\hbar}\right)^6 +
22 \frac{k_F^9}{\pi^2} \left(\frac{p}{\hbar}\right)^4 +
\frac{55}{9} \frac{k_F^{11}}{\pi^2} \left(\frac{p}{\hbar}\right)^2 +
\frac{1}{3}\frac{k_F^{13}}{\pi^2}
\right].
\end{aligned}
\end{equation}

The nucleon effective mass is used to characterize the momentum dependence of the single-nucleon potential, and in nonrelativistic models, it can be expressed as \cite{Jaminon:1989wj,Li:2018lpy}
\begin{equation}
\frac{m_{\tau}^{\ast}(\rho,\delta)}{m}=\left[1+\left.\frac{m}{p} \frac{d U_\tau(\rho, \delta,p)}{d p}\right|_{p=p_{F_{\tau}}}\right]^{-1}.
\end{equation}
The isoscalar nucleon effective mass $m_{s}^{\ast}$ is the nucleon effective mass in SNM, and the isovector nucleon effective mass $m_{v}^{\ast}$ is the effective mass of proton (neutron) in pure neutron (proton) matter, and their expressions are provided in Appendix~\ref{sec:App_quantities}.
Additionally, a subscript ``0" denotes that the nucleon effective mass is defined at the saturation density $\rho_0$, e.g., $m_{s,0}^{\ast}$ and $m_{v,0}^{\ast}$.
The nucleon effective mass splitting, denoted as $\mspl(\rho,\delta) \equiv \left[m_{n}^{\ast}(\rho,\delta)-m_{p}^{\ast}(\rho,\delta) \right]/m$, is extensively used in nuclear physics.
$\mspl(\rho,\delta)$ can be expanded as a power series in $\delta$, i.e.,
\begin{equation}
\label{eq:spl_coes}
\mspl(\rho,\delta) =\sum_{n=1}^{\infty}
\Delta m_{2 n-1}^{\ast}(\rho) \delta^{2 n-1},
\end{equation}
where $\Delta m_{2 n-1}^{\ast}(\rho)$ are the isospin splitting coefficients (of the nucleon effective mass).
The first coefficient $\Delta m_{1}^{\ast}(\rho)$ is usually referred to as the linear isospin splitting coefficient, whose expression is presented in Appendix~\ref{sec:App_quantities}.

\end{widetext}

\section{Fitting strategy and new interactions}
\label{sec:fitting}
The gradient terms make no contribution to the properties of uniform nuclear matter, thus the parameters $E^{[n]}$ and $F^{[n]}$ ($n=2,4,6,8,10$) are irrelevant to the following discussions about the nuclear matter.
However, the gradient terms are important for transport model and nuclear structure, and the values of $E^{[n]}$ and $F^{[n]}$ could be determined by the finite nuclei calculation, which is beyond the scope of this work.

The construction of interactions with the N3LO model has been discussed in detail in Ref.~\cite{Wang:2023zcj}, and we briefly outline the similar process carried out in this work for completeness.
To clearly demonstrate the fitting process, $U_{0}(\rho_0,p)$ [Eq.~(\ref{eq:U0_N5LO})] and $U_{\mathrm{sym}}(\rho_0,p)$ [Eq.~(\ref{eq:Usym})] can be expressed as
\begin{equation}
\begin{aligned}
\label{eq:U0_a}
U_{0}(\rho_0,p) \equiv& a_{0}
+ a_{2} \left(\frac{p}{\hbar}\right)^{2}
+ a_{4} \left(\frac{p}{\hbar}\right)^{4} \\
&+ a_{6} \left(\frac{p}{\hbar}\right)^{6}
+ a_{8} \left(\frac{p}{\hbar}\right)^{8}
+ a_{10} \left(\frac{p}{\hbar}\right)^{10},
\end{aligned}
\end{equation}
and
\begin{equation}
\begin{aligned}
\label{eq:Usym_b}
U_{\mathrm{sym}}(\rho_0,p) \equiv & b_{0}
+ b_{2} \left(\frac{p}{\hbar}\right)^{2}
+ b_{4} \left(\frac{p}{\hbar}\right)^{4} \\
&+ b_{6} \left(\frac{p}{\hbar}\right)^{6}
+ b_{8} \left(\frac{p}{\hbar}\right)^{8}
+ b_{10} \left(\frac{p}{\hbar}\right)^{10},
\end{aligned}
\end{equation}
where $a_{n}$ and $b_{n}$ ($n=0,2,4,6,8,10$) take the following forms
\begin{equation}
\begin{aligned}
a_{0} =&
\frac{3}{4}t_{0}\rho_{0} + \sum_{n=1}^{5} \frac{t_3^{[2n-1]}}{16} \left( \frac{2n-1}{3}+2 \right) \rho_{0}^{\frac{2n-1}{3}+1} \\
&+ \frac{k_{F_{0}}^5}{40\pi^2}\left(2 C^{[2]}+D^{[2]}\right)
+ \frac{k_{F_{0}}^7}{112\pi^2}\left(2 C^{[4]}+D^{[4]}\right) \\
&+ \frac{k_{F_{0}}^9}{72\pi^2}\left(2 C^{[6]}+D^{[6]}\right)
+ \frac{k_{F_{0}}^{11}}{704\pi^2}\left(2 C^{[8]}+D^{[8]}\right) \\
&+ \frac{k_{F_{0}}^{13}}{1664\pi^2}\left(2 C^{[10]}+D^{[10]}\right),
\end{aligned}
\end{equation}
\begin{equation}
\begin{aligned}
a_{2} =&
\frac{k_{F_{0}}^{3}}{24 \pi^{2}} \left(2 C^{[2]}+D^{[2]}\right)
+ \frac{k_{F_{0}}^{5}}{24 \pi^{2}} \left(2 C^{[4]}+D^{[4]}\right) \\
&+ \frac{k_{F_{0}}^{7}}{8 \pi^{2}} \left(2 C^{[6]}+D^{[6]}\right)
+ \frac{k_{F_{0}}^{9}}{48 \pi^{2}} \left(2 C^{[8]}+D^{[8]}\right) \\
&+ \frac{5k_{F_{0}}^{11}}{384 \pi^{2}} \left(2 C^{[10]}+D^{[10]}\right),
\end{aligned}
\end{equation}
\begin{equation}
\begin{aligned}
a_{4} =& \frac{k_{F_{0}}^{3}}{48 \pi^{2}} \left(2 C^{[4]}+D^{[4]}\right)
+ \frac{7k_{F_{0}}^{5}}{40 \pi^{2}} \left(2 C^{[6]}+D^{[6]}\right) \\
&+ \frac{9k_{F_{0}}^{7}}{160 \pi^{2}} \left(2 C^{[8]}+D^{[8]}\right)
+ \frac{11k_{F_{0}}^{9}}{192 \pi^{2}} \left(2 C^{[10]}+D^{[10]}\right),
\end{aligned}
\end{equation}
\begin{equation}
\begin{aligned}
a_{6} =& \frac{k_{F_{0}}^{3}}{24 \pi^{2}} \left(2 C^{[6]}+D^{[6]}\right)
+ \frac{3k_{F_{0}}^{5}}{80 \pi^{2}} \left(2 C^{[8]}+D^{[8]}\right) \\
&+ \frac{33k_{F_{0}}^{7}}{448 \pi^{2}} \left(2 C^{[10]}+D^{[10]}\right),
\end{aligned}
\end{equation}
\begin{equation}
a_{8} = \frac{k_{F_{0}}^{3}}{192 \pi^{2}} \left(2 C^{[8]}+D^{[8]}\right)
+ \frac{11k_{F_{0}}^{5}}{384 \pi^{2}} \left(2 C^{[10]}+D^{[10]}\right),
\end{equation}
\begin{equation}
a_{10} = \frac{k_{F_{0}}^{3}}{384 \pi^{2}} \left(2 C^{[10]}+D^{[10]}\right),
\end{equation}
and
\begin{equation}
\begin{aligned}
b_{0} = & - \frac{1}{4} t_{0} (2x_{0}+1) \rho_{0} \\
&- \sum_{n=1}^{5} \frac{1}{24} t_{3}^{[2n-1]} \left( 2x_{3}^{[2n-1]} + 1 \right) \rho_{0}^{\frac{2n-1}{3}+1} \\
& + \frac{k_{F_{0}}^{5}}{24 \pi^2} D^{[2]} + \frac{k_{F_{0}}^{7}}{48 \pi^2} D^{[4]} + \frac{k_{F_{0}}^{9}}{24 \pi^2} D^{[6]} \\
&+ \frac{k_{F_{0}}^{11}}{192 \pi^2} D^{[8]} + \frac{k_{F_{0}}^{13}}{384 \pi^2} D^{[10]},
\end{aligned}
\end{equation}
\begin{equation}
\begin{aligned}
b_{2} =& \frac{k_{F_{0}}^{3}}{24 \pi^2} D^{[2]} + \frac{5 k_{F_{0}}^{5}}{72 \pi^2} D^{[4]} + \frac{7 k_{F_{0}}^{7}}{24 \pi^2} D^{[6]} \\
&+ \frac{k_{F_{0}}^{9}}{16 \pi^2} D^{[8]} + \frac{55 k_{F_{0}}^{11}}{1152 \pi^2} D^{[10]},
\end{aligned}
\end{equation}
\begin{equation}
\begin{aligned}
b_{4}=& \frac{k_{F_{0}}^{3}}{48 \pi^2} D^{[4]} + \frac{7 k_{F_{0}}^{5}}{24 \pi^2} D^{[6]} + \frac{21 k_{F_{0}}^{7}}{160 \pi^2} D^{[8]} \\
&+ \frac{11 k_{F_{0}}^{9}}{64 \pi^2} D^{[10]} ,
\end{aligned}
\end{equation}
\begin{equation}
b_{6} = \frac{k_{F_{0}}^{3}}{24 \pi^2} D^{[6]} + \frac{k_{F_{0}}^{5}}{16 \pi^2} D^{[8]} + \frac{11 k_{F_{0}}^{7}}{64 \pi^2} D^{[10]}
\end{equation}
\begin{equation}
b_{8} = \frac{k_{F_{0}}^{3}}{192 \pi^2} D^{[8]} + \frac{55 k_{F_{0}}^{5}}{1152 \pi^2} D^{[10]} ,
\end{equation}
\begin{equation}
b_{10} = \frac{k_{F_{0}}^{3}}{384 \pi^2} D^{[10]} .
\end{equation}
In Table~\ref{tab:Models_compare}, we list the independent model parameters and the adjustable quantities in the N3LO, N4LO and N5LO models, respectively.
The number of parameters equals to the number of characteristic quantities, which are also listed in Table~\ref{tab:Models_compare}.
The values of all adjustable parameters must be provided in order to fully construct an interaction, i.e., determine all the model parameters.
It is worth emphasizing again that these parameters and quantities are related to the properties of uniform nuclear matter.
Only when the values of $E^{[n]}$ and $F^{[n]}$ are determined through finite nuclei calculations can all the Skyrme parameters in Eqs.~(\ref{eq:vN3})-(\ref{eq:vN5}) be completely obtained.
\begin{table}
\caption{
\label{tab:Models_compare}
The independent model parameters, the adjustable quantities and the total number (t.n.) of them for different models.
}
\begin{tabular}{|c|c|c|c|}
\hline
Models & Parameters & Quantities & t.n. \\ \hline
       & $t_0$, $x_0$, $t_{3}^{[1]}$, $x_{3}^{[1]}$,$t_3^{[3]}$, $x_{3}^{[3]}$,
       & $\rho_0$, $E_0(\rho_0)$, $K_0$, $J_0$,& \\
N3LO   & $t_3^{[5]}$, $x_{3}^{[5]}$, $C^{[2]}$, $D^{[2]}$,
       & $E_{{\mathrm{sym}}}(\rho_0)$, $L$, $K_{{\mathrm{sym}}}$,
         & 14 \\
       & $C^{[4]}$,$D^{[4]}$,$C^{[6]}$,$D^{[6]}$
       & $J_{{\mathrm{sym}}}$, $a_2$,$a_4$,$a_6$,$b_2$,$b_4$,$b_6$ & \\ \hline
N4LO   & N3LO $+$ &  N3LO $+$ & 18 \\
       & $t_3^{[7]}$, $x_{3}^{[7]}$,$C^{[8]}$,$D^{[8]}$
       & $I_0$, $I_{{\mathrm{sym}}}$, $a_8$, $b_8$ & \\ \hline
N5LO   & N4LO $+$ &  N4LO $+$ & 22 \\
       & $t_3^{[9]}$, $x_{3}^{[9]}$,$C^{[10]}$,$D^{[10]}$
       & $H_0$, $H_{{\mathrm{sym}}}$, $a_{10}$, $b_{10}$ &  \\ \hline
\end{tabular}
\end{table}

In N3LO, N4LO, and N5LO models, we set the values of $\rho_0$, $E_{0}(\rho_0)$ and $K_0$ to be $0.16\,\mathrm{fm}^{-3}$, $-16\,\mathrm{MeV}$, and $230 \,\mathrm{MeV}$, respectively.
Next, we use GEKKO optimization suite \cite{beal2018gekko} to minimize the weighted squared difference $\chi^{2}$ between $U_{0}$ in Eq.~(\ref{eq:U0_N5LO}) and the nucleon optical potential data $U_{\mathrm{opt}}$ \cite{Hama:1990vr,Cooper:1993nx} and its extrapolation above $1$ GeV:
\begin{equation}
\chi^{2} = \sum_{i=1}^{N_{d}} \left( \frac{U_{0,i}-U_{\mathrm{opt},i}}{\sigma_{i}} \right)^{2},
\end{equation}
with constraint of the HVH theorem \cite{Hugenholtz:1958zz,SATPATHY199985}, i.e.,
\begin{equation}
E_{0}\left(\rho_0\right) = \frac{p_{F_0}^{2}}{2m} +U_{0} \left(\rho_0,p_{F_0}\right).
\end{equation}
The $N_d$ is the number of the experimental data points.
Since there are actually no practical errors $\sigma_i$ here, we assign equal weights to each data point within the range of the nucleon momentum up to $1.5\,\mathrm{GeV}/c$, $2.0\,\mathrm{GeV}/c$, and $2.5\,\mathrm{GeV}/c$ (approximately corresponding to nucleon kinetic energy of $1$ GeV, $1.5$ GeV and $2$ GeV) for N3LO, N4LO and N5LO models, respectively.
The values of $a_n$ parameters and the isoscalar nucleon effective mass at saturation density $m^{\ast}_{s,0}$ are presented in Table~\ref{tab:an_compare}.
\begin{table}
\caption{
\label{tab:an_compare}
The isoscalar nucleon effective mass at saturation density ($m^{\ast}_{s,0}$) and the $a_{n}$ parameters related to the momentum dependence of the single-nucleon potential.
}
\begin{ruledtabular}
\begin{tabular}{cccc}
    & N3LO & N4LO & N5LO \\ \hline
$m^{\ast}_{s,0}/m$ & $0.773$ & $0.761$ & $0.760$ \\
$a_{0}$ ($\mathrm{MeV}$) & $-64.03$ & $-64.88$ & $-64.97$ \\
$a_{2}$ ($\mathrm{MeV}\,\mathrm{fm}^{2}$)
& $6.518$ & $7.044$ & $7.104$ \\
$a_{4}$ ($\mathrm{MeV}\,\mathrm{fm}^{4}$)
& $-0.1260$ & $-0.1545$ & $-0.1628$ \\
$a_{6}$ ($\mathrm{MeV}\,\mathrm{fm}^{6}$)
& $8.133\times10^{-4}$ & $1.428\times10^{-3}$ & $1.731\times10^{-3}$ \\
$a_{8}$ ($\mathrm{MeV}\,\mathrm{fm}^{8}$)
& - & $-4.703\times10^{-6}$ & $-8.614\times10^{-6}$ \\
$a_{10}$ ($\mathrm{MeV}\,\mathrm{fm}^{10}$)
& - & - & $1.621\times10^{-8}$
\end{tabular}
\end{ruledtabular}
\end{table}
In N3LO model, the last independent quantity related to SNM, $J_0$, is constrained by the flow data in HICs \cite{Danielewicz:2002pu}, and it is set to be the maximum allowed value, i.e., $-383\,\mathrm{MeV}$.
In N4LO (N5LO) model, $I_0$ (as well as $H_0$) are adjustable, and we set them to their corresponding values calculated from N3LO model, i.e., $I_{0}=1818.9\,\mathrm{MeV}$ and $H_{0}=-12065\,\mathrm{MeV}$.

Considering its significant uncertainty, we construct eight kinds of symmetry potential (i.e., 8 different $b_n$ parameter sets), with $\Delta m_{1}^{\ast}(\rho_0)$ values of $-0.7$, $-0.5$, $-0.3$, $-0.1$, $0.1$, $0.3$, $0.5$, and $0.7$.
To avoid too many degrees of freedom and ensure that symmetry potential behaves well within the momentum range up to $2$ GeV/$c$, the values of $b_n$ are given by analogy the Taylor expansion of the cosine function, i.e.,
\begin{equation}
\label{eq:bn_taylor}
b_{n} = A \frac{(-1)^{n/2}}{n!}\left(\frac{\pi}{10} \right)^{n}
\quad (\mathrm{MeV}\;\mathrm{fm}^{n}),
\end{equation}
where $A$ is the only adjustable parameter to provide the corresponding $\Delta m_{1}^{\ast}(\rho_0)$ value, and $n=2,4,6$ for N3LO model, $n=2,4,6,8$ for N4LO model, and $n=2,4,6,8,10$ for N5LO model.
The values of $A$ corresponding to 8 different $\Delta m_{1}^{\ast}(\rho_0)$ values (8 different momentum dependencies of the symmetry potential) are listed in Table.~\ref{tab:A_m1}, and the corresponding $b_{n}$ can be obtained through Eq.~(\ref{eq:bn_taylor}).
\begin{table}
\caption{
\label{tab:A_m1}
The values of $A$ in Eq.~(\ref{eq:bn_taylor}) used in the parametrization of the symmetry potential, corresponding to different models and different $\Delta m_{1}^{\ast}(\rho_0)$.
The values of $b_{n}$ can be calculated through Eq.~(\ref{eq:bn_taylor}).}
\begin{ruledtabular}
\begin{tabular}{cccc}
$\Delta m_{1}^{\ast}(\rho_0)$
& $A$ (N3LO) & $A$ (N4LO) & $A$ (N5LO) \\ \hline
$0.1$ & $30.17$ & $30.08$ & $29.84$ \\
$-0.1$ & $-42.19$ & $-44.61$ & $-45.04$ \\
$0.3$ & $102.5$ & $104.8$ & $104.7$ \\
$-0.3$ & $-114.6$ & $-119.3$ & $-119.9$ \\
$0.5$ & $174.9$ & $179.5$ & $179.6$ \\
$-0.5$ & $-186.9$ & $-194.0$ & $-194.8$ \\
$0.7$ & $247.2$ & $254.2$ & $254.5$ \\
$-0.7$ & $-259.3$ & $-268.7$ & $-269.7$
\end{tabular}
\end{ruledtabular}
\end{table}
It is worth noting that the value of $b_0$ is determined by the theorem of the symmetry energy decomposition \cite{Brueckner:1964zz,Dabrowski:1972mbb,Dabrowski:1973zz,Xu:2010fh,Xu:2010kf,Chen:2011ag}:
\begin{equation}
\label{eq:Esym_decom}
E_{\mathrm{sym}}(\rho_0)=\frac{1}{3} \frac{p_{F_0}^{2}}{2m^{\ast}_{s,0}}+\frac{1}{2} U_{\mathrm{sym}}\left(\rho_0,p_{F_0}\right),
\end{equation}
once the value of $E_{\mathrm{sym}}(\rho_0)$ is given.
The quantities related to the symmetry energy in N3LO model have been discussed in detail in Ref.~\cite{Wang:2023zcj}.
We set $E_{\mathrm{sym}}(\rho_0)$, $L$, $K_{\mathrm{sym}}$, and $J_{\mathrm{sym}}$ to their values in parameter set ``SP6L45" from Ref.~\cite{Wang:2023zcj}, namely $E_{\mathrm{sym}}(\rho_0)=30\,\mathrm{MeV}$, $L=45\,\mathrm{MeV}$, $K_{\mathrm{sym}}=-110\,\mathrm{MeV}$, and $J_{\mathrm{sym}}=700\,\mathrm{MeV}$, to simultaneously satisfy the theoretical prediction of the EOS of PNM and various neutron star observations.
The values of $I_{\mathrm{sym}}$ in N4LO models as well as $H_{\mathrm{sym}}$ in N5LO model are set to their corresponding values calculated from N3LO model, which will vary due to the differences in $\Delta m_{1}^{\ast}(\rho_0)$.

Finally, the combination of the 3 different models namely, N$3$LO, N$4$LO and N$5$LO and 8 different symmetry potentials forms a parameter set family consisting of 24 parameter sets.
These parameters have similar density behavior of the SNM EOS and the symmetry energy by construction.
We name these parameter sets as SPNL$45$X, where:
``SP'' indicates the framework of the Skyrme pseudopotential, and N represents the highest power of momentum, with N taking values of 6, 8, and 10 for N3LO, N4LO, and N5LO models, respectively;
``L45'' indicates that their $L$ value is 45 MeV;
``X'' denotes the value of $\Delta m_{1}^{\ast}(\rho_0)$, i.e., X=Dm07, Dm05, Dm03, Dm01, D01, D03, D05, and D07 respectively indicate that the values of $\Delta m_{1}^{\ast}(\rho_0)$ is $-0.7$, $-0.5$, $-0.3$, $-0.1$, $0.1$, $0.3$, $0.5$, and $0.7$.
In Tables~\ref{tab:SP6}-\ref{tab:SP10} at Appendix~\ref{sec:Skyrme_paras}, we list the Skyrme parameters for these 24 new interactions.

\begin{table}
\caption{
\label{tab:SNM_paras}
The values of characteristic quantities related to SNM, and non-adjustable quantities are enclosed in `` ''.
}
\begin{ruledtabular}
\begin{tabular}{cccc}
    & SP6 & SP8 & SP10 \\ \hline
$\rho_0\,(\rm{fm}^{-3})$  & $0.160$ & $0.160$ & $0.160$ \\
$E_0(\rho_0)\,(\rm{MeV})$
& $-16.0$ & $-16.0$ & $-16.0$ \\
$K_0\,(\rm{MeV})$
& $230.0$ & $230.0$ & $230.0$ \\
$J_0\,(\rm{MeV})$
& $-383.0$ & $-383.0$ & $-383.0$ \\
$I_0\,(\rm{MeV})$
& ``$1818.9$'' & $1818.9$ & $1818.9$ \\
$H_0\,(\rm{MeV})$
& ``$-12065$'' & ``$-12090$'' & $-12065$
\end{tabular}
\end{ruledtabular}
\end{table}

\begin{figure}[ht]
    \centering
    \includegraphics[width=\linewidth]{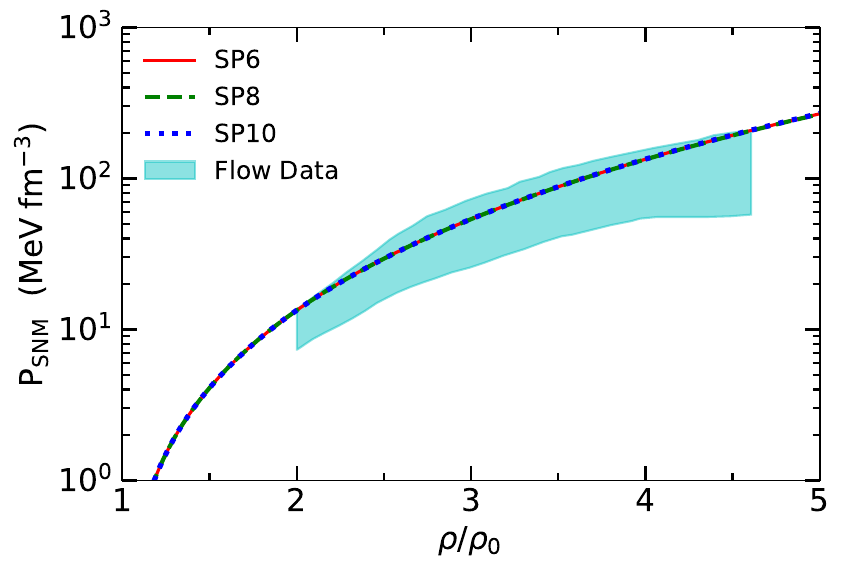}
    \caption{The pressure of SNM [$P_{\mathrm{SNM}}(\rho)$] as a function of nucleon density given by the SP6, SP8 and SP10 models. Also included are the constraints from flow data in HICs \cite{Danielewicz:2002pu}.}
    \label{fig:Psnm}
\end{figure}

\begin{table*}
\caption{
\label{tab:Esym_paras}
The values of high-order characteristic quantities of symmetry energy, including the kurtosis coefficient ($I_{\mathrm{sym}}$) and the hyper-skewness coefficient ($H_{\mathrm{sym}}$) corresponding to different linear isospin splitting coefficients [$\Delta m_{1}^{\ast}(\rho_0)$] for different models.
Also listed are the values of the isovector effective mass ($m^{\ast}_{v,0}$) and the fourth-order symmetry energy [$E_{\mathrm{sym},4}(\rho_0)$].
Note that all the models share the common lower-order symmetry energy parameters  of $E_{\mathrm{sym}}(\rho_0)=30~\mathrm{MeV}$, $L=45~\mathrm{MeV}$, $K_{\mathrm{sym}}=-110~\mathrm{MeV}$ and $J_{\mathrm{sym}}=700~\mathrm{MeV}$.
}
\begin{ruledtabular}
\begin{tabular}{ccccccccc}
SP6L45  & Dm07 & Dm05 & Dm03 & Dm01 & D01 & D03 & D05 & D07 \\
$\Delta m_{1}^{\ast}(\rho_0)$ &
$-0.7$ & $-0.5$ & $-0.3$ & $-0.1$ &
$0.1$ & $0.3$ & $0.5$ & $0.7$ \\
$I_{\mathrm{sym}}$ (MeV) &
$-2241.3$ & $-2284.7$ & $-2328.2$ & $-2371.6$ &
$-2415.1$ & $-2458.5$ & $-2502.0$ & $-2545.5$ \\
$H_{\mathrm{sym}}$ (MeV) &
$18845$ & $19011$ & $19177$ & $19343$ &
$19509$ & $19676$ & $19842$ & $20008$ \\
$m^{\ast}_{v,0}/m$ &
$1.464$ & $1.169$ & $0.972$ & $0.833$ &
$0.728$ & $0.647$ & $0.582$ & $0.529$ \\
$E_{\mathrm{sym},4}(\rho_0)$ (MeV) &
$-0.333$ & $-0.0916$ & $0.150$ & $0.392$ &
$0.633$  & $0.875$   & $1.117$ & $1.358$ \\ \hline
SP8L45  & Dm07 & Dm05 & Dm03 & Dm01 & D01 & D03 & D05 & D07 \\
$\Delta m_{1}^{\ast}(\rho_0)$ &
$-0.7$ & $-0.5$ & $-0.3$ & $-0.1$ &
$0.1$ & $0.3$ & $0.5$ & $0.7$ \\
$I_{\mathrm{sym}}$ (MeV) &
$-2241.3$ & $-2284.7$ & $-2328.2$ & $-2371.6$ & $-2415.1$
& $-2458.5$ & $-2502.0$ & $-2545.5$ \\
$H_{\mathrm{sym}}$ (MeV) &
$18799$ & $18972$ & $19145$ &  $19318$ &
$19491$ & $19663$ & $19836$ & $20009$  \\
$m^{\ast}_{v,0}/m$ &
$1.464$ & $1.161$ & $0.962$ & $0.812$ &
$0.716$ & $0.635$ & $0.571$ & $0.518$ \\
$E_{\mathrm{sym},4}(\rho_0)$ (MeV) &
$-0.366$ & $-0.117$ & $0.133$ & $0.382$ &
$0.632$  & $0.881$  & $1.131$ & $1.380$ \\ \hline
SP10L45  & Dm07 & Dm05 & Dm03 & Dm01 & D01 & D03 & D05 & D07 \\
$\Delta m_{1}^{\ast}(\rho_0)$ &
$-0.7$ & $-0.5$ & $-0.3$ & $-0.1$ &
$0.1$ & $0.3$ & $0.5$ & $0.7$ \\
$I_{\mathrm{sym}}$ (MeV) &
$-2241.3$ & $-2284.7$ & $-2328.2$ & $-2371.6$ &
$-2415.1$ & $-2458.5$ & $-2502.0$ & $-2545.5$ \\
$H_{\mathrm{sym}}$ (MeV) &
$18845$ & $19011$ & $19177$ & $19343$ &
$19509$ & $19676$ & $19842$ & $20008$ \\
$m^{\ast}_{v,0}/m$ &
$1.465$ & $1.161$ & $0.962$ & $0.821$ &
$0.716$ & $0.635$ & $0.570$ & $0.517$ \\
$E_{\mathrm{sym},4}(\rho_0)$ (MeV) &
$-0.372$ & $-0.121$ & $0.129$ & $0.379$ &
$0.628$  & $0.879$  & $1.129$ & $1.379$ \\
\end{tabular}
\end{ruledtabular}
\end{table*}

\begin{figure*}[ht]
    \centering
    \includegraphics[width=\linewidth]{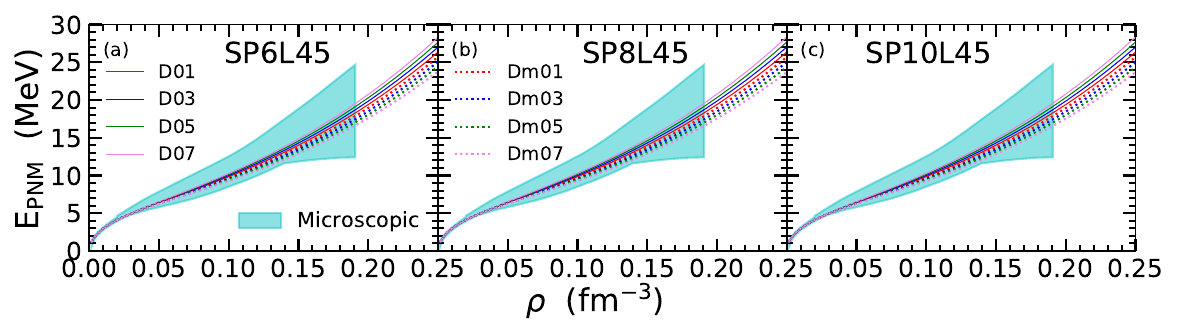}
    \caption{The EOS of PNM predicted by different interactions, categorized according to different models.
    The band represents the microscopic calculation results \cite{Zhang:2022bni} (see text for the details).}
    \label{fig:Epnm}
\end{figure*}
\begin{figure*}[ht]
    \centering
    \includegraphics[width=\linewidth]{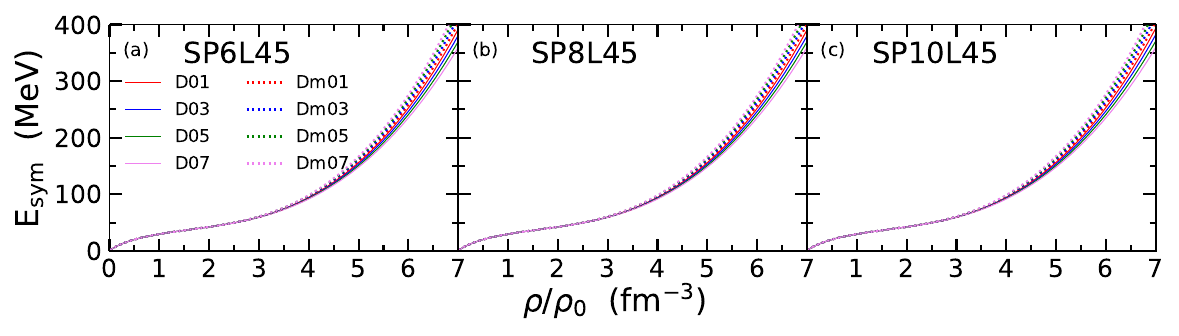}
    \caption{The energy dependence of the symmetry energy predicted by different interactions, categorized according to different models (see text for the details).}
    \label{fig:Esym}
\end{figure*}

\section{The properties of cold nuclear matter}
\label{sec:properties}
\subsection{Bulk properties of cold nuclear matter}
The values of characteristic quantities related to SNM of the three models are listed in Table~\ref{tab:SNM_paras}, and non-adjustable quantities are denoted by `` ''. In Fig.~\ref{fig:Psnm}, we present the pressure of SNM [$P_{\mathrm{SNM}}(\rho)$] as a function of nucleon density given by SP6, SP8, and SP10 models, respectively.
Also shown in Fig.~\ref{fig:Psnm} are the constraints from flow data in HICs \cite{Danielewicz:2002pu}.
It can be seen that the SNM pressure given by the three models is almost the same by the construction, and all conform to the constraints given by the flow data.

For all interactions, the values of $E_{\mathrm{sym}}(\rho_0)$, $L$, $K_{\mathrm{sym}}$ and $J_{\mathrm{sym}}$ are taken as $30\,\mathrm{MeV}$, $45\,\mathrm{MeV}$, $-110\,\mathrm{MeV}$ and $700\,\mathrm{MeV}$, respectively.
Once these quantities are fixed:
the values of $I_{\mathrm{sym}}$ and $H_{\mathrm{sym}}$ are determined in SP6 model;
the values of $I_{\mathrm{sym}}$ remains adjustable, but $H_{\mathrm{sym}}$ is not independent in SP8 model;
both $I_{\mathrm{sym}}$ and $H_{\mathrm{sym}}$ remain free to vary in SP10 model.
To minimize the influence of the symmetry energy and thereby focus on the effects of different models (different momentum behaviors of the single-nucleon potential at high energies) and different symmetry potentials (different effective mass splitting), we set the values of $I_{\mathrm{sym}}$ in SP8 model, as well as $I_{\mathrm{sym}}$ and $H_{\mathrm{sym}}$ in SP10 model, to the values calculated in SP6 model.
Table.~\ref{tab:Esym_paras} presents the values of $I_{\mathrm{sym}}$ and $H_{\mathrm{sym}}$ for different interactions corresponding to various $\Delta m_{1}^{\ast}(\rho_0)$.
Also listed in Table~\ref{tab:Esym_paras} are the values of the isovector nucleon effective mass $m^{\ast}_{v,0}$ and the fourth-order symmetry energy $E_{\mathrm{sym},4}(\rho_0)$.

Shown in Fig.~\ref{fig:Epnm}
is the EOS of PNM as a function of density predicted by different interactions and the result from combined microscopic calculations (see Ref.~\cite{Zhang:2022bni} and its references for details).
It can be seen that all interactions are consistent with the microscopic theoretical calculations.
Shown in Fig.~\ref{fig:Esym}
is the density dependence of the symmetry energy predicted by different interactions.
It can be observed from each panel in Fig.~\ref{fig:Esym} that the impact of different $\Delta m_{1}^{\ast}(\rho_0)$ on the symmetry energy becomes apparent only at high densities.
Moreover, for the same $\Delta m_{1}^{\ast}(\rho_0)$, the symmetry energy predicted by the three models are almost identical by construction.

\begin{figure}[ht]
    \centering
    \includegraphics[width=\linewidth]{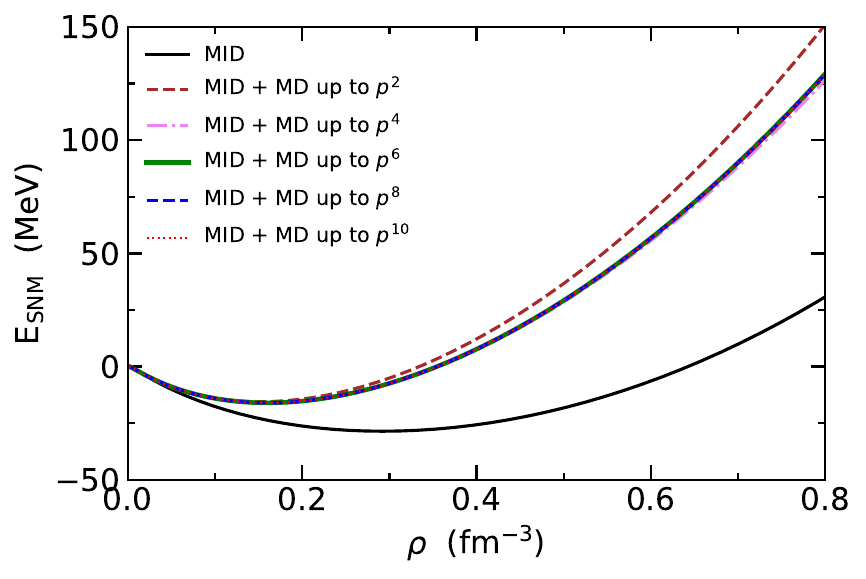}
    \caption{The EOS of SNM as a function of nucleon density from the MID part as well from the sum of the MID part and the MD part up to different orders in the SP10 model.}
    \label{fig:Esnm_pw}
\end{figure}

\begin{figure}[ht]
    \centering
    \includegraphics[width=\linewidth]{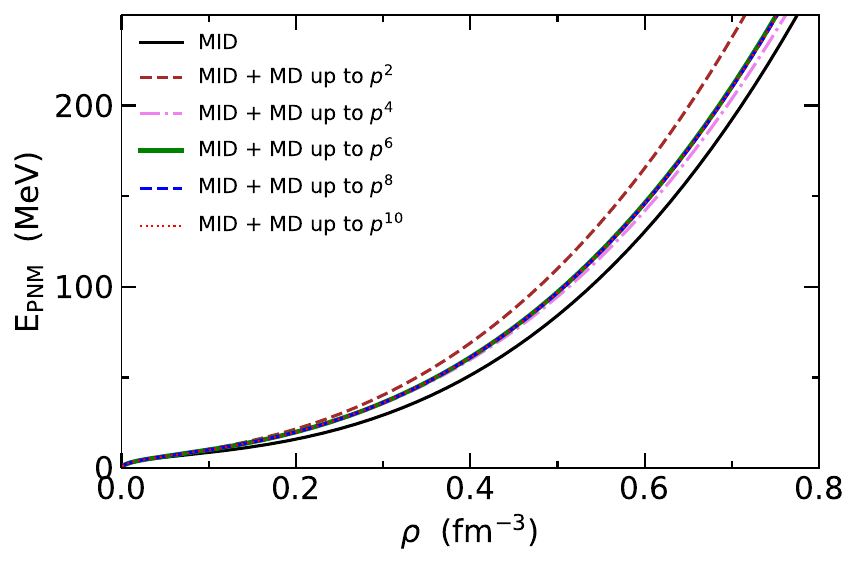}
    \caption{The EOS of PNM as a function of nucleon density from the MID part as well from the sum of the MID part and the MD part up to different orders in the SP10L45D03 model.}
    \label{fig:Epnm_pw}
\end{figure}

To demonstrate the convergence of the high-order derivatives, we further present the cumulative contributions of the MD term for the nuclear matter bulk properties.
In Fig.~\ref{fig:Esnm_pw}, we display the SNM EOS from the momentum-independent (MID) part as well as from the sum of the MID part and the MD part up to different orders in the SP10 model.
The result for the PNM EOS are presented in Fig.~\ref{fig:Epnm_pw}, using the SP10L45D03 model as an example.
From Figs.~\ref{fig:Esnm_pw} and \ref{fig:Epnm_pw}, it can be seen that the SNM EOS converges up to $p^{4}$, while the PNM EOS converges up to $p^{6}$, with higher-order terms contributing negligibly.
This is consistent with the conclusions in Ref.~\cite{Davesne:2016fqg}.

\subsection{Single-nucleon potential and symmetry potential}
\begin{figure}[ht]
    \centering
    \includegraphics[width=\linewidth]{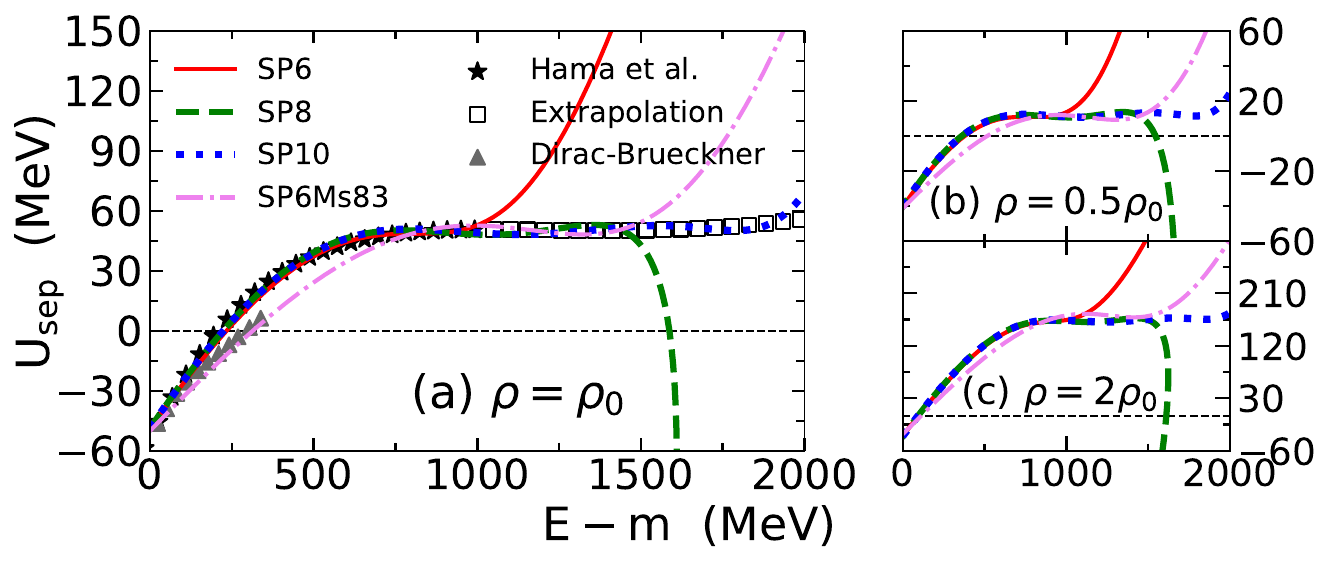}
    \caption{
    The energy dependence of the single-nucleon potential in cold SNM predicted by SP6, SP8, SP10 and SP6Ms83.
    Also shown are the nucleon optical potential (Schr\"{o}dinger equivalent potential, $\mathrm{U}_{\mathrm{sep}}$) in SNM at $\rho_0$ obtained by Hama \textit{et al.}~\cite{Hama:1990vr,Cooper:1993nx} and the Dirac-Brueckner calculation \cite{TerHaar:1986xpv}.
    }
    \label{fig:U0_compare}
\end{figure}

The momentum dependence of single-nucleon potential is characterized by isoscalar nucleon effective mass ($m^{\ast}_{s}$), and it is also
represented by parameters $a_n$ in the three models.
The values of $m^{\ast}_{s,0}$ and $a_n$ parameters are presented in Table~\ref{tab:an_compare}.
Shown in Fig.~\ref{fig:U0_compare} is the single-nucleon potential [$U_{0}(\rho,p)$] in cold SNM, predicted by SP6, SP8, and SP10 models, as a function of nucleon kinetic energy $E-m=\sqrt{p^2+m^2}+U_{0}(\rho,p)-m$, at $\rho=0.5\rho_0$, $\rho_0$ and $2\rho_0$, respectively.
Also shown in Fig.~\ref{fig:U0_compare}(a) is the real part of nucleon optical potential (Schr\"{o}dinger equivalent potential) in SNM at $\rho_0$ at the energy range up to $1$ GeV obtained by Hama \textit{et al.} \cite{Hama:1990vr,Cooper:1993nx}, from Dirac phenomenology of nucleon-nucleus scattering data.
The extrapolation of Hama's data are also plotted in Fig.~\ref{fig:U0_compare}(a).
It can be seen from Fig.~\ref{fig:U0_compare}(a) that through the optimization process we have performed in Sec.~\ref{sec:fitting}, $U_{0}(\rho,p)$ conforms rather well to the empirical nucleon optical potential (and its extrapolation) for nucleon kinetic energy up to $1$ GeV, $1.5$ GeV, and $2$ GeV, with SP6, SP8, and SP10 models, respectively.

\begin{figure}[ht]
    \centering
    \includegraphics[width=\linewidth]{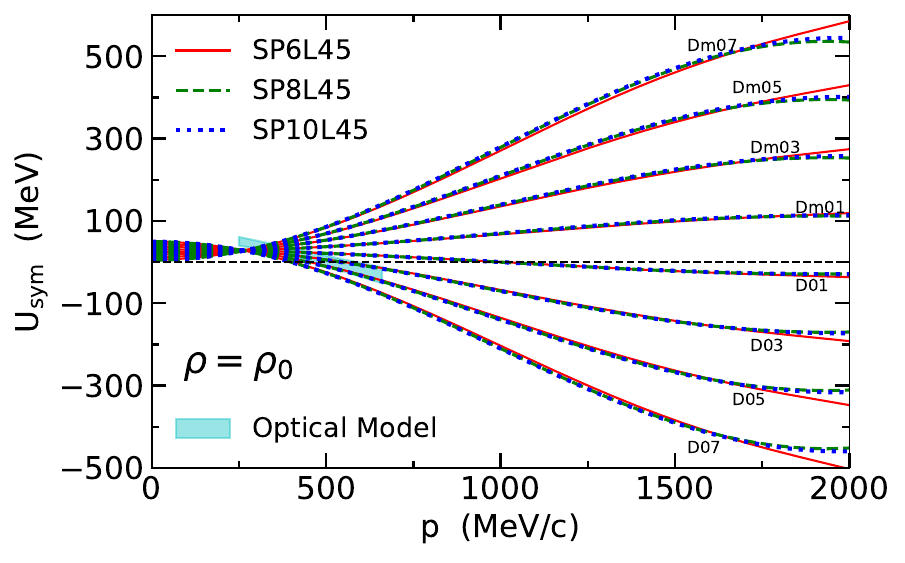}
    \caption{The momentum dependence of the symmetry potential at saturation density with various interactions.
    Also included are the results from global optical model analyses \cite{Li:2014qta,Xu:2010fh}.}
    \label{fig:Usym_compare}
\end{figure}

The momentum dependence of the symmetry potential is determined by $b_n$ parameters, whose values can be obtained through the parametrization in Eq.~(\ref{eq:bn_taylor}) and the $A$ values listed in Table~\ref{tab:A_m1}.
Figure~\ref{fig:Usym_compare} displays the symmetry potential $U_{\mathrm{sym}}(\rho_0,p)$ as a function of momentum for different interactions.
For the same $\Delta m_{1}^{\ast}(\rho_0)$, $U_{\mathrm{sym}}(\rho_0,p)$ obtained from SP6, SP8 and SP10 models are almost consistent within the momentum range up to $2$ $\mathrm{GeV}/c$.
Also shown in Fig.~\ref{fig:Usym_compare} is the momentum dependence of $U_{\mathrm{sym}}(\rho_0,p)$ obtained from a global optical model analyses of the nucleon-nucleus scattering data at beam energies from $0.05$ to $200$~MeV \cite{Li:2014qta,Xu:2010fh}, and the value of $\mspl(\rho_0,\delta)$ is predicted to be $(0.41\pm0.15)\delta$.
It can be seen from Fig.~\ref{fig:Usym_compare} that, through this parametrization method, the predicted $U_{\mathrm{sym}}(\rho_0,p)$ for interactions with $\Delta m_{1}^{\ast}(\rho_0)$ values being $0.3$ and $0.5$, i.e., SP6L45D03, SP8L45D03, SP10L45D03 and SP6L45D05, SP8L45D05, SP10L45D05 are consistent with the results from the optical model.
The isovector nucleon effective mass $m^{\ast}_{v}$ is commonly used to describe the momentum dependence of $U_{\mathrm{sym}}(\rho,p)$.
In the conventional Skyrme interaction, the linear isospin splitting coefficient $\Delta m_{1}^{\ast}(\rho)$ and the fourth-order symmetry energy $E_{\mathrm{sym},4}(\rho)$ have straightforward and interesting relations with $m^{\ast}_{s}(\rho)$ and $m^{\ast}_{v}(\rho)$ \cite{Zhang:2015qdp,Pu:2017kjx}.
However, in the N3LO (as well as N4LO and N5LO) Skyrme pseudopotential, since $m_{s}(\rho,p)$ and $m_{v}(\rho,p)$ are momentum-dependent, these relations become more complex \cite{Wang:2023zcj}.
In Table~\ref{tab:Esym_paras}, we present the values of $m^{\ast}_{v,0}$ and $E_{\mathrm{sym},4}(\rho_0)$ of the different interactions.

\begin{figure}[ht]
    \centering
    \includegraphics[width=\linewidth]{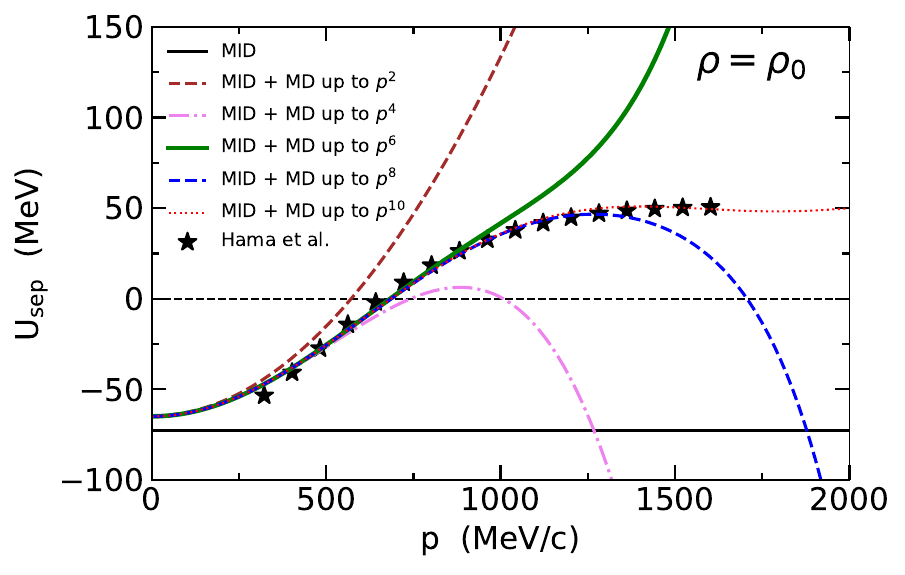}
    \caption{
    The momentum dependence of the single-nucleon potential at saturation density from the MID part as well from the sum of the MID part and the MD part up to different orders in the SP10 model.
    Also shown are the nucleon optical potential obtained by Hama \textit{et al.}~\cite{Hama:1990vr,Cooper:1993nx}.
    }
    \label{fig:U0_pw}
\end{figure}

\begin{figure}[ht]
    \centering
    \includegraphics[width=\linewidth]{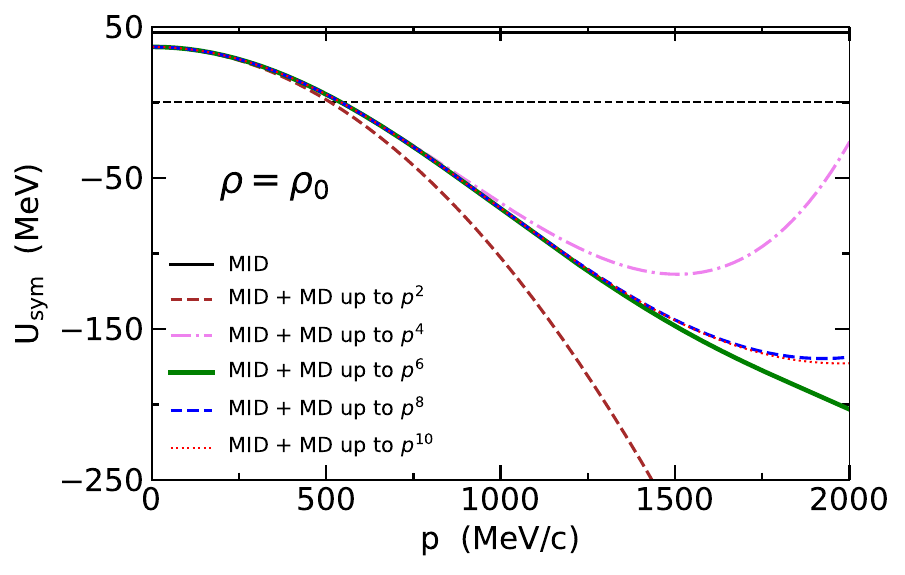}
    \caption{
    The momentum dependence of the symmetry potential at saturation density from the MID part as well from the sum of the MID part and the MD part up to different orders in the SP10L45D03 model.}
    \label{fig:Usym_pw}
\end{figure}

Shown in Fig.~\ref{fig:U0_pw} are the cumulative contributions from the MD terms up to different orders for the single-nucleon potential in the SP10 model.
Also included in Fig.~\ref{fig:U0_pw} are the nucleon optical potential obtained by Hama \textit{et al.}~\cite{Hama:1990vr,Cooper:1993nx}.
It can be seen from Fig.~\ref{fig:U0_pw} that the contribution from $p^{10}$ becomes significant when the nucleon momentum exceeds $1.5$~GeV/$c$.
The results for the symmetry potential with the SP10L45D03 model are presented in Fig.~\ref{fig:Usym_pw}, where it can be observed that $p^8$ is necessary for the symmetry potential to converge up to $2$~GeV/$c$.
It can be seen from Figs~\ref{fig:U0_pw} and \ref{fig:Usym_pw} that the higher-order momentum-dependence is crucial for the single-nucleon potential of high-momentum nucleons, although their contributions to the nuclear matter bulk properties around the saturation density are negligible.


\subsection{Neutron star properties}
In the present work, we assume the core of neutron stars consists of free neutrons, protons, electrons and possible muons ($npe\mu$ matter) without phase transition and other degrees of freedom at high densities.
The EOS of the $\beta$-equilibrium and charge-neutral neutron star core is derived based on the different interactions previously constructed in this paper.
The core-crust transition density $\rho_t$ is obtained self-consistently using the dynamical methods \cite{Xu:2009vi}, and the critical density separating the inner and the outer crust is taken to be $\rho_{\mathrm{out}}=2.46\times10^{-4}\,\mathrm{fm}^{-3}$ \cite{Carriere:2002bx,Xu:2008vz,Xu:2009vi}.
For the outer crust, where $\rho<\rho_{\mathrm{out}}$, we use the EOS of BPS (FMT) \cite{1971ApJ...170..299B};
for the inner crust, where $\rho_{\mathrm{out}}<\rho<\rho_t$, we construct the EOS by interpolation with the form \cite{Carriere:2002bx,Xu:2008vz,Xu:2009vi}:
\begin{equation}
\label{eq:innerCrust}
P=a+b \epsilon^{4/3} .
\end{equation}

\begin{figure*}[ht]
    \centering
    \includegraphics[width=\linewidth]{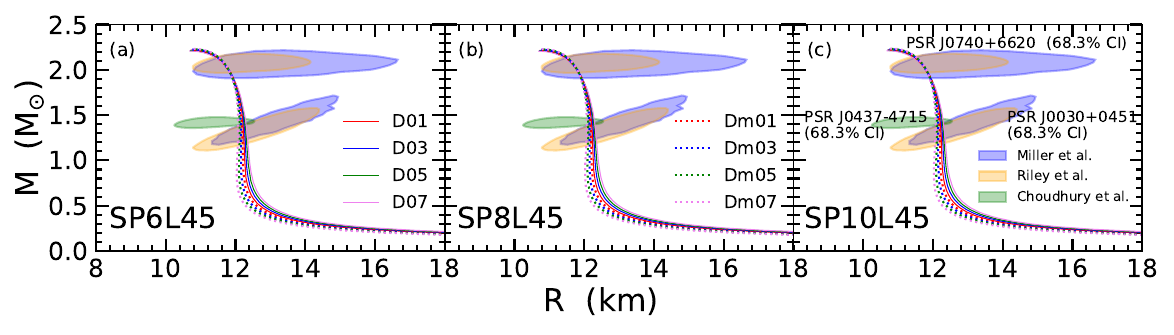}
    \caption{M-R relation for static neutron stars from different interactions, categorized according to different models (see text for the details).
    The NICER (XMM-Newton) constraints for PSR J0030+0451 \cite{Miller:2019cac,Riley:2019yda}, PSR J0740+6620 \cite{Miller:2021qha,Riley:2021pdl} and PSR J0437-4715 \cite{Choudhury:2024xbk} are also included for comparison.
    All contours are plotted for 68.3\% CI.}
    \label{fig:MR}
\end{figure*}

With the EOS of neutron star matter $P(\epsilon)$, one can obtain the mass-radius (M-R) relation of static neutron stars by solving the famous Tolman-Oppenheimer-Volkoff (TOV) equation \cite{Tolman:1939jz,Oppenheimer:1939ne}.
Shown in Fig.~\ref{fig:MR} are the mass-radius relations of neutron stars obtained using different interactions.
The simultaneous mass-radius determinations for PSR J0030+0451 \cite{Miller:2019cac,Riley:2019yda} and PSR J0437-4715 \cite{Choudhury:2024xbk}, both with a mass around $1.4\,M_{\odot}$, as well as for PSR J0740+6620 \cite{Miller:2021qha,Riley:2021pdl} with a mass around $2.0\,M_{\odot}$, obtained from NICER (XMM-Newton) are also plotted for $68.3\%$ credible intervals (CI) in Fig.~\ref{fig:MR} for comparison.
As shown in Fig.~\ref{fig:MR}, all of the 24 interactions are compatible with the constraint for PSR J0030+0451, PSR J0437-4715 and PSR J0740+6620, falling within the $68.3\%$ CI.
From Figs.~\ref{fig:Epnm} and \ref{fig:MR}, it can be observed that interactions with smaller $\Delta m_{1}^{\ast}(\rho_0)$ predict smaller radii for low-mass neutron stars due to their smaller symmetry energy around and below saturation density.

Table.~\ref{tab:NS} summarizes the core-crust transition density $\rho_t$, the central density $\rho_{\mathrm{cen}}^{\mathrm{TOV}}$ and mass $M_{\mathrm{TOV}}$ of the maximum mass neutron-star configuration and the dimensionless tidal deformability $\Lambda_{1.4}$ of $1.4M_{\odot}$ neutron stars obtained with different interactions.
It can be seen from Table.~\ref{tab:NS} that within the range of $-0.7$ to $0.7$, the nucleon effective mass splitting have little impact on these properties of neutron stars, and $\Lambda_{1.4}$ in all interactions complies with the limit of $\Lambda_{1.4} \leqslant 580$ from the gravitationalwave signal GW170817 \cite{LIGOScientific:2018cki}.
\begin{table*}
\caption{\label{tab:NS}
    Core-curst transition density ($\rho_t$), the central density ($\rho_{\mathrm{cen}}^{\mathrm{TOV}}$) and mass ($M_{\mathrm{TOV}}$) of the maximum mass neutron star configuration and the dimensionless tidal deformability ($\Lambda_{1.4}$) of $1.4M_{\odot}$ neutron star for different interactions.
}
\begin{ruledtabular}
\begin{tabular}{ccccccccc}
SP6L45&Dm07&Dm05&Dm03&Dm01&D01&D03&D05&D07 \\
$\rho_t$ ($\mathrm{fm}^{-3}$) &
$0.0821$&$0.0817$&$0.0813$&$0.0809$&$0.0806$&$0.0802$&$0.0799$&$0.0796$\\
$\rho_{\mathrm{cen}}^{\mathrm{TOV}}$ ($\mathrm{fm}^{-3}$) &
$1.03$&$1.03$&$1.03$&$1.04$&$1.04$&$1.04$&$1.05$&$1.05$\\
$M_{\mathrm{TOV}}/M_{\odot}$&
$2.23$&$2.23$&$2.22$&$2.22$&$2.22$&$2.21$&$2.21$&$2.21$\\
$\Lambda_{1.4}$&
$380.4$&$383.2$&$385.8$&$388.0$&$390.0$&$391.8$&$393.4$&$394.8$\\ \hline
SP8L45&Dm07&Dm05&Dm03&Dm01&D01&D03&D05&D07 \\
$\rho_t$ ($\mathrm{fm}^{-3}$) &
$0.0822$&$0.0817$&$0.0813$&$0.0809$&$0.0805$&$0.0802$&$0.0798$&$0.0795$\\
$\rho_{\mathrm{cen}}^{\mathrm{TOV}}$ ($\mathrm{fm}^{-3}$) &
$1.03$&$1.03$&$1.03$&$1.04$&$1.04$&$1.05$&$1.05$&$1.05$\\
$M_{\mathrm{TOV}}/M_{\odot}$&
$2.23$&$2.23$&$2.22$&$2.22$&$2.21$&$2.21$&$2.21$&$2.20$\\
$\Lambda_{1.4}$&
$379.4$&$382.5$&$385.2$&$387.6$&$389.7$&$391.6$&$393.2$&$394.8$\\ \hline
SP10L45&Dm07&Dm05&Dm03&Dm01&D01&D03&D05&D07 \\
$\rho_t$ ($\mathrm{fm}^{-3}$) &
$0.0822$&$0.0817$&$0.0813$&$0.0809$&$0.0805$&$0.0802$&$0.0798$&$0.0795$\\
$\rho_{\mathrm{cen}}^{\mathrm{TOV}}$ ($\mathrm{fm}^{-3}$) &
$1.03$&$1.03$&$1.03$&$1.04$&$1.04$&$1.05$&$1.05$&$1.05$\\
$M_{\mathrm{TOV}}/M_{\odot}$&
$2.23$&$2.23$&$2.23$&$2.22$&$2.22$&$2.22$&$2.21$&$2.21$\\
$\Lambda_{1.4}$&
$379.6$&$382.7$&$385.4$&$387.8$&$389.9$&$391.7$&$393.4$&$394.9$\\
\end{tabular}
\end{ruledtabular}
\end{table*}

\begin{figure*}[ht]
    \centering
    \includegraphics[width=\linewidth]{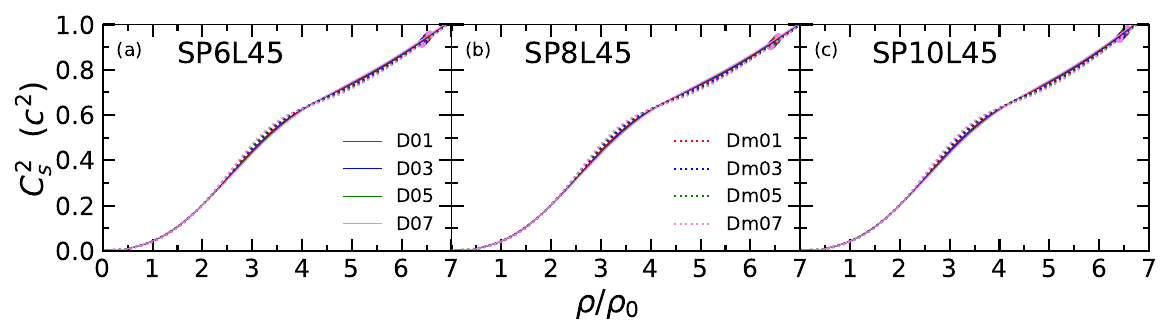}
    \caption{The squared sound speed ($C_{s}^{2}\equiv dP/d\epsilon$) of neutron star matter as a function of nucleon density predicted by different interactions, categorized according to different models (see text for the details).
    The central density of maximum mass neutron star configuration is indicated by ``$\bullet$''.}
    \label{fig:Cs2}
\end{figure*}

The sound speed has widely used to characterize the properties of dense matter.
Figure~\ref{fig:Cs2} shows the squared sound speed $C_{s}^{2}\equiv dP/d\epsilon$ of the neutron star matter as a function of nucleon density for the different interactions.
Also shown in Fig.~\ref{fig:Cs2} is the central density of maximum mass neutron star configuration with different interactions, and it can be seen that the causality condition $C_{s}^{2}\leq c^2 $ is satisfied by all the interactions.
From Fig.~\ref{fig:Cs2}, it can be observed that the nucleon effective mass splitting has negligible impact on the $C_{s}^{2}\equiv dP/d\epsilon$ of the neutron star matter, and the effect of different single-nucleon potentials is even more insignificant.

\section{Microscopic transport model simulations of HICs}
\label{sec:LBUU}
These new constructed N$n$LO interactions can be employed into BUU equations to study the HICs.
In this section, we briefly outline the methods we adopt to solve the BUU equation, including the lattice Hamiltonian method to deal with the mean-field evolution and the stochastic approach to handle the collision term.
As a benchmark, we select a series of interactions to simulate the fixed-target Au+Au collisions at $E_{\mathrm{beam}}=1.23$~GeV/nucleon (corresponding to $\sqrt{s_{N N}}=2.4 \,\mathrm{GeV}$), and compare the predicted proton flows with the experimental data measured by the HADES collaboration.
We would like to mention that a series of studies has been conducted recently to simulate the HADES experiment using various transport models~\cite{Hillmann:2019wlt,Parfenov:2022brq,Du:2023ype,Fang:2023sna,Li:2023ydi,Liu:2023rlm,Mohs:2024gyc,Lan:2024flu,Steinheimer:2024eha,Kireyeu:2024hjo,Reichert:2024ayg}.

\subsection{The lattice BUU transport model}
The time evolution of the one-body phase-space distribution function (Wigner function) $f_{\tau}$ $=$ $f_{\tau}(\vec{r},\vec{p},t)$ satisfies the BUU equation.
In the present work, we solve the following BUU equation with a momentum-dependent mean-field potential, i.e.,
\begin{equation}
(\partial_t+\vec{\nabla}_p\epsilon_{\tau} \cdot\vec{\nabla}_{r}-\vec{\nabla}_{r}\epsilon_{\tau}\cdot \vec{\nabla}_{p})f_{\tau} = I^{\rm coll}_{\tau}[f_n,f_p,f_{\Delta},f_{\pi}],
\label{eq:BUU}
\end{equation}
where $\tau$ represent different particle species, i.e., neutrons $n$, protons $p$, $\Delta$-resonances and $\pi$-mesons.
The $\epsilon_{\tau}$ is the single-particle energy of the particle species $\tau$, and it contains the kinetic part and MD mean-field potential $U(\vec{r},\vec{p})$.
Note $U(\vec{r},\vec{p})$ should be treated as a functional of $f_{\tau}$.
The $I_{\tau}^{\rm coll}$ is the collision integral, which takes into account the effect of quantum statistics, i.e., Pauli blocking for Fermions and Bose enhancement for Bosons.
It consists of the gain term ($<$) and the loss term ($>$), i.e.,
\begin{equation}
    I_{\tau}^{\rm coll} = {\cal K}_{\tau}^{<}[f_n,f_p,f_{\Delta},f_{\pi}](1\pm f_{\tau}) - {\cal K}_{\tau}^{>}[f_n,f_p,f_{\Delta},f_{\pi}]f_{\tau}.
\end{equation}
The factor $1\pm f_{\tau}$ in the gain term represents the effect from quantum statistics, and the plus sign is for Bosons while the minus sign for Fermions.
In practice, since the $f_{\tau}$ for Bosons is usually very small, we omit the effect of Bose enhancement.
The gain term and loss term contain contributions from different scatterings, namely, two-body elastic scatterings, $NN$ $\leftrightarrow$ $N\Delta$ and $\Delta(N^*,\Delta^*)$ $\leftrightarrow$ $N\pi$, whose scattering matrix can be deduced from their measured cross sections.

Based on the test particle method \cite{Wong:1982zzb}, the $f$ can be mimicked by $A \times N_{\mathrm{E}}$ test particles with a form factor $S$ in the coordinate space, i.e.,
\begin{equation}
f_{\tau}\left(\vec{r}, \vec{p}, t\right) = \frac{(2 \pi \hbar)^{3}}{N_{\mathrm{E}}} \sum_{i}^{AN_{\rm{E}}\atop\tau} S\left[\vec{r}_{i}(t)-\vec{r}\right] \delta\left[\vec{p}_{i}(t)-\vec{p}\right],
\label{eq:testP}
\end{equation}
where $A$ is the mass number of the system and $N_{\mathrm{E}}$ is the number of ensembles or number of test particles.
The sum in the above expression runs over all test particles with isospin $\tau$.

The mean-field evolution of the BUU equation is solved by the lattice Hamiltonian method \cite{Lenk:1989zz,Wang:2019ghr}, where the total Hamiltonian $H$ can be approximated by the lattice Hamiltonian $H_{L}$, i.e.,
\begin{equation}
\label{eq:H_Lattice}	
H=\int \mathcal{H}(\vec{r}) \mathrm{d} \vec{r} \approx l^{3} \sum_{\alpha} \mathcal{H}\left(\vec{r}_{\alpha}\right) \equiv H_{L},
\end{equation}
where $\vec{r}_{\alpha}$ represents the coordinate of certain lattice site $\alpha$, and $l$ is the lattice spacing.
In the above expression, $\mathcal{H}$ is the Hamiltonian density including both nuclear interaction [Eq.~(\ref{eq:Hdensity})] and Coulomb interaction.
By substituting Eq.~(\ref{eq:testP}) into Eq.~(\ref{eq:Hdensity}) and then into Eq.~(\ref{eq:H_Lattice}), the coordinates $\vec{r}_i$ and momenta $\vec{p}_i$ of test particles can be treated as canonical variables of the lattice Hamiltonian $H_L$, and the time evolution of $\vec{r}_i(t)$ and $\vec{p}_i(t)$ are then governed by the Hamilton equation of total lattice Hamiltonian of all ensembles.
It therefore ensures the conservation of energy during the dynamic process.
With Eq.~(\ref{eq:testP}), the time evolution of $\vec{r}_i(t)$ and $\vec{p}_i(t)$ then give the time evolution of $f(\vec{r},\vec{p},t)$, from which one obtains expectations of certain quantities.

Within the test particle method, the integral when calculating the MD part of the single-particle potential, i.e., Eqs.~(\ref{E:UMD1}) and (\ref{E:UMDt1}), can be converted to a summation over test particles, for $U^{\mathrm{md}[2n]}(\vec{r},\vec{p})$ one has
\begin{equation}
\begin{split}
U^{\mathrm{md}[2n]}(\vec{r},\vec{p}) & = \int \frac{{\rm d}^3 p^{\prime}}{(2\pi\hbar)^3} \left(\vec{p}-\vec{p}\,{}^{\prime}\right)^{2n} f(\vec{r},\vec{p}\,{}^{\prime})\\
 & = \frac{1}{N_E}\sum_i^{AN_E}(\vec{p}-\vec{p}_i)^{2n}S(\vec{r}_i-\vec{r}),
\end{split}
\end{equation}
and for $U_{\tau}^{\mathrm{md}[2n]}(\vec{r},\vec{p})$ the summation only applies to test particles of the given $\tau$.
One of the advantages of the polynomial form of the MD part of the single-particle potential with N$n$LO, compared with the logarithm form (see, e.g., Refs.~\cite{Aichelin:1987ti,Zhang:2006vb,Feng:2011pu,Wang:2014rva,Liu:2020jbg}) or Lorentzian form (see, e.g., Refs.~\cite{Das:2002fr,Chen:2004si,Isse:2005nk,Rizzo:2008zz,Cozma:2013sja,Su:2016iif,Ikeno:2023cyc}) used in various transport models, is that its $\vec{p}$-dependence can be factored out from the integral or summation, as shown in Appendix~\ref{S:A1}.
This feature significantly reduces the computational complexity of the transport model when calculating the MD part of single-particle potentials, if a very large $N_E$ is adopted.
Note that a sufficiently large $N_E$ is essential when calculating certain quantities, e.g. the width of giant dipole resonance~\cite{Wang:2020xgk}.
It should be noted that, also in the QMD models, the single-particle potential with forms as Eqs.~(\ref{E:UMD1}) and (\ref{E:UMDt1}) helps to reduce the computational complexity compared with the logarithm form and Lorentzian form \cite{Yang:2023umm}.

We adopt the stochastic approach \cite{Danielewicz:1991dh,Wang:2020ixf} to deal with the collision terms $I^{\rm coll}_{\tau}$.
The collision probability $P_{ij}$ of the $i$-th and $j$-th test particles during a time interval $\Delta t$ can be derived directly from the lost term of $I^{\rm coll}_{\tau}$.
For example, for nucleon-nucleon elastic scatterings, one has
\begin{equation}
\label{eq:P_ij}
P_{ij}=v_{\mathrm{rel}} \sigma_{\mathrm{NN}}^{\ast} S\left(\vec{r}_{i}-\vec{r}_{\alpha}\right) S\left(\vec{r}_{j}-\vec{r}_{\alpha}\right) l^{3} \Delta t ,
\end{equation}
where $v_{\mathrm{rel}}$ is the relative velocity of the test particles and $\sigma_{\mathrm{NN}}^{\ast}$ is the in-medium nucleon-nucleon cross-section.
The factor $\left[1-f\left(\vec{r}_{\alpha}, \vec{p}_{i}{}^{\prime}\right)\right]\left[1-f\left(\vec{r}_{\alpha}, \vec{p}_{j}{}^{\prime}\right)\right]$ is calculated according to their final state $\vec{p}_{i}{}^{\prime}$ and $\vec{p}_{j}{}^{\prime}$ to determine whether the collision is blocked by the Pauli principle.

The Thomas-Fermi initialization is applied to obtain the ground state of the nuclei (see Refs.~\cite{Wang:2019ghr,Wang:2020ixf} and the references therein for the details).
The gradient parameter $E^{[2]}$ is modified for each interaction to reproduce the experiment binding energy of the ground state nuclei.
For simplicity, we omit the higher-order $E^{[n]}$ parameters and all $F^{[n]}$ parameters.
The Thomas-Fermi method actually provides a static solution of the BUU equation, it therefore ensures the stability of the ground state evolution.

\subsection{Comparisons to HADES collective flow data}
As mentioned above, we extend the central term of the Skyrme effective interactions to provide additional momentum/energy dependence of the single-nucleon potential, which is displayed in Fig.~\ref{fig:U0_compare}.
To focus on the effects of the momentum dependence of the single-nucleon potential, we fix $\Delta m_{1}^{\ast}(\rho_{0})=0.3$ (i.e., fix the shape of the symmetry potential in Fig.~\ref{fig:Usym_compare}) for the interactions used in the simulations, namely we choose SP6L45D03, SP8L45D03 and SP10L45D03.
For brevity, in the following text and figures, these interactions will be referred to SP6, SP8 and SP10, respectively.
The values of $m_{s,0}^{\ast}$ given by these three interactions are all approximately $0.77m$ (see Table~\ref{tab:an_compare}), because their single-nucleon potentials have similar behaviors below $1$ GeV.
Additionally, to investigate the effect of $m_{s,0}^{\ast}$, we also employ another single-nucleon potential constructed based on the N3LO model from Ref.~\cite{Wang:2023zcj}, for which the value of $m^{\ast}_{s,0}$ equals to $0.83m$, and we label the corresponding interaction as SP6Ms83.
The $a_{n}$ parameters of SP6Ms83 are listed in Table~\uppercase\expandafter{\romannumeral 4} in Ref.~\cite{Wang:2023zcj}, while other adjustable quantities are the same as those in the SP6 interaction.
The single-nucleon potential of SP6Ms83 is shown in Fig.~\ref{fig:U0_compare}, and it can be seen that SP6Ms83 predicts a weaker energy dependence than other three interactions and Hama's data, while aligning with the Dirac-Brueckner calculations \cite{TerHaar:1986xpv}.

To evaluate the validity and establish a benchmark, we solve the BUU equation with those four interactions to simulate the fixed-target Au+Au collisions at $E_{\mathrm{beam}}=1.23$~GeV/nucleon (corresponding to $\sqrt{s_{N N}}=2.4 \,\mathrm{GeV}$) conducted by HADES collaboration \cite{HADES:2020lob,HADES:2022osk}.
Some details of the simulations are as follow:
the values of $E^{[2]}$ for SP6, SP8, SP10 and SP6Ms83 are $-305$, $-300$, $-310$ and $-310 \, \mathrm{MeV} \, \mathrm{fm}^{5}$, respectively, obtained by fitting the experimental binding energy of $^{197}$Au;
the number of test particles is set to be $N_{E} = 100000$;
the factor $S(\vec{r}_{i}-\vec{r})$ in Eq.~(\ref{eq:testP}) is chosen to be triangle form \cite{Wang:2019ghr,Wang:2020ixf};
the end of the time evolution is set to $60$ fm/$c$, with a time step of $0.2$ fm/$c$;
a fixed impact parameter $b=7.4 \, \mathrm{fm}$ is adopted for the centrality class of $20\%$-$30\%$ \cite{HADES:2017def};
the free nucleon-nucleon cross section $\sigma_{\mathrm{NN}}^{\mathrm{free}}$ is adopted based on the parametrization of experimental nucleon-nucleon scattering data as in Ref.\cite{Cugnon:1996kh}, and we have tested that the in-medium correction to $\sigma_{\mathrm{NN}}$ has very limited impact on the simulation results;
the nucleon resonances and $\Delta$ resonances are included up to $N(1720)$ and $\Delta(1950)$, respectively;
the scatterings related to $\Delta$ resonances and pions, i.e., $N N \leftrightarrow N \Delta$ and $\Delta \leftrightarrow N \pi$, as well as $N^*(\Delta^*)$ $\leftrightarrow$ $N\pi$ are included with the standard cross sections and decay width taken from Refs.~\cite{SMASH:2016zqf,TMEP:2023ifw}.
When solving the BUU equation, we omit the single-particle potential of $\pi$ and all other resonances except $\Delta$.
It should be noted that the isospin dependence of the $\Delta$ single-particle potentials is still very elusive~\cite{Drago:2014oja,Cai:2015hya,Cozma:2014yna,Li:2015hfa}, and there are two popular forms in the transport model simulations for heavy-ion collisions, with one proposed in Ref.~\cite{UmaMaheswari:1997mc} and the other in Refs.~\cite{Li:2002qx,Li:2002yda}, and both are expressed as a liner scaling from those of $n$ and $p$.
In this work, we assume the former form for the single-particle potentials of $\Delta$ ~\cite{UmaMaheswari:1997mc}, i.e., $U_{\Delta^{++}}= -U_n + 2U_p$, $U_{\Delta^{+}} = U_p$, $U_{\Delta^{0}} =  U_n$ and $U_{\Delta^{-}} = 2U_n - U_p$.
Note that the above $\Delta$ potentials do not cause a difference in the potential energy between the initial and final states of the scattering $NN$ $\leftrightarrow$ $N\Delta$, while their effects on the energy conservation of the scattering $\Delta$ $\leftrightarrow$ $N\pi$, i.e., the threshold effects~\cite{Ferini:2005del,Song:2015hua,Zhang:2017nck,Ikeno:2023cyc}, should be taken into account when solving the BUU equation.

The particles transverse momentum $p_{\mathrm{t}}$ anisotropy is resulted from the pressure anisotropy of the compressed matter formed during the noncentral HICs, and is thus sensitive to the EOS of dense nuclear matter from the collision products.
The anisotropic flows $v_{n}$ of particles are the Fourier coefficients in the decomposition of their $p_{\mathrm{t}}$ spectra in the azimuthal angle $\phi$ with respect to the reaction plane \cite{Poskanzer:1998yz}, i.e.,
\begin{equation}
\label{eq:flow_n}
E \frac{d^3 N}{d p^3}= \frac{1}{2\pi}\frac{d^{2} N}{p_{\mathrm{t}} d p_{\mathrm{t}} d y } \left[ 1 + \sum_{n=1}^{\infty} 2 v_{n} (p_{\mathrm{t}},y) \cos{(n \phi)} \right].
\end{equation}
The anisotropic flows $v_{n}$ generally depend on particle transverse momentum $p_{\mathrm{t}}$ as well as rapidity $y$, and for a given $y$ the anisotropic flows at $p_{\mathrm{t}}$ can be evaluated according to
\begin{equation}
\label{eq:flow_average}
v_{n} (p_{\mathrm{t}}) = \left \langle \cos{(n \phi)} \right \rangle,
\end{equation}
where $\langle \cdots \rangle$ denotes average over the azimuthal distribution of particles with transverse momentum $p_{\mathrm{t}}$.
The anisotropic flows $v_{n}$ can further be expressed in terms of the single-particle averages \cite{Chen:2004dv,Chen:2004vha}:
\begin{align}
\label{eq:flow_expressions}
v_{1}(p_{\mathrm{t}}) &= \left \langle \frac{p_{x}}{p_{\mathrm{t}}} \right \rangle, \\
v_{2}(p_{\mathrm{t}}) &= \left \langle \frac{p_{x}^{2}-p_{y}^{2}}{p_{\mathrm{t}}^{2}} \right \rangle, \\
v_{3}(p_{\mathrm{t}}) &= \left \langle \frac{p_{x}^{3}- 3 p_{x} p_{y}^{2}}{p_{\mathrm{t}}^{3}} \right \rangle, \\
v_{4}(p_{\mathrm{t}}) &= \left \langle \frac{p_{x}^{4}- 6 p_{x}^{2} p_{y}^{2} + p_{y}^{4} }{p_{\mathrm{t}}^{4}} \right \rangle,
\end{align}
where $p_{x}$ and $p_{y}$ are, respectively, the projections of particle momentum in and perpendicular to the reaction plane.

\begin{figure}[ht]
    \raggedleft
    \includegraphics[width=\linewidth]{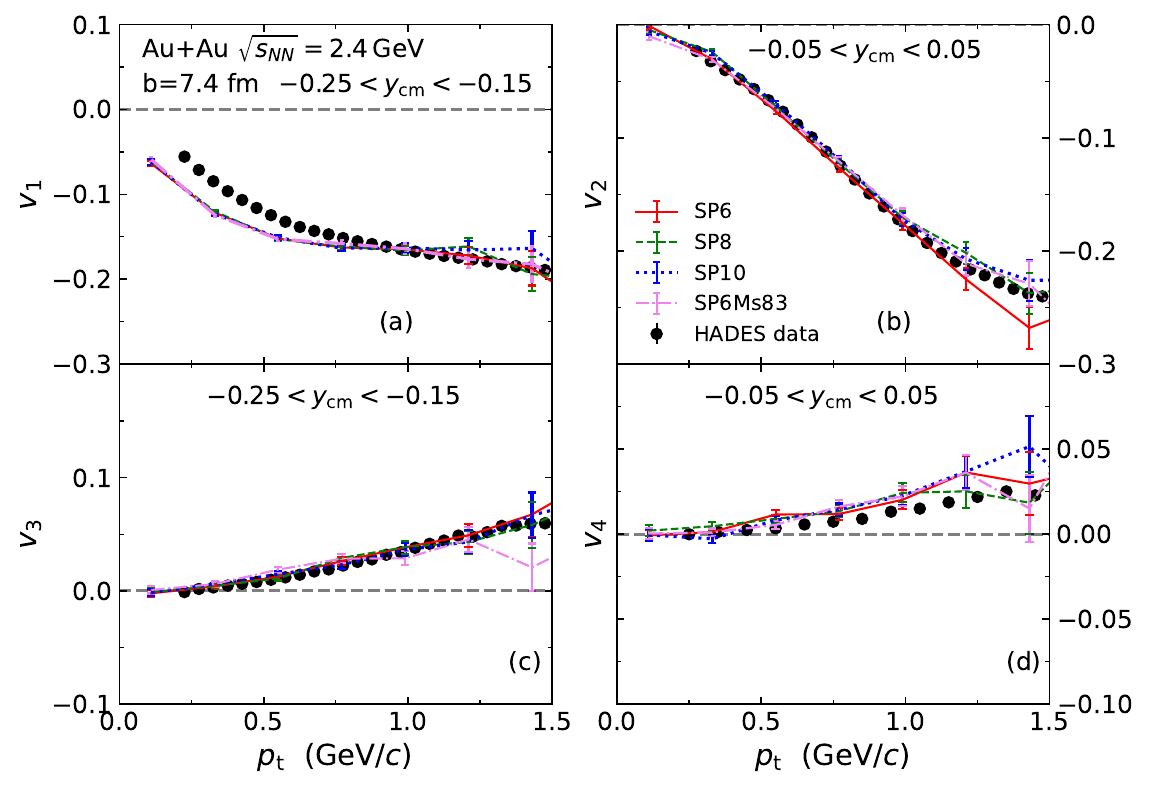}
    \caption{
    Directed ($v_1$), elliptic ($v_2$), triangular ($v_3$) and quadrangular ($v_4$) flows as function of transverse momentum ($p_{\mathrm{t}}$) for free protons in Au+Au collisions at $E_\mathrm{beam}=1.23$~GeV/nucleon (corresponding to $\sqrt{s_{N N}}=2.4 \,\mathrm{GeV}$) predicted by lattice BUU model with SP6, SP8, SP10 and SP6Ms83 interactions.
    Also included are the HADES data \cite{HADES:2020lob,HADES:2022osk}.
    }
	\label{fig:ptall}
\end{figure}

\begin{figure}[ht]
	\raggedright
	\includegraphics[width=\linewidth]{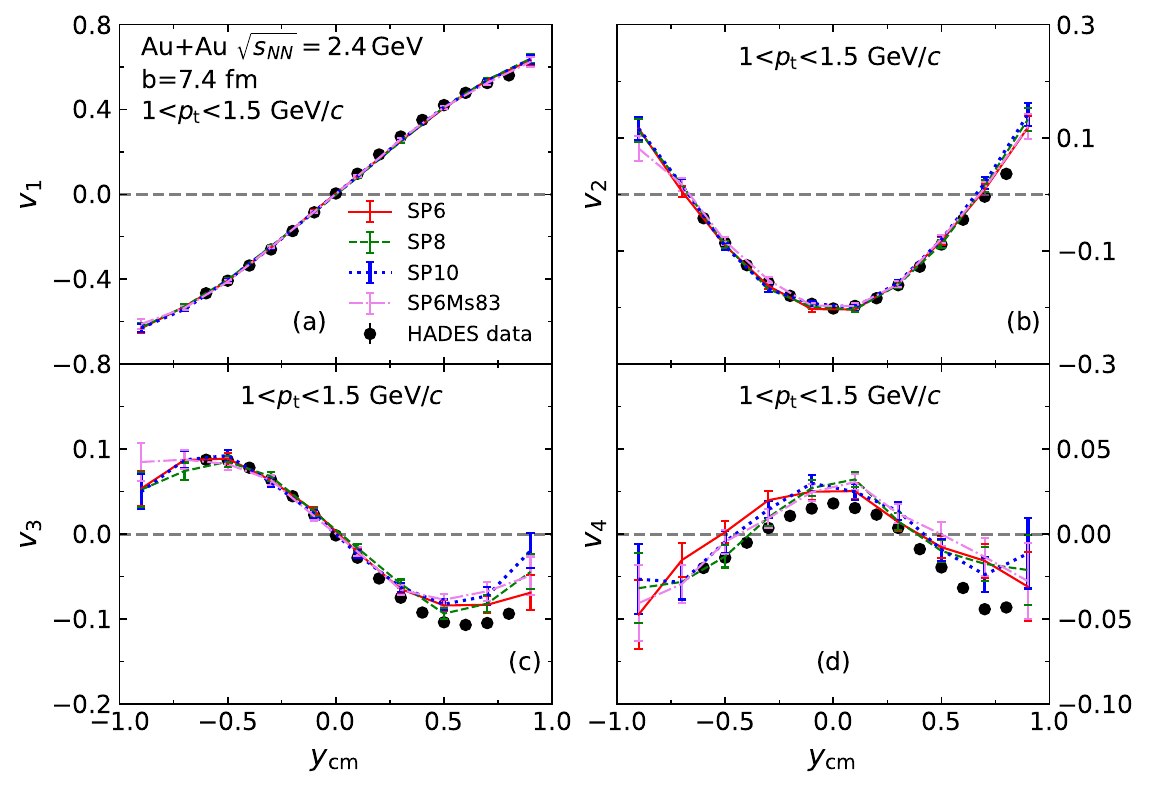}
	\caption{
    Directed ($v_1$), elliptic ($v_2$), triangular ($v_3$) and quadrangular ($v_4$) flows as function of center-of-mass rapidity ($y_{\mathrm{cm}}$) for free protons in Au+Au collisions at $E_\mathrm{beam}=1.23$~GeV/nucleon (corresponding to $\sqrt{s_{N N}}=2.4 \,\mathrm{GeV}$) predicted by lattice BUU model with SP6, SP8, SP10 and SP6Ms83 interactions.
    Also included are the HADES data~\cite{HADES:2020lob,HADES:2022osk}.
    }
	\label{fig:rapall}
\end{figure}

We present in Fig.~\ref{fig:ptall} the lattice BUU results for directed flow $v_1$, elliptic flow $v_2$, triangular flow $v_3$ and quadrangular flow $v_4$ for protons as function of transverse momentum $p_{\mathrm{t}}$.
All of these predictions are generally in good agreement with the HADES data.
It is seen from Fig.~\ref{fig:ptall}(b) that SP6Ms83 predicts a smaller magnitude of $v_2$, whereas SP6 predicts a larger magnitude of $v_2$, while SP8 and SP10 are nearly identical.
This difference is due to the different energy dependencies of the single-nucleon potential.
It is seen from Fig.~\ref{fig:U0_compare} that SP6Ms83 predicts a larger range of negative single-nucleon potential, and the attractive potential will weaken the $v_2$.
In contrast, SP6 shows a rapid increase in the single-nucleon potential above $1$ GeV, leading to a stronger repulsive potential for high-energy particles and, consequently, a larger magnitude of $v_2$ at high $p_{\mathrm{t}}$.
In Fig.~\ref{fig:rapall}, we show the predicted proton flows, including $v_1$, $v_2$, $v_3$ and $v_4$, as a function of center-of-mass rapidity $y_{\mathrm{cm}}$, and all of them are generally in good agreement with the HADES data.
It is seen from Fig.~\ref{fig:rapall}(b) that at mid-rapidity, SP6Ms83 predicts the smallest magnitude of $v_2$, SP6 predicts the largest, and the predictions of SP8 and SP10 are almost identical and fall in between.
This is consistent with the results presented in Fig.~\ref{fig:ptall}(b).

It should be noted that these results serve as a benchmark and broadly illustrate the impact of the single-nucleon potential.
In fact, the predicted proton flows can also be affected by the nuclear EOS, which is fixed in these interactions by construction.
Particularly, the stiffness of the SNM EOS could directly affect the matter densities produced during collisions, thereby influencing the single-nucleon potential in the dynamic process.
Considering the joint effects of both nuclear EOS and the single-nucleon potential, extracting them from HICs will rely on future large-scale simulations.

Finally, we would like to mention that in the LBUU transport model, the momentum-dependent Skyrme pseudopotential is non-relativistic,
although the kinematics in Hamiltonian equations of motion is treated relativistically.
To the best of our knowledge, for transport model simulations for heavy-ion collisions at energies up to about $2$~GeV/nucleon, the non-relativistic mean-field potentials are still widely used in some popular transport model codes, e.g., IBUU~\cite{Li:2002yda,Du:2023ype,Wei:2024ncz}, updated ART~\cite{Yong:2022pyb}, UrQMD~\cite{LI:2005zi,Ye:2018vbc,Wang:2020dru,Liu:2020jbg,Liu:2025pzr}, QMD~\cite{Yang:2023umm}, and LQMD~\cite{Feng:2018emx,Liu:2023rlm}.
Furthermore, in the present LBUU model,
instead of the commonly used geometric method, the stochastic collision method is implemented for the two-body collision term in the BUU equation, and in this way the collision probability of two test nucleons can be derived directly from the two-body collision term in the BUU equation, and it thus significantly weakens the relativistic covariance effects which are very important in the geometric method where accurate distance of two scattering particles is necessary and the mean free path of a test particle should be much larger than the interaction length between the two scattering test particles.
As shown in this paper,
the LBUU transport model with empirical parameters can successfully reproduce the HADES experimental data, suggesting that at HADES energy ($1.23$~GeV/nucleon), the relativistic correction to the mean-field potential may not be so significant.
Of course, a detailed and quantitative check on the relativistic effects of mean-field potential will be definitely interesting and important, although it is not trivial at all (see, e.g., Ref.~\cite{Isse:2005nk}.

\section{SUMMARY AND OUTLOOK}
\label{sec:summary}
Based on a general approach conforming to the basic symmetry principles, we have extended the central terms of the conventional Skyrme interactions by incorporating specific higher-order derivative terms and obtain the general form of the Skyrme pseudopotential up to N$n$LO.
The corresponding expressions of Hamiltonian density and single-nucleon potential are derived within the Hartree-Fock approach under general non-equilibrium conditions.
The higher-order derivative terms can provide additional momentum dependence for the single-nucleon potential, especially affecting its behavior at high energies.
In particular, for the Skyrme pseudopotential up to N3LO, N4LO and N5LO, the corresponding single-nucleon potential can describe the empirical nucleon optical potential up to the energy of $1$ GeV, $1.5$ GeV and $2$ GeV, respectively.
The density-dependent term is also extended in the spirit of Fermi momentum expansion, enabling significant freedom to adjust the density dependence of the symmetry energy as well as the high-density behavior of the isospin symmetric nuclear matter.
Consequently, the corresponding nuclear EOS can simultaneously fit the empirical properties of symmetric nuclear matter, the microscopic calculations of pure neutron matter and the properties of neutron stars from astrophysical observations.
It is worth noting that the density behavior of the EOS corresponding to N3LO Skyrme pseudopotential is sufficiently flexible to satisfy the current constraints, and the introduction of higher-order characteristic parameters may help to extract the nuclear EOS at suprasaturation densities based on new available experimental probes in the future.
Specifically, for the Skyrme pseudopotential up to N5LO, the adjustable characteristic parameters will be extended up to $H_{0}$ and $H_{\mathrm{sym}}$, which correspond to the fifth-order expansion coefficients of $E_{0}(\rho)$ and $E_{\mathrm{sym}}(\rho)$, respectively.
Considering the large uncertainty in the symmetry potential, we construct eight symmetry potentials with different momentum dependencies, corresponding to the values of the linear isospin splitting coefficient $\Delta m_{1}^{\ast}(\rho_{0})$ ranging from $-0.7$ to $0.7$.
Combining three single-nucleon potential featuring different high-momentum behaviors obtained from the Skyrme pseudopotential up to N3LO, N4LO and N5LO, respectively, with the eight symmetry potentials, we have constructed a parameter sets, by which we study the properties of nuclear matter and neutron stars. Our results indicate that these interactions can well describe the neutron star mass-radius relation from NICER and the tidal deformability extracted from gravitational wave signal GW170817.

Furthermore, these interactions with $\Delta m_{1}^{\ast}(\rho_{0})=0.3$ have been applied in the lattice BUU model to simulate the Au+Au collisions at $1.23$~GeV/nucleon conducted by HADES collaboration.
Another interaction with larger $m^{\ast}_{s,0}$ has been also used for the simulation.
We find that these interactions can provide good predictions of the proton collective flows in the HADES experiments.
Moreover, we find that the proton elliptic flow $v_2$ is sensitive to the single-nucleon potential as well as $m^{\ast}_{s,0}$.
A larger $m^{\ast}_{s,0}$, which indicates a more attractive potential at low energies, will weaken the $v_2$ effects, while a stronger repulsive potential at high energies will result in a larger magnitude of $v_2$ at high $p_{\mathrm{t}}$.

In future, we will conduct large-scale simulations to extract the single-nucleon potential as well as the nuclear matter EOS at suprasaturation densities.
Considering that the hyperons production may be significant in heavy-in collisions at higher energies, we will incorporate the hyperon degrees of freedom in the lattice BUU model in the next step.

\begin{acknowledgments}
This work was supported in part by the National Natural Science Foundation of China under Grant No 12235010, the National SKA Program of China (Grant No. 2020SKA0120300), and the Science and Technology Commission of Shanghai Municipality (Grant No. 23JC1402700).
\end{acknowledgments}

\appendix

\begin{widetext}
\section{THE SKYRME PSEUDOPOTENTIAL UP TO N$n$LO}
\label{sec:App_NnLO}
\subsection{Hamiltonian density with the extended Skyrme interaction in one-body transport model of HICs}
In the transport model, considering the system being out of equilibrium, the Hamiltonian density needs to be expressed as a function of the Wigner function $f(\vec{r},\vec{p})$, which can be though as the quantum analogy of classical phase-space distribution function.
The Wigner function $f(\vec{r},\vec{p})$ is defined as the Fourier transform of the density matrix, i.e.,
\begin{equation}
\label{eq:Wingerf_r}
f(\vec{r}, \vec{p})=\int \exp \left(-i \frac{\vec{p}}{\hbar} \cdot \vec{s}\right) \rho(\vec{r}+\vec{s} / 2, \vec{r}-\vec{s} / 2) d^{3} s
\end{equation}
in coordinate configuration, and
\begin{equation}
\label{eq:Wingerf_p}
f(\vec{r}, \vec{p})= \int \exp \left(i \frac{\vec{q}}{\hbar} \cdot \vec{r}\right) g(\vec{p}+\vec{q} / 2, \vec{p}-\vec{q} / 2) d^{3} q
\end{equation}
in momentum configuration, where $\rho(\vec{r}+\vec{s} / 2, \vec{r}-\vec{s} / 2)$ and $g(\vec{p}+\vec{q} / 2, \vec{p}-\vec{q} / 2)$ are the matrix elements of the density matrices in coordinate and momentum representations, respectively.
With HF approximation, the density matrix of the system is expressed as $\hat{\rho}=\sum_{i}\left|\phi_i\right\rangle\left\langle\phi_i\right|$, and thus we have
\begin{align}
\label{eq:DM_r}
\rho(\vec{r}+\vec{s} / 2, \vec{r}-\vec{s} / 2)&=\langle\vec{r}+\vec{s} / 2 \left| \hat{\rho} \right| \vec{r}-\vec{s}/2 \rangle=\sum_{i} \phi_{i}^{\ast}(\vec{r}+\vec{s} / 2) \phi_{i}(\vec{r}-\vec{s} / 2), \\
\label{eq:DM_p}
g(\vec{p}+\vec{q} / 2, \vec{p}-\vec{q}/2)&= \langle\vec{p}+\vec{q} / 2 \left|\hat{\rho}\right| \vec{p}-\vec{q}/2 \rangle=\sum_{i} \phi_{i}^{\ast}(\vec{p}+\vec{q} / 2) \phi_{i}(\vec{p}-\vec{q} / 2),
\end{align}
where $\phi_{i}(\vec{r})$ and $\phi_{i}(\vec{p})$ are the one particle wave functions in coordinate and momentum configuration, respectively.

The central part of the standard Skyrme interaction is written as
\begin{equation}
\label{eq:NLO}
\begin{aligned}
v_{sk}  = & \,
t_0\left(1+x_0 \hat{P}_\sigma\right)  \delta \left( \vec{r}_1 -\vec{r}_2  \right) \\
& + t_1^{[2]}\left(1+x_1^{[2]} \hat{P}_\sigma\right) \frac{1}{2}\left[\hat{\vec{k}}^{\prime 2} \delta \left( \vec{r}_1 -\vec{r}_2  \right) + \delta \left( \vec{r}_1 -\vec{r}_2  \right) \hat{\vec{k}}^2\right]
+t_2^{[2]}\left(1+x_2^{[2]} \hat{P}_\sigma\right) \hat{\vec{k}}^{\prime} \cdot \delta \left( \vec{r}_1 -\vec{r}_2  \right) \hat{\vec{k}},
\end{aligned}
\end{equation}
which is also recognized as pseudopotential up to NLO, since it includes not only the local term ($t_0$ term), but also the spatial derivative operators up to the first order, i.e., $\hat{A}=\frac{1}{2}\left[\hat{\vec{k}}^{\prime}{}^2 \delta \left( \vec{r}_1 -\vec{r}_2  \right) + \delta \left( \vec{r}_1 -\vec{r}_2  \right) \hat{\vec{k}}{}^2\right]$ and $\hat{B}=\hat{\vec{k}}^{\prime} \cdot \delta \left( \vec{r}_1 -\vec{r}_2  \right) \hat{\vec{k}}$.
The HF calculations of the expectation values of $\hat{A}$ and $\hat{B}$ are considered as the starting point, which can be calculated as (detailed derivation can be found in Ref.~\cite{Wang:2018yce})
\begin{equation}
\label{eq:block1}
\begin{aligned}
 \sum_{i,j} \langle ij| \hA | ij \rangle
= & \sum_{i,j} \langle ij| \frac{1}{2} \left[
\vec{k}^{\prime}{}^2 \delta(\vec{r}_{1}-\vec{r}_{2}) + \delta(\vec{r}_{1}-\vec{r}_{2}) \vec{k}{}^2
\right] | ij \rangle \\
= &   \int d^{3}r \int \frac{{\rm d}^{3} p}{(2\pi\hbar)^3}\frac{{\rm d}^{3} p'}{(2\pi\hbar)^3}\frac{{\rm d}^{3} q}{(2\pi\hbar)^3}\frac{{\rm d}^{3} q'}{(2\pi\hbar)^3}\left( \frac{1}{2\hbar}  \right)^2 \left[
\left( \vec{p}-\vec{p}\,{}^{\prime} \right)^2  { \bm{+} } \left( \frac{\vec{q}-\vec{q}\,{}^{\prime}}{2}  \right) ^2 \right] \\
& \times \exp{\left( \frac{ \mathrm{i} \vec{q} \cdot \vec{r} }{\hbar}   \right)} \exp{\left( \frac{ \mathrm{i} \vec{q} \, {}^{\prime} \cdot \vec{r} }{\hbar}   \right)}\left[ g ( \vec{p} + \frac{\vec{q}}{2} , \vec{p} - \frac{\vec{q}}{2} ) \, g ( \vec{p}\,{}^{\prime} + \frac{\vec{q}\,{}^{\prime}}{2} , \vec{p}\,{}^{\prime} - \frac{\vec{q}\,{}^{\prime}}{2} )  \right]
\end{aligned}
\end{equation}
being the even parity term (even order of $\hat{B}$), and
\begin{equation}
\label{eq:block2}
\begin{aligned}
 \sum_{i,j} \langle ij| \hB | ij \rangle
= &  \sum_{i,j} \langle ij|  \left[
\vec{k}^{\prime} \cdot \delta(\vec{r}_{1}-\vec{r}_{2}) \vec{k}
\right] | ij \rangle \\
= &  \int d^{3}r \int \frac{{\rm d}^{3} p}{(2\pi\hbar)^3}\frac{{\rm d}^{3} p'}{(2\pi\hbar)^3}\frac{{\rm d}^{3} q}{(2\pi\hbar)^3}\frac{{\rm d}^{3} q'}{(2\pi\hbar)^3}\left( \frac{1}{2\hbar}  \right)^2 \left[
\left( \vec{p}-\vec{p}\,{}^{\prime} \right)^2  { \bm{-} } \left( \frac{\vec{q}-\vec{q}\,{}^{\prime}}{2}  \right) ^2 \right] \\
& \times \exp{\left( \frac{ \mathrm{i} \vec{q} \cdot \vec{r} }{\hbar}   \right)} \exp{\left( \frac{ \mathrm{i} \vec{q} \, {}^{\prime} \cdot \vec{r} }{\hbar}   \right)}\left[ g ( \vec{p} + \frac{\vec{q}}{2} , \vec{p} - \frac{\vec{q}}{2} ) \, g ( \vec{p}\,{}^{\prime} + \frac{\vec{q}\,{}^{\prime}}{2} , \vec{p}\,{}^{\prime} - \frac{\vec{q}\,{}^{\prime}}{2} )  \right]
\end{aligned}
\end{equation}
being with odd parity.
The integral of $\frac{1}{4\hbar^{2}} \left( \vec{p} - \vec{p}\,{}^{\prime} \right)^2$ becomes the MD terms, i.e.,
\begin{equation}
\frac{1}{4\hbar^{2}} \int d^{3}r \int \frac{{\rm d}^{3} p}{(2\pi\hbar)^3}\frac{{\rm d}^{3} p'}{(2\pi\hbar)^3}\left( \vec{p} - \vec{p}\,{}^{\prime} \right)^2  f (\vec{r},\vec{p}) \, f (\vec{r},\vec{p}\,^{\prime}),
\end{equation}
while the integral of $\frac{1}{4\hbar^{2}}\left( \frac{\vec{q}-\vec{q}\,{}^{\prime}}{2}  \right) ^2$ will become the gradient (GD) terms (see Ref.~\cite{Wang:2018yce} for details), i.e.,
\begin{align}
\label{eq:intq1}
\frac{1}{4\hbar^{2}} \frac{\vec{q}\,{}^2 + \vec{q}\,{}^{\prime}{}^2}{4} \implies &  \frac{1}{4} \left(\frac{\mathrm{i}}{2} \right)^2  \left[ 2 \rho \nabla^{2} \rho \right], \\
\label{eq:intq2}
\frac{1}{4\hbar^{2}} \frac{-2 \, \vec{q} \cdot  \vec{q}\,{}^{\prime}}{4} \implies & \frac{1}{4} \left(\frac{\mathrm{i}}{2} \right)^2  \left[ -2   \left( \nabla \rho \right)^2\right],
\end{align}
where $\mathrm{i}$ is the imaginary unit.

The Skyrme pseudopotential up to N3LO (including derivative terms up to the sixth-order) is constructed by combining $\hat{A}$, $\hat{B}$ as well as the spin part through the so-called LS-coupling \cite{Carlsson:2008gm,Raimondi:2011pz}, and the corresponding expressions in Cartesian representation have been derived in Refs.~\cite{Davesne:2014wya,Davesne:2014rva}.
It is seen that this potential is invariant under space rotation, space and time reversal and Hermitian conjugation by construction, and the coupling constants can be further constrained by the Galilean symmetry and the gauge symmetry.
In the present work, we only keep the central term and ignore the spin-dependent component since they do not contribute to spin-averaged quantities, on which we are focusing here.
We assume that the 2$n$th-order derivative term in the central term of the Skyrme pseudopotential can be constructed by using the binomial expansion of $\left(\hat{A} + \hat{B}    \right)^{n}$ (one can see $\left[ \hat{A},\hat{B} \right] =0 $, and the Dirac delta function $\DiracrR$ is omitted here for clarity), i.e.,
\begin{equation}
\label{eq:AB_binom}
(\hA+\hB)^n = \sum_{m=0}^{n} \binom{n}{m} \hA^{n-m} \hB^{m}.
\end{equation}
This assumption complies with the aforementioned symmetry principles and is consistent with corresponding forms in N3LO model Ref.~\cite{Davesne:2014wya,Davesne:2014rva}.

We partition Eq.~(\ref{eq:AB_binom}) into two groups according to the parity of the powers of $\hat{B}=\Bhat$ with the corresponding Skyrme parameters $\Skypat{1}{2n}$ and $\Skypat{2}{2n}$, and the N$n$LO pseudopotential central term $v^{[2n]}$ can be written as:
\begin{equation}
\label{eq:def_Vn}
v^{[2n]} = \Skypat{1}{2n} \left( 1 + \Skypax{1}{2n} \Psig \right) v^{[2n]}_{\mathrm{even}} +  \Skypat{2}{2n} \left( 1 + \Skypax{2}{2n} \Psig \right) v^{[2n]}_{\mathrm{odd}},
\end{equation}
where the even-parity term $v^{[2n]}_{\mathrm{even}}$ and odd-parity term $v^{[2n]}_{\mathrm{odd}}$ take the following forms:
\begin{equation}
\label{eq:VP_2n}
v^{[2n]}_{\mathrm{P}} =  \sum_{
\substack{
m = 0  \\ (m \in  \mathrm{P})
}
}^{n} \binom{n}{m} \left( \frac{\vec{k}^{\prime}{}^2+\vec{k}{}^2}{2} \right)^{n-m} \left( \vec{k}^{\prime} \cdot \vec{k} \right)^{m},
\end{equation}
with P being even or odd according to the parity.
And the contribution of the 2$n$th-order derivative terms with certain parity P can be calculated through HF approximation $\frac{1}{2} \sum_{i,j} \langle ij| \, v^{[2n]}_{\mathrm{P}} \left( 1 - \Psig \Ptau \Pm \right) | ij \rangle$.

One can see that for arbitrary natural numbers $n$ and $m$,
\begin{equation}
\label{eq:lemma0}
\begin{aligned}
\langle ij | \left( \frac{\vec{k}^{\prime}{}^2+\vec{k}{}^2}{2} \right) ^{n} \left( \vec{k}^{\prime} \cdot \vec{k} \right)^{m} | ij \rangle
= & \int d^{3}r \int \frac{{\rm d}^{3}p_{1}}{(2\pi\hbar)^3} \frac{{\rm d}^{3}p_{2}}{(2\pi\hbar)^3} \dots \frac{{\rm d}^{3}p_{2n+1}}{(2\pi\hbar)^3} \frac{{\rm d}^{3}p_{2n+2}}{(2\pi\hbar)^3} \dots \frac{{\rm d}^{3}p_{2(n+m)+1}}{(2\pi\hbar)^3} \frac{{\rm d}^{3}p_{2(n+m)+2}}{(2\pi\hbar)^3} \\
& \times \langle ij | p_{1} p_{2} \rangle \exp{\left[ \frac{- \mathrm{i} \left( \vec{p}_{1} + \vec{p}_{2} -\vec{p}_{3} - \vec{p}_{4}  \right)  }{\hbar} \cdot \vec{r} \right]}
\langle p_{1} p_{2} | \left( \frac{  \vec{k}^{\prime}{}^2+\vec{k}{}^2}{2} \right)_{1}  | p_{3} p_{4}  \rangle \\
& \times \dots \dots \\
& \times \exp{\left[ \frac{- \mathrm{i} \left( \vec{p}_{2n-1} + \vec{p}_{2n} -\vec{p}_{2n+1} - \vec{p}_{2n+2}  \right)  }{\hbar} \cdot \vec{r} \right]}
\langle p_{2n-1} p_{2n} | \left( \frac{  \vec{k}^{\prime}{}^2+\vec{k}{}^2}{2} \right)_{n}  | p_{2n+1} p_{2n+2}  \rangle \\
& \times \exp{\left[ \frac{- \mathrm{i} \left( \vec{p}_{2n+1} + \vec{p}_{2n+2} -\vec{p}_{2n+3} - \vec{p}_{2n+4}  \right)  }{\hbar} \cdot \vec{r} \right]}
\langle p_{2n+1} p_{2n+2} | \left( \vec{k}^{\prime} \cdot \vec{k}  \right)_{1}  | p_{2n+3} p_{2n+4}  \rangle \\
& \times \dots \dots \\
& \times \exp{\left[ \frac{- \mathrm{i} \left( \vec{p}_{2(n+m)-1} + \vec{p}_{2(n+m)} -\vec{p}_{2(n+m)+1} - \vec{p}_{2(n+m)+2}  \right)  }{\hbar} \cdot \vec{r} \right]} \\
& \times \langle p_{2(n+m)-1} p_{2(n+m)} | \left( \vec{k}^{\prime} \cdot \vec{k}  \right)_{m}  | p_{2(n+m)+1} p_{2(n+m)+2}  \rangle \, \langle p_{2(n+m)+1} p_{2(n+m)+2} | i j \rangle .
\end{aligned}
\end{equation}
Based on Eq.~(\ref{eq:block1}), Eq.~(\ref{eq:block2}), Eq.~(\ref{eq:VP_2n}) and Eq.~(\ref{eq:lemma0}), we can obtain $V ^{[2n]}_{\mathrm{P}}$, i.e., the expectation value of $v^{[2n]}_{\mathrm{P}}$ without the exchange term, as
\begin{equation}
\label{eq:V2n_P_1}
\begin{aligned}
V ^{[2n]}_{\mathrm{P}}= \frac{1}{2} \sum_{i,j} \langle ij| \, v^{[2n]}_{\mathrm{P}} \, | ij \rangle
= &  \, \frac{1}{2} \frac{1}{\left( 2 \hbar \right)^{2n}} \int d^{3}r \int \frac{{\rm d}^{3} p}{(2\pi\hbar)^3}\frac{{\rm d}^{3} p'}{(2\pi\hbar)^3}\frac{{\rm d}^{3} q}{(2\pi\hbar)^3}\frac{{\rm d}^{3} q'}{(2\pi\hbar)^3}\exp{\left( \frac{ \mathrm{i} \vec{q} \cdot \vec{r} }{\hbar}   \right)} \exp{\left( \frac{ \mathrm{i} \vec{q} \, {}^{\prime} \cdot \vec{r} }{\hbar}   \right)} \\
& \times\left[ g ( \vec{p} + \frac{\vec{q}}{2} , \vec{p} - \frac{\vec{q}}{2} ) \, g ( \vec{p}\,{}^{\prime} + \frac{\vec{q}\,{}^{\prime}}{2} , \vec{p}\,{}^{\prime} - \frac{\vec{q}\,{}^{\prime}}{2} )  \right] \\
& \times \sum_{ \substack{ m = 0  \\ (m \in  \mathrm{P})}}^{n} \binom{n}{m} \left[ \left( \vec{p}-\vec{p}\,{}^{\prime} \right)^2  { \bm{+} } \left( \frac{\vec{q}-\vec{q}\,{}^{\prime}}{2}  \right) ^2 \right] ^{n-m} \left[ \left( \vec{p}-\vec{p}\,{}^{\prime} \right)^2  { \bm{-} } \left( \frac{\vec{q}-\vec{q}\,{}^{\prime}}{2}  \right) ^2 \right] ^{m}.
\end{aligned}
\end{equation}
With the two identities:
\begin{align}
\label{eq:ID_even}
\sum_{ \substack{ m = 0  \\ (m \in  \mathrm{even})}}^{n} \binom{n}{m} \left( a + b \right)^{n-m} \left( a-b \right) ^{m} & = 2^{n-1} \left( a^{n} + b^{n}  \right) , \\
\label{eq:ID_odd}
\sum_{ \substack{ m = 0  \\ (m \in  \mathrm{odd})}}^{n} \binom{n}{m} \left( a + b \right)^{n-m} \left( a-b \right) ^{m} & = 2^{n-1} \left( a^{n} - b^{n}  \right) ,
\end{align}
Eq.~(\ref{eq:V2n_P_1}) becomes:
\begin{equation}
\label{eq:V2n_P_2}
\begin{aligned}
V ^{[2n]}_{\mathrm{P}} = & \frac{1}{2} \frac{1}{\left( 2 \hbar \right)^{2n}}  \int d^{3}r \int \frac{{\rm d}^{3} p}{(2\pi\hbar)^3}\frac{{\rm d}^{3} p'}{(2\pi\hbar)^3}\frac{{\rm d}^{3} q}{(2\pi\hbar)^3}\frac{{\rm d}^{3} q'}{(2\pi\hbar)^3}\exp{\left( \frac{ \mathrm{i} \vec{q} \cdot \vec{r} }{\hbar}   \right)} \exp{\left( \frac{ \mathrm{i} \vec{q} \, {}^{\prime} \cdot \vec{r} }{\hbar}   \right)} \\
& \times\left[ g ( \vec{p} + \frac{\vec{q}}{2} , \vec{p} - \frac{\vec{q}}{2} ) \, g ( \vec{p}\,{}^{\prime} + \frac{\vec{q}\,{}^{\prime}}{2} , \vec{p}\,{}^{\prime} - \frac{\vec{q}\,{}^{\prime}}{2} )  \right]2^{n-1}  \left[
\left( \vec{p}-\vec{p}\,{}^{\prime} \right)^{2n}  { \bm{\pm} } \left( \frac{\vec{q}-\vec{q}\,{}^{\prime}}{2}  \right) ^{2n} \right],
\end{aligned}
\end{equation}
where the even (odd) parity term take the $+$ ($-$) sign.
One can see that there are no cross terms in Eq.~(\ref{eq:V2n_P_2}), and the integrals of $\left( \vec{p}-\vec{p}\,{}^{\prime} \right)^{2n}$ and $\left( \frac{\vec{q}-\vec{q}\,{}^{\prime}}{2}  \right) ^{2n}$ become the MD term and GD term, respectively.
Next, we consider the contribution of the exchange term, i.e.,
\begin{equation}
\label{eq:V2n_iso}
\frac{1}{2} \sum_{i,j} \langle ij| \, v^{[2n]}_{\mathrm{P}} \left( - \Psig \Ptau \Pm \right) | ij \rangle .
\end{equation}
The Majorana operator $\Pm$ can be replaced by $\hat{1}$ ($-\hat{1}$) for even-parity (odd-parity) terms.
Assuming that there is no isospin mixing of the HF states, the isospin exchange operator $\Ptau$ can be replaced by a Kronecker delta function $\delta_{\tau_{i}\tau_{j}}$ with $\tau_{i}$ being the isospin of the $i$th state.
We assume the collision system is spin-averaged, and thus the spin exchange operator $\Psig$ is simply replaced by a factor of $\frac{1}{2}$.
By incorporating the exchange terms and summing the contributions from the odd and even terms, we can obtain the Hamiltonian density of the $n$th-order MD term $\mathcal{H}_{[2n]}^{\mathrm{MD}}$ as
\begin{equation}
\label{eq:HMD2n}
\begin{aligned}
\mathcal{H}_{[2n]}^{\mathrm{MD}} ( \vec{r} ) = & \, \frac{2^n}{8} \frac{1}{\left( 2 \hbar \right)^{2n}} \left[ t_{1}^{[2n]}\left( 2 + x_{1}^{[2n]} \right)  + t_{2}^{[2n]}\left( 2 + x_{2}^{[2n]} \right)  \right]  \mathcal{H}^{\mathrm{md}[n]}(\vec{r})  \\
& + \frac{2^n}{8} \frac{1}{\left( 2 \hbar \right)^{2n}} \left[ -t_{1}^{[2n]}\left( 2x_{1}^{[2n]} + 1 \right)  + t_{2}^{[2n]}\left( 2x_{2}^{[2n]} + 1 \right)  \right] \sum_{\tau=n,p} \mathcal{H}_\tau^{\mathrm{md}[n]}(\vec{r}),
\end{aligned}
\end{equation}
where
\begin{align}
\label{eq:Hmd_2n}
& \mathcal{H}^{\mathrm{md}[2n]}(\vec{r})=\int \frac{{\rm d}^3 p}{(2\pi\hbar)^3} \frac{{\rm d}^3 p'}{(2\pi\hbar)^3}\left(\vec{p}-\vec{p}\,{}^{\prime}\right)^{2n} f(\vec{r}, \vec{p}) f\left(\vec{r}, \vec{p}\,{}^{\prime}\right), \\
\label{eq:Hmdtau_2n}
& \mathcal{H}_\tau^{\mathrm{md}[2n]}(\vec{r})=\int \frac{{\rm d}^3 p}{(2\pi\hbar)^3} \frac{{\rm d}^3 p'}{(2\pi\hbar)^3}\left(\vec{p}-\vec{p}\,{}^{\prime}\right)^{2n} f_\tau(\vec{r}, \vec{p}) f_\tau\left(\vec{r}, \vec{p}\,{}^{\prime}\right).
\end{align}

On the other side, notice that $ \vec{q}\,{}^{2n-m} \cdot \vec{q}\,^{\prime}{}^{m}$ and $\vec{q}\,{}^{m} \cdot \vec{q}\,^{\prime}{}^{2n-m}$ contribute equally in the integral of $\left( \frac{\vec{q}-\vec{q}\,{}^{\prime}}{2}  \right) ^{2n}$ in Eq.~(\ref{eq:V2n_P_2}), i.e., both are $\left( \frac{\hbar}{\mathrm{i}} \right)^{2n} \nabla^{2n-m}\rho \nabla^{m}\rho$ (and $\nabla^{0}\rho = \rho$), the Hamiltonian density of the $n$th-order GD term can be expressed as:
\begin{equation}
\label{eq:HGD2n}
\begin{aligned}
\mathcal{H}_{[2n]}^{\mathrm{GD}} ( \vec{r} ) = & \, \frac{2^n}{8} \frac{1}{\left( 2 \hbar \right)^{2n}} \left[ t_{1}^{[2n]}\left( 2 + x_{1}^{[2n]} \right)  - t_{2}^{[2n]}\left( 2 + x_{2}^{[2n]} \right)  \right]  \left( \frac{\mathrm{i}\hbar}{2} \right) ^{2n}  \mathcal{H}^{\mathrm{gd}[2n]} ( \vec{r} ) \\
& - \frac{2^n}{8} \frac{1}{\left( 2 \hbar \right)^{2n}} \left[ t_{1}^{[2n]}\left( 2x_{1}^{[2n]} + 1 \right)  +  t_{2}^{[2n]}\left( 2x_{2}^{[2n]} + 1 \right)  \right]  \left( \frac{\mathrm{i}\hbar}{2} \right) ^{2n} \sum_{\tau=n,p} \mathcal{H}_{\tau}^{\mathrm{gd}[2n]} ( \vec{r} ),
\end{aligned}
\end{equation}
where
\begin{equation}
\label{eq:H_gd2n}
\begin{aligned}
\mathcal{H}^{\mathrm{gd}[2n]} ( \vec{r} ) = &  \sum_{m=0}^{2n} (-1)^{m} \binom{2n}{m} \nabla^{m}\rho(\vec{r}) \, \nabla^{2n-m}\rho(\vec{r}) \\
=& 2 \sum_{m=0}^{n-1} \left[ (-1)^{m} \binom{2n}{m} \nabla^{m}\rho(\vec{r}) \, \nabla^{2n-m}\rho(\vec{r})  \right] + (-1)^{n} \binom{2n}{n} \left[ \nabla^{n} \rho (\vec{r}) \right]^{2},
\end{aligned}
\end{equation}
and
\begin{equation}
\label{eq:Htau_gd2n}
\begin{aligned}
 \mathcal{H}_{\tau}^{\mathrm{gd}[2n]} ( \vec{r} ) = &  \sum_{m=0}^{2n} (-1)^{m} \binom{2n}{m} \nabla^{m}\rhotr \, \nabla^{2n-m}\rhotr \\
=& 2 \sum_{m=0}^{n-1} \left[ (-1)^{m} \binom{2n}{m} \nabla^{m}\rho_{\tau}(\vec{r}) \, \nabla^{2n-m}\rho_{\tau}(\vec{r})  \right] + (-1)^{n} \binom{2n}{n} \left[ \nabla^{n} \rho_{\tau} (\vec{r}) \right]^{2}.
\end{aligned}
\end{equation}

For convenience, we can recombine the Skyrme parameters in Eqs.~(\ref{eq:HMD2n}) and (\ref{eq:HGD2n}), namely, $t_{1}^{[2n]}$, $t_{2}^{[2n]}$, $x_{1}^{[2n]}$ and $x_{2}^{[2n]}$, into the parameters $C^{[2n]}$, $D^{[2n]}$, $E^{[2n]}$ and $F^{[2n]}$, i.e.,
\begin{align}
\label{eq:C2n}
& C^{[2n]} = t_{1}^{[2n]}\left( 2 + x_{1}^{[2n]} \right)  + t_{2}^{[2n]}\left( 2 + x_{2}^{[2n]} \right), \\
\label{eq:D2n}
& D^{[2n]} = -t_{1}^{[2n]}\left( 2x_{1}^{[2n]} + 1 \right)  + t_{2}^{[2n]}\left( 2x_{2}^{[2n]} + 1 \right), \\
\label{eq:E2n}
& E^{[2n]} = \left( \frac{\mathrm{i}}{2} \right)^{2n} \left[
t_{1}^{[2n]}\left( 2 + x_{1}^{[2n]} \right)  - t_{2}^{[2n]}\left( 2 + x_{2}^{[2n]} \right) \right], \\
\label{eq:F2n}
& F^{[2n]} = - \left( \frac{\mathrm{i}}{2} \right)^{2n} \left[
t_{1}^{[2n]}\left( 2x_{1}^{[2n]} + 1 \right)  + t_{2}^{[2n]}\left( 2x_{2}^{[2n]} + 1 \right) \right].
\end{align}
Or we can obtain that
\begin{align}
& t_{1}^{[2n]} = \frac{1}{6} \left[ \left( 2C^{[2n]} + D^{[2n]} \right) + \left( 2\mathrm{i} \right)^{2n}\left( 2E^{[2n]} + F^{[2n]}  \right) \right], \\
& t_{2}^{[2n]} = \frac{1}{6} \left[ \left( 2C^{[2n]} - D^{[2n]} \right) - \left( 2\mathrm{i} \right)^{2n}\left( 2E^{[2n]} - F^{[2n]}  \right) \right], \\
& x_{1}^{[2n]} = -\frac{\left( C^{[2n]} + 2D^{[2n]} \right) + \left( 2\mathrm{i} \right)^{2n}\left( E^{[2n]} + 2F^{[2n]}  \right) }{\left( 2C^{[2n]} + D^{[2n]} \right) + \left( 2\mathrm{i} \right)^{2n}\left( 2E^{[2n]} + F^{[2n]}  \right) }, \\
& x_{2}^{[2n]} = -\frac{\left( C^{[2n]} - 2D^{[2n]} \right) - \left( 2\mathrm{i} \right)^{2n}\left( E^{[2n]} - 2F^{[2n]}  \right) }{\left( 2C^{[2n]} - D^{[2n]} \right) - \left( 2\mathrm{i} \right)^{2n}\left( 2E^{[2n]} - F^{[2n]}  \right) },
\end{align}
once $C^{[2n]}$, $D^{[2n]}$, $E^{[2n]}$ and $F^{[2n]}$ are given.
The units for $t_{1}^{[2n]}$, $t_{2}^{[2n]}$, $C^{[2n]}$, $D^{[2n]}$, $E^{[2n]}$ and $F^{[2n]}$ are $\mathrm{MeV}\,\mathrm{fm}^{2n+3}$, while $x_{1}^{[2n]}$ and $x_{2}^{[2n]}$ are dimensionless.

Finally, for the N$n$LO Skyrme pseudopotential central term defined in Eq.~(\ref{eq:def_Vn}), we obtain its Hamiltonian density as
\begin{equation}
\label{eq:H_2n}
\begin{aligned}
\mathcal{H}_{[2n]} ( \vec{r} ) = & \, \frac{1}{2} \sum_{i,j} \langle i j | v^{[2n]} \left( 1 - \Psig \Ptau \Pm     \right) | i j \rangle =  \mathcal{H}_{[2n]}^{\mathrm{MD}} ( \vec{r} ) + \mathcal{H}_{[2n]}^{\mathrm{GD}} ( \vec{r} ) \\
= & \, \frac{1}{2^{n+3}} \frac{C^{[2n]}}{\hbar^{2n}}  \mathcal{H}^{\mathrm{md}[2n]} (\vec{r}) + \frac{1}{2^{n+3}} \frac{D^{[2n]}}{ \hbar^{2n}}  \sum_{\tau=n,p} \mathcal{H}_\tau^{\mathrm{md}[2n]}(\vec{r}) \\
& + \frac{1}{2^{n+3}} E^{[2n]} \mathcal{H}^{\mathrm{gd}[2n]} ( \vec{r} ) + \frac{1}{2^{n+3}} F^{[2n]} \sum_{\tau=n,p}  \mathcal{H}_{\tau}^{\mathrm{gd}[2n]} ( \vec{r} ) .
\end{aligned}
\end{equation}

\subsection{Single-nucleon potential with the extended Skyrme interaction}
\label{S:A1}
Within the framework of Landau Fermi liquid theory, the $n$th-order single-nucleon potential $U_{\tau}^{[2n]}$ can be expressed as:
\begin{equation}
\label{eq:def_U2n}
U_{\tau}^{[2n]}(\vec{r},\vec{p}) = \frac{ \delta \int \mathcal{H}_{[2n]} ( \vec{r} ) d\vec{r} }{ \delta f_{\tau}(\vec{r},\vec{p})}
= \frac{\delta \int \mathcal{H}^{\mathrm{MD}}_{[2n]} ( \vec{r} ) d\vec{r}}{\delta f_{\tau}(\vec{r},\vec{p})} + \sum_{m=0}^{2n} (-1)^{m} \nabla^{m} \frac{\partial \mathcal{H}_{[2n]}^{\mathrm{GD}} ( \vec{r} ) }{ \partial \left[ \nabla^{m} \rhotr  \right] } .
\end{equation}
The variation of the $n$th-order MD term becomes:
\begin{equation}
\frac{\delta \int \mathcal{H}^{\mathrm{MD}}_{[2n]} ( \vec{r} ) d\vec{r}}{\delta f_{\tau}(\vec{r},\vec{p})} = \frac{1}{2^{n+2}} \frac{C^{[2n]}}{\hbar^{2n}}  U^{\mathrm{md}[2n]} (\vec{r},\vec{p}) + \frac{1}{2^{n+2}} \frac{D^{[2n]}}{ \hbar^{2n}}  U_\tau^{\mathrm{md}[2n]}(\vec{r},\vec{p}),
\end{equation}
where
\begin{equation}
U^{\mathrm{md}[2n]}(\vec{r},\vec{p}) = \int \frac{{\rm d}^3 p'}{(2\pi\hbar)^3}\left(\vec{p}-\vec{p}\,{}^{\prime}\right)^{2n} f\left(\vec{r}, \vec{p}\,{}^{\prime}\right),
\label{E:UMD}
\end{equation}
\begin{equation}
U_{\tau}^{\mathrm{md}[2n]}(\vec{r},\vec{p}) = \int \frac{{\rm d}^3 p'}{(2\pi\hbar)^3}\left(\vec{p}-\vec{p}\,{}^{\prime}\right)^{2n} f_{\tau} \left(\vec{r}, \vec{p}\,{}^{\prime}\right).
\label{E:UMDt}
\end{equation}

Substitute Eq.~(\ref{eq:H_gd2n}) and Eq.~(\ref{eq:Htau_gd2n}) into Eq.~(\ref{eq:def_U2n}), and note that every term in Eq.~(\ref{eq:H_gd2n}) and Eq.~(\ref{eq:Htau_gd2n}) (take $\nabla^{k} \rhor \nabla^{2n-k} \rhor$ as an example) will contribute twice with the partial derivative $\partial/\partial \left[ \nabla^{m} \rhotr  \right]$ (when $m=k$ and $m=2n-k$), i.e.,
\begin{equation}
\begin{aligned}
& \sum_{m=0}^{2n} (-1)^{m} \nabla^{m} \frac{\partial \left[ (-1)^{k} \binom{2n}{k} \nabla^{k} \rhor \nabla^{2n-k} \rhor   \right] }{ \partial \left[ \nabla^{m} \rhotr  \right] } \\
= & \, (-1)^{2k} \binom{2n}{k} \nabla^{k} \left[ \nabla^{2n-k} \rhor \right] + (-1)^{2n} \binom{2n}{k} \nabla^{2n-k} \left[ \nabla^{k} \rhor \right] \\
= & \, 2 \binom{2n}{k} \nabla^{2n} \rhor .
\end{aligned}
\end{equation}
Then we can obtain that
\begin{equation}
\begin{aligned}
\sum_{m=0}^{2n} (-1)^{m} \nabla^{m} \frac{\partial \mathcal{H}_{[2n]}^{\mathrm{GD}} ( \vec{r} ) }{ \partial \left[ \nabla^{m} \rhotr  \right] } & = \frac{1}{2^{n+3}} E^{[2n]} 2^{2n+1} \nabla^{2n} \rhor + \frac{1}{2^{n+3}} F^{[2n]} 2^{2n+1}  \nabla^{2n} \rhotr \\
& = 2^{n-2} E^{[2n]} \nabla^{2n} \rhor + 2^{n-2} F^{[2n]} \nabla^{2n} \rhotr,
\end{aligned}
\end{equation}
and here the identity $\sum_{m=0}^{2n}\binom{2n}{m}=2^{2n}$ is used.
Finally, the $n$th-order single-nucleon potential can be expressed as
\begin{equation}
\begin{aligned}
\label{eq:Utau_2n}
U_{\tau}^{[2n]}(\vec{r},\vec{p}) = & \, \frac{1}{2^{n+2}} \frac{C^{[2n]}}{\hbar^{2n}}  U^{\mathrm{md}[2n]} (\vec{r},\vec{p}) + \frac{1}{2^{n+2}} \frac{D^{[2n]}}{ \hbar^{2n}}  U_\tau^{\mathrm{md}[2n]}(\vec{r},\vec{p}) \\
& + 2^{n-2} E^{[2n]} \nabla^{2n} \rhor + 2^{n-2} F^{[2n]} \nabla^{2n} \rhotr .
\end{aligned}
\end{equation}

As mentioned in the main text, one of the advantages of the polynomial form of the MD term in Eqs.~(\ref{E:UMD}) and (\ref{E:UMDt}) [also for Eqs.~(\ref{eq:Hmd_2n}) and (\ref{eq:Hmdtau_2n})] is that their $\vec{p}$ dependence can be factored out from the integral.
For Eq.~(\ref{E:UMD}) one has
\begin{equation}
\begin{split}
U^{\mathrm{md}[2n]}(\vec{r},\vec{p}) = & \int \frac{{\rm d}^3 p'}{(2\pi\hbar)^3} \left(\vec{p}-\vec{p}\,{}^{\prime}\right)^{2n} f(\vec{r},\vec{p}\,{}^{\prime})\\
=&\int \frac{{\rm d}^3 p'}{(2\pi\hbar)^3} (p^2-2p_xp'_x-2p_yp'_y-2p_zp'_z+p'^2)^n f(\vec{r},\vec{p}\,{}^{\prime})\\
 = &\sum_{n_1+n_2+\cdots+n_5=n\atop n_1,n_2,\cdots,n_5\geqslant0}\Big({n\atop n_1,n_2,\cdots,n_5}\Big)(p^2)^{n_1}(-2p_x)^{n_2}(-2p_y)^{n_3}(-2p_z)^{n_4}\Big[\int\frac{{\rm d}^3 p'}{(2\pi\hbar)^3}p'_xf(\vec{r},\vec{p}\,{}^{\prime})\Big]^{n_2}\\
 &\times\Big[\int\frac{{\rm d}^3 p'}{(2\pi\hbar)^3}p'_yf(\vec{r},\vec{p}\,{}^{\prime})\Big]^{n_3}\Big[\int\frac{{\rm d}^3 p'}{(2\pi\hbar)^3}p'_zf(\vec{r},\vec{p}\,{}^{\prime})\Big]^{n_4}\Big[\int\frac{{\rm d}^3 p'}{(2\pi\hbar)^3}p'^2f(\vec{r},\vec{p}\,{}^{\prime})\Big]^{n_5}\\
 =&\sum_{n_1+n_2+\cdots+n_5=n\atop n_1,n_2,\cdots,n_5\geqslant0}\Big({n\atop n_1,n_2,\cdots,n_5}\Big)(p^2)^{n_1}(-2p_x)^{n_2}(-2p_y)^{n_3}(-2p_z)^{n_4}\langle p_x\rangle_{\vec{r}}^{n_2}\langle p_y\rangle_{\vec{r}}^{n_3}\langle p_z\rangle_{\vec{r}}^{n_4}\langle p^2\rangle_{\vec{r}}^{n_5}.
\end{split}
\label{E:ME}
\end{equation}
Similarly one has
\begin{equation}
\begin{split}
U_{\tau}^{\mathrm{md}[2n]}(\vec{r},\vec{p}) = & \int \frac{{\rm d}^3 p'}{(2\pi\hbar)^3}\left(\vec{p}-\vec{p}\,{}^{\prime}\right)^{2n} f_{\tau}(\vec{r},\vec{p}\,{}^{\prime})\\
 =&\sum_{n_1+n_2+\cdots+n_5=n\atop n_1,n_2,\cdots,n_5\geqslant0}\Big({n\atop n_1,n_2,\cdots,n_5}\Big)(p^2)^{n_1}(-2p_x)^{n_2}(-2p_y)^{n_3}(-2p_z)^{n_4}\langle p_x\rangle_{\vec{r},\tau}^{n_2}\langle p_y\rangle_{\vec{r},\tau}^{n_3}\langle p_z\rangle_{\vec{r},\tau}^{n_4}\langle p^2\rangle_{\vec{r},\tau}^{n_5}.
\end{split}
\label{E:MEt}
\end{equation}
In the above equations,
\begin{equation}
    \Big({n\atop n_1,n_2,\cdots,n_5}\Big) = \frac{n!}{n_1!n_2!\cdots n_5!}
\end{equation}
is the the multinomial coefficient, and we have defined the mean value of a quantity at $\vec{r}$, i.e.,
\begin{eqnarray}
    \langle A\rangle_{\vec{r}} = \int\frac{{\rm d}^3 p}{(2\pi\hbar)^3}A(\vec{p})f(\vec{r},\vec{p}),\\
    \langle A\rangle_{\vec{r},\tau} = \int\frac{{\rm d}^3 p}{(2\pi\hbar)^3}A(\vec{p})f_{\tau}(\vec{r},\vec{p}).
\end{eqnarray}
We notice from Eqs.~(\ref{E:ME}) and (\ref{E:MEt}) that the $\vec{p}$ dependence of $U^{\mathrm{md}[2n]}(\vec{r},\vec{p})$ and $U_{\tau}^{\mathrm{md}[2n]}(\vec{r},\vec{p})$ are factored out from the integral over $\vec{p}\,^{\prime}$.
It thus enables us to eliminate a loop over test particles when calculating the momentum-dependent part of single-particle potentials in the transport model, and reduces significantly the computational complexity of the transport model if a very large $N_E$ is adopted.

\section{THE CHARACTERISTIC QUANTITIES OF THE SNM EOS AND THE SYMMETRY ENERGY, THE FOURTH-ORDER SYMMETRY ENERGY, AND THE NUCLEON EFFECTIVE MASSES}
\label{sec:App_quantities}
The SNM EOS $E_{0}(\rho)$, the symmetry energy $E_{\mathrm{sym}}(\rho)$ along with their expansion coefficients defined in Eqs.~(\ref{eq:L_def})-(\ref{eq:H_def}) are used to characterize the density behavior of the nuclear matter, which are expressed as
\begin{equation}
\label{eq:E0}
\begin{aligned}
E_{0}(\rho) =& \frac{3 a^2 \hbar ^2}{10 m} \rho^{2/3}
+ \frac{3 t_{0}}{8} \rho^{3/3}
+ \frac{t_{3}^{[1]}}{16} \rho^{4/3}
+ \frac{3 a^{2}}{80} \left(2C^{[2]}+D^{[2]} \right) \rho^{5/3}
+ \frac{t_{3}^{[3]}}{16} \rho^{6/3}
+ \frac{9 a^{4}}{280} \left(2C^{[4]}+D^{[4]} \right) \rho^{7/3} \\
& + \frac{t_{3}^{[5]}}{16} \rho^{8/3}
+ \frac{2 a^{6}}{15} \left(2C^{[6]}+D^{[6]} \right) \rho^{9/3}
+ \frac{t_{3}^{[7]}}{16} \rho^{10/3}
+ \frac{3 a^{8}}{77} \left(2C^{[8]}+D^{[8]} \right) \rho^{11/3}
+ \frac{t_{3}^{[9]}}{16} \rho^{12/3}\\
&+ \frac{9 a^{10}}{182} \left(2C^{[10]}+D^{[10]} \right) \rho^{13/3},
\end{aligned}
\end{equation}
\begin{equation}
\label{eq:L0}
\begin{aligned}
L_{0}(\rho) =& \frac{3 a^2 \hbar ^2}{5 m} \rho^{2/3}
+ \frac{9 t_{0}}{8} \rho^{3/3}
+ \frac{t_{3}^{[1]}}{4} \rho^{4/3}
+ \frac{3 a^{2}}{16} \left(2C^{[2]}+D^{[2]} \right) \rho^{5/3}
+ \frac{3 t_{3}^{[3]}}{8} \rho^{6/3}
+ \frac{9 a^{4}}{40} \left(2C^{[4]}+D^{[4]} \right) \rho^{7/3} \\
&+ \frac{t_{3}^{[5]}}{2} \rho^{8/3}
+ \frac{6 a^{6}}{5} \left(2C^{[6]}+D^{[6]} \right) \rho^{9/3}
+ \frac{5 t_{3}^{[7]}}{8} \rho^{10/3}
+ \frac{3 a^{8}}{7} \left(2C^{[8]}+D^{[8]} \right) \rho^{11/3}
+ \frac{3 t_{3}^{[9]}}{4} \rho^{12/3} \\
&+ \frac{9 a^{10}}{14} \left(2C^{[10]}+D^{[10]} \right) \rho^{13/3},
\end{aligned}
\end{equation}
\begin{equation}
\label{eq:K0}
\begin{aligned}
K_{0}(\rho) =& -\frac{3 a^{2} \hbar^2}{5m} \rho^{2/3}
+ \frac{t_{3}^{[1]}}{4} \rho^{4/3}
+ \frac{3 a^{2}}{8} \left(2C^{[2]}+D^{[2]} \right) \rho^{5/3}
+ \frac{9 t_{3}^{[3]}}{8} \rho^{6/3}
+ \frac{9 a^{4}}{10} \left(2C^{[4]}+D^{[4]} \right) \rho^{7/3}
+ \frac{5 t_{3}^{[5]}}{2} \rho^{8/3} \\
&+ \frac{36 a^{6}}{5} \left(2C^{[6]}+D^{[6]} \right) \rho^{9/3}
+ \frac{35 t_{3}^{[7]}}{8} \rho^{10/3}
+ \frac{24 a^{8}}{7} \left(2C^{[8]}+D^{[8]} \right) \rho^{11/3}
+ \frac{27 t_{3}^{[9]}}{4} \rho^{12/3}\\
&+ \frac{45 a^{10}}{7} \left(2C^{[10]}+D^{[10]} \right) \rho^{13/3},
\end{aligned}
\end{equation}
\begin{equation}
\label{eq:J0}
\begin{aligned}
J_{0}(\rho) =& \frac{12 a^{2} \hbar^2}{5 m} \rho^{2/3}
- \frac{t_{3}^{[1]}}{2} \rho^{4/3}
- \frac{3 a^{2}}{8} \left(2C^{[2]}+D^{[2]} \right) \rho^{5/3}
+ \frac{9 a^{4}}{10} \left(2C^{[4]}+D^{[4]} \right) \rho^{7/3}
+ 5 t_{3}^{[5]} \rho^{8/3} \\
&+ \frac{108 a^{6}}{5} \left(2C^{[6]}+D^{[6]} \right) \rho^{9/3}
+ \frac{35 t_{3}^{[7]}}{2} \rho^{10/3}
+ \frac{120 a^{8}}{7} \left(2C^{[8]}+D^{[8]} \right) \rho^{11/3}
+ \frac{81 t_{3}^{[9]}}{2} \rho^{12/3} \\
&+ 45 a^{10} \left(2C^{[10]}+D^{[10]} \right) \rho^{13/3},
\end{aligned}
\end{equation}
\begin{equation}
\label{eq:I0}
\begin{aligned}
I_{0}(\rho) =& -\frac{84 a^2 \hbar ^2}{5 m} \rho^{2/3}
+ \frac{5 t_{3}^{[1]}}{2} \rho^{4/3}
+ \frac{3 a^{2}}{2} \left(2C^{[2]}+D^{[2]} \right) \rho^{5/3}
- \frac{9 a^{4}}{5} \left(2C^{[4]}+D^{[4]} \right) \rho^{7/3}
- 5 t_{3}^{[5]} \rho^{8/3} \\
&+ \frac{35 t_{3}^{[7]}}{2} \rho^{10/3}
+ \frac{240 a^{8}}{7} \left(2C^{[8]}+D^{[8]} \right) \rho^{11/3}
+ \frac{243 t_{3}^{[9]}}{2} \rho^{12/3}
+ 180 a^{10} \left(2C^{[10]}+D^{[10]} \right) \rho^{13/3},
\end{aligned}
\end{equation}
\begin{equation}
\label{eq:H0}
\begin{aligned}
H_{0}(\rho) =& \frac{168 a^2 \hbar ^2}{m} \rho^{2/3}
- 20 t_{3}^{[1]} \rho^{4/3}
- \frac{21 a^{2}}{2} \left(2C^{[2]}+D^{[2]} \right) \rho^{5/3}
+ 9 a^{4} \left(2C^{[4]}+D^{[4]} \right) \rho^{7/3} \\
&+ 20 t_{3}^{[5]} \rho^{8/3}
- 35 t_{3}^{[7]} \rho^{10/3}
- \frac{240 a^{8}}{7} \left(2C^{[8]}+D^{[8]} \right) \rho^{11/3}
+ 180 a^{10} \left(2C^{[10]}+D^{[10]} \right) \rho^{13/3},
\end{aligned}
\end{equation}
and
\begin{equation}
\label{eq:Esym}
\begin{aligned}
E_{\mathrm{sym}}(\rho) =& \frac{a^2 \hbar ^2}{6 m} \rho^{2/3}
- \frac{1}{8} t_{0} (2 x_{0} +1) \rho^{3/3}
- \frac{1}{48} t_{3}^{[1]} \left(2 x_{3}^{[1]} +1 \right) \rho^{4/3}
+ \frac{a^2}{24} \left(C^{[2]}+2 D^{[2]} \right) \rho^{5/3} \\
&- \frac{1}{48} t_{3}^{[3]} \left(2 x_{3}^{[3]} +1 \right) \rho^{6/3}
+ \frac{a^4}{24} \left(2C^{[4]}+3 D^{[4]} \right) \rho^{7/3}
- \frac{1}{48} t_{3}^{[5]} \left(2 x_{3}^{[5]} +1 \right) \rho^{8/3}\\
&+ \frac{a^6}{5} \left(3C^{[6]}+ 4D^{[6]} \right) \rho^{9/3}
- \frac{1}{48} t_{3}^{[7]} \left(2 x_{3}^{[7]} +1 \right) \rho^{10/3}
+ \frac{a^8}{15} \left(4C^{[8]}+ 5D^{[8]} \right) \rho^{11/3} \\
&- \frac{1}{48} t_{3}^{[9]} \left(2 x_{3}^{[9]} +1 \right) \rho^{12/3}
+ \frac{2a^{10}}{21} \left(5C^{[10]}+ 6D^{[10]} \right) \rho^{13/3},
\end{aligned}
\end{equation}
\begin{equation}
\label{eq:L}
\begin{aligned}
L(\rho) =& \frac{a^2 \hbar ^2}{3 m} \rho^{2/3}
- \frac{3}{8} t_{0} (2 x_{0} +1) \rho^{3/3}
- \frac{1}{12} t_{3}^{[1]} \left(2 x_{3}^{[1]} +1 \right) \rho^{4/3}
+ \frac{5 a^{2}}{24} \left(C^{[2]}+2 D^{[2]} \right) \rho^{5/3} \\
&- \frac{1}{8} t_{3}^{[3]} \left(2 x_{3}^{[3]} +1 \right) \rho^{6/3}
+ \frac{7 a^{4}}{24} \left(2C^{[4]}+ 3D^{[4]} \right) \rho^{7/3}
- \frac{1}{6} t_{3}^{[5]} \left(2 x_{3}^{[5]} +1 \right) \rho^{8/3} \\
& + \frac{9 a^{6}}{5} \left(3C^{[6]}+ 4D^{[6]} \right) \rho^{9/3}
- \frac{5}{24} t_{3}^{[7]} \left(2 x_{3}^{[7]} +1 \right) \rho^{10/3}
+ \frac{11 a^{8}}{15} \left(4C^{[8]}+ 5D^{[8]} \right) \rho^{11/3} \\
&- \frac{1}{4} t_{3}^{[9]} \left(2 x_{3}^{[9]} +1 \right) \rho^{12/3}
+ \frac{26 a^{10}}{21} \left(5C^{[10]}+ 6D^{[10]} \right) \rho^{13/3} ,
\end{aligned}
\end{equation}
\begin{equation}
\label{eq:Ksym}
\begin{aligned}
K_{\mathrm{sym}}(\rho) =& -\frac{a^2 \hbar ^2}{3m} \rho^{2/3}
- \frac{1}{12} t_{3}^{[1]} \left(2 x_{3}^{[1]} +1 \right) \rho^{4/3}
+ \frac{5 a^{2}}{12} \left(C^{[2]}+2 D^{[2]} \right) \rho^{5/3}
- \frac{3}{8} t_{3}^{[3]} \left(2 x_{3}^{[3]} +1 \right) \rho^{6/3} \\
&+ \frac{7 a^{4}}{6} \left(2C^{[4]}+ 3D^{[4]} \right) \rho^{7/3}
- \frac{5}{6} t_{3}^{[5]} \left(2 x_{3}^{[5]} +1 \right) \rho^{8/3}
+ \frac{54 a^{6}}{5} \left(3C^{[6]}+ 4D^{[6]} \right) \rho^{9/3} \\
&- \frac{35}{24} t_{3}^{[7]} \left(2 x_{3}^{[7]} +1 \right) \rho^{10/3}
+ \frac{88 a^{8}}{15} \left(4C^{[8]}+ 5D^{[8]} \right) \rho^{11/3}
- \frac{9}{4} t_{3}^{[9]} \left(2 x_{3}^{[9]} +1 \right) \rho^{12/3} \\
&+ \frac{260 a^{10}}{21} \left(5C^{[10]}+ 6D^{[10]} \right) \rho^{13/3} ,
\end{aligned}
\end{equation}
\begin{equation}
\label{eq:Jsym}
\begin{aligned}
J_{\mathrm{sym}}(\rho) =& \frac{4 a^2 \hbar ^2}{3m} \rho^{2/3}
+ \frac{1}{6} t_{3}^{[1]} \left(2 x_{3}^{[1]} +1 \right) \rho^{4/3}
- \frac{5 a^{2}}{12} \left(C^{[2]}+2 D^{[2]} \right) \rho^{5/3}
+ \frac{7 a^{4}}{6} \left(2C^{[4]}+ 3D^{[4]} \right) \rho^{7/3} \\
&- \frac{5}{3} t_{3}^{[5]} \left(2 x_{3}^{[5]} +1 \right) \rho^{8/3}
+ \frac{162 a^{6}}{5} \left(3C^{[6]}+ 4D^{[6]} \right) \rho^{9/3}
- \frac{35}{6} t_{3}^{[7]} \left(2 x_{3}^{[7]} +1 \right) \rho^{10/3} \\
&+ \frac{88 a^{8}}{3} \left(4C^{[8]}+ 5D^{[8]} \right) \rho^{11/3}
- \frac{27}{2} t_{3}^{[9]} \left(2 x_{3}^{[9]} +1 \right) \rho^{12/3}
+ \frac{260 a^{10}}{3} \left(5C^{[10]}+ 6D^{[10]} \right) \rho^{13/3} ,
\end{aligned}
\end{equation}
\begin{equation}
\label{eq:Isym}
\begin{aligned}
I_{\mathrm{sym}}(\rho) =& -\frac{28 a^2 \hbar^2}{3m} \rho^{2/3}
- \frac{5}{6} t_{3}^{[1]} \left(2 x_{3}^{[1]} +1 \right) \rho^{4/3}
+ \frac{5 a^{2}}{3} \left(C^{[2]}+2 D^{[2]} \right) \rho^{5/3}
- \frac{7 a^{4}}{3} \left(2C^{[4]}+ 3D^{[4]} \right) \rho^{7/3} \\
&+ \frac{5}{3} t_{3}^{[5]} \left(2 x_{3}^{[5]} +1 \right) \rho^{8/3}
- \frac{35}{6} t_{3}^{[7]} \left(2 x_{3}^{[7]} +1 \right) \rho^{10/3}
+ \frac{176 a^{8}}{3} \left(4C^{[8]}+ 5D^{[8]} \right) \rho^{11/3} \\
&- \frac{81}{2} t_{3}^{[9]} \left(2 x_{3}^{[9]} +1 \right) \rho^{12/3}
+ \frac{1040 a^{10}}{3} \left(5C^{[10]}+ 6D^{[10]} \right) \rho^{13/3} ,
\end{aligned}
\end{equation}
\begin{equation}
\label{eq:Hsym}
\begin{aligned}
H_{\mathrm{sym}}(\rho) =& \frac{280 a^2 \hbar ^2}{3m} \rho^{2/3}
+ \frac{20}{3} t_{3}^{[1]} \left(2 x_{3}^{[1]} +1 \right) \rho^{4/3}
- \frac{35 a^{2}}{3} \left(C^{[2]}+2 D^{[2]} \right) \rho^{5/3}
+ \frac{35 a^{4}}{3} \left(2C^{[4]}+ 3D^{[4]} \right) \rho^{7/3} \\
&- \frac{20}{3} t_{3}^{[5]} \left(2 x_{3}^{[5]} +1 \right) \rho^{8/3}
+ \frac{35}{3} t_{3}^{[7]} \left(2 x_{3}^{[7]} +1 \right) \rho^{10/3}
- \frac{176 a^{8}}{3} \left(4C^{[8]}+ 5D^{[8]} \right) \rho^{11/3} \\
&+ \frac{1040 a^{10}}{3} \left(5C^{[10]}+ 6D^{[10]} \right) \rho^{13/3} .
\end{aligned}
\end{equation}
And the fourth-order symmetry energy can be expressed as
\begin{equation}
\label{eq:Esym4}
\begin{aligned}
E_{\mathrm{sym},4}(\rho) \equiv& \left. \frac{1}{4!} \frac{\partial^{4} E(\rho,\delta)}{\partial \delta^{4}} \right|_{\delta=0}
=\frac{a^{2} \hbar ^2}{162m} \rho^{2/3}
+ \frac{a^{2}}{648} \left(C^{[2]} -  D^{[2]} \right) \rho^{5/3}
+ \frac{a^{4}}{648} \left(8C^{[4]} + 3D^{[4]} \right) \rho^{7/3} \\
&+ \frac{2 a^{6}}{135} \left(13C^{[6]} + 9D^{[6]} \right) \rho^{9/3}
+ \frac{a^{8}}{405} \left(61C^{[8]} + 50D^{[8]} \right) \rho^{11/3}
+ \frac{2 a^{10}}{81} \left(17C^{[10]}+ 15D^{[10]} \right) \rho^{13/3}.
\end{aligned}
\end{equation}

The isoscalar nucleon effective mass $m_{s}^{\ast}$ and isovector nucleon effective mass $m_{v}^{\ast}$ are momentum dependent in our models.
We define $\Tilde{M}_{s} \equiv m/m_{s}$ and $\Tilde{M}_{v} \equiv m/m_{v}$.
Thus, we have
\begin{equation}
\label{eq:Ms_exp}
\small
\begin{aligned}
\Tilde{M}_{s}(\rho,p) = &  1+ \frac{m}{p} \frac{d U_0(\rho,p)}{d p} \\
 = & 1 + \frac{m}{8 \hbar^2} \rho \left( 2 C^{[2]} + D^{[2]}  \right)
+ \frac{m}{8 \hbar^2} a^2 \rho^{5/3} \left( 2 C^{[4]} + D^{[4]}  \right)
+ \frac{3 m}{8 \hbar^2} a^4 \rho^{7/3} \left( 2 C^{[6]} + D^{[6]}  \right) \\
&+ \frac{m}{16 \hbar^2} a^6 \rho^{3} \left( 2 C^{[8]} + D^{[8]}  \right)
+\frac{5 m}{128 \hbar^2} a^8 \rho^{11/3} \left( 2 C^{[10]} + D^{[10]}  \right)
\\
& + \frac{p^2}{\hbar^2}
\left[
\frac{m}{8 \hbar^2} \rho \left( 2 C^{[4]} + D^{[4]} \right)
+  \frac{21 m}{20 \hbar^2} a^2 \rho^{5/3} \left( 2 C^{[6]} + D^{[6]}  \right)
+  \frac{27 m}{80 \hbar^2} a^4 \rho^{7/3} \left( 2 C^{[8]} + D^{[8]}  \right)
+  \frac{11 m}{32 \hbar^2} a^6 \rho^{3} \left( 2 C^{[10]} + D^{[10]}  \right)
\right] \\
& + \frac{p^4}{\hbar^4}
\left[
\frac{3 m}{8 \hbar^2} \rho \left( 2 C^{[6]} + D^{[6]}  \right)
+ \frac{27 m}{80 \hbar^2} a^2 \rho^{5/3} \left( 2 C^{[8]} + D^{[8]}  \right)
+ \frac{297 m}{448 \hbar^2} a^4 \rho^{7/3} \left( 2 C^{[8]} + D^{[8]}  \right)
\right] \\
& + \frac{p^6}{\hbar^6}
\left[
\frac{m}{16 \hbar^2} \rho \left( 2 C^{[8]} + D^{[8]}  \right)
+ \frac{11 m}{32 \hbar^2} a^2 \rho^{5/3} \left( 2 C^{[10]} + D^{[10]}  \right)
\right]
+ \frac{p^8}{\hbar^8}
\left[
\frac{5 m}{128 \hbar^2} \rho \left( 2 C^{[10]} + D^{[10]}  \right)
\right] ,
\end{aligned}
\end{equation}
and
\begin{equation}
\begin{aligned}
\label{eq:Mv_exp}
\Tilde{M}_{v}(\rho,p) = &   1+ \frac{m}{p} \frac{d U_{\tau}(\rho,-\tau,p)}{d p} \\
=& 1 + \frac{m}{4 \hbar^2} \rho C^{[2]}
+ \frac{m}{4 \hbar^2} 2^{2/3} a^2 \rho^{5/3} C^{[4]}
+ \frac{m}{2 \hbar^2} 2^{1/3} a^4 \rho^{7/3} C^{[6]}
+ \frac{m}{2 \hbar^2} a^6 \rho^{3} C^{[8]}
+ \frac{5 m}{16 \hbar^2} 2^{1/3} a^8 \rho^{11/3} C^{[10]}
\\
& + \frac{p^2}{\hbar^2}
\left[
\frac{m}{4 \hbar^2} \rho C^{[4]}
+ \frac{21 m}{20 \hbar^2} 2^{2/3} a^2 \rho^{5/3} C^{[6]}
+ \frac{27 m}{20 \hbar^2} 2^{1/3} a^4 \rho^{7/3} C^{[8]}
+ \frac{11 m}{4 \hbar^2} a^6 \rho^{3} C^{[10]}
\right] \\
&+ \frac{p^4}{\hbar^4} \left[
\frac{3 m}{4 \hbar^2} \rho C^{[6]}
+ \frac{27 m}{40 \hbar^2} 2^{2/3} a^2 \rho^{5/3} C^{[8]}
+ \frac{297 m}{112 \hbar^2} 2^{1/3} a^4 \rho^{7/3} C^{[10]}
\right] \\
&+ \frac{p^6}{\hbar^6} \left[
\frac{m}{8 \hbar^2} \rho C^{[8]}
+ \frac{11 m}{16 \hbar^2} 2^{1/3} a^2 \rho^{5/3} C^{[10]}
\right]
+ \frac{p^8}{\hbar^8} \left[
\frac{5 m}{64 \hbar^2} \rho C^{[10]}
\right],
\end{aligned}
\end{equation}
with $\tau = 1\, [-1]$ for neutron [proton].
And the linear isospin splitting coefficient can be expressed as
\begin{equation}
\label{eq:Dm1}
\Delta m_{1}^{\ast}(\rho)  \equiv \left. \frac{\partial \mspl(\rho,\delta)}{\partial \delta} \right|_{\delta=0}
= - \frac{ \mathcal{A}(\rho) }{ \mathcal{B}(\rho) },
\end{equation}
where
\begin{equation}
\label{eq:Dm1A}
\begin{aligned}
\mathcal{A}(\rho) =& \, 80 m \hbar^{2} \left[ 15 D^{[2]} \rho^{3/3}
+ 10a^{2}\left(2C^{[4]} + 5D^{[4]} \right) \rho^{5/3}
+ 72a^{4}\left( 4C^{[6]} + 7D^{[6]} \right) \rho^{7/3}  \right. \\
&+ \left. 96a^{6}\left( 2C^{[8]} + 3D^{[8]} \right) \rho^{9/3}
+ \frac{400}{7}a^{8}\left( 8C^{[10]} + 11D^{[10]} \right) \rho^{11/3}
\right],
\end{aligned}
\end{equation}
and
\begin{equation}
\label{eq:Dm1B}
\begin{aligned}
\mathcal{B}(\rho) =& \, 3 m^{2} \left[  \frac{40 \hbar^{2}}{m}
+ 5 \left(2C^{[2]} + D^{[2]} \right) \rho^{3/3}
+ 10 a^{2} \left(2C^{[4]} + D^{[4]} \right) \rho^{5/3}
+ 72 a^{4} \left(2C^{[6]} + D^{[6]} \right) \rho^{7/3} \right. \\
&+ \left. 32 a^{6} \left(2C^{[8]} + D^{[8]} \right) \rho^{9/3}
+ \frac{400}{7} a^{8} \left(2C^{[10]} + D^{[10]} \right) \rho^{11/3}
\right]^{2}.
\end{aligned}
\end{equation}

Ignoring $C^{[10]}$, $D^{[10]}$, $E^{[10]}$, $F^{[10]}$ and $t_3^{[9]}$ terms (or as well as the $C^{[8]}$, $D^{[8]}$, $E^{[8]}$, $F^{[8]}$ and $t_3^{[7]}$ terms), these expressions in N5LO model will reduce to their corresponding forms in the N4LO (N3LO) model.

\section{PARAMETERS OF THE SKYRME INTERACTIONS}
\label{sec:Skyrme_paras}
In Table~\ref{tab:SP6}, \ref{tab:SP8} and \ref{tab:SP10}, we list the Skyrme parameters of the 24 interactions.
\begin{table*}
\caption{\label{tab:SP6}
The 14 parameters of the interactions SP6L45X.
Here the recombination of Skyrme parameters defined in Eq.~(\ref{eq:C2n})
and Eq.~(\ref{eq:D2n}) are used.
The units of parameters:
$t_0$: $\mathrm{MeV}\, \mathrm{fm}^3$; $t_{3}^{[n]}$ ($n=1,3,5$), $C^{[n]}$ and $D^{[n]}$ ($n=2,4,6$): $\mathrm{MeV}\, \mathrm{fm}^{n+3}$; $x_0$ and $x_{3}^{[n]}$ ($n=1,3,5$) are dimensionless.
}
\begin{ruledtabular}
\begin{tabular}{ccccccccc}
X&Dm07&Dm05&Dm03&Dm01&D01&D03&D05&D07 \\ \hline
$t_0$ &
$-1840.45$ & $-1840.45$ & $-1840.45$ & $-1840.45$ &
$-1840.45$ & $-1840.45$ & $-1840.45$ & $-1840.45$ \\
$t_{3}^{[1]}$ &
$13010.2$  & $13010.2$  & $13010.2$  & $13010.2$  &
$13010.2$  & $13010.2$  & $13010.2$  & $13010.2$  \\
$t_{3}^{[3]}$ &
$-4036.41$ & $-4036.41$ & $-4036.41$ & $-4036.41$ &
$-4036.41$ & $-4036.41$ & $-4036.41$ & $-4036.41$ \\
$t_{3}^{[5]}$ &
$2386.36$  & $2386.36$  & $2386.36$  & $2386.36$  &
$2386.36$  & $2386.36$  & $2386.36$  & $2386.36$  \\
$x_0$ &
$0.328136$ & $0.309276$ & $0.290417$ & $0.271557$ &
$0.252697$ & $0.233838$ & $0.214978$ & $0.196119$ \\
$x_{3}^{[1]}$ &
$0.720796$ & $0.611729$ & $0.502661$ & $0.393593$ &
$0.284524$ & $0.175457$ & $0.066389$ &$-0.042677$ \\
$x_{3}^{[3]}$ &
$-4.69172$ & $-3.98025$ & $-3.26878$ & $-2.55731$ &
$-1.84583$ & $-1.13437$ & $-0.422901$& $0.288567$ \\
$x_{3}^{[5]}$&
$-8.11908$ & $-7.58510$ & $-7.05111$ & $-6.51713$ &
$-5.98314$ & $-5.44915$ & $-4.91517$ & $-4.38119$ \\
$C^{[2]}$   &
$-322.800$ & $-135.318$ & $52.1646$  & $239.648$ &
$427.132$  & $614.614$  & $802.097$  & $989.580$ \\
$D^{[2]}$ &
$1343.54$  & $968.578$  & $593.612$  & $218.645$ &
$-156.323$ & $-531.287$ & $-906.253$ & $-1281.21$ \\
$C^{[4]}$  &
$-2.25298$ & $-5.30996$ & $-8.36694$ & $-11.4239$ &
$-14.4809$ & $-17.5379$ & $-20.5948$ & $-23.6518$ \\
$D^{[4]}$  &
$-21.9069$ & $-15.7930$ & $-9.67908$ & $-3.56509$ &
$2.54890$  & $8.66285$  & $14.7768$  & $20.8907$  \\
$C^{[6]}$  &
$0.023356$ & $0.028186$ & $0.033017$ & $0.037848$ &
$0.042679$ & $0.047510$ & $0.052341$ & $0.057172$ \\
$D^{[6]}$  &
$0.034618$ & $0.024957$ & $0.015295$ & $0.005633$ &
$-0.004027$& $-0.013689$& $-0.023351$& $-0.033012$
\end{tabular}
\end{ruledtabular}
\end{table*}
\begin{table*}
\caption{\label{tab:SP8}
The 18 parameters of the interactions SP8L45X.
Here the recombination of Skyrme parameters defined in Eq.~(\ref{eq:C2n})
and Eq.~(\ref{eq:D2n}) are used.
The units of parameters:
$t_0$: $\mathrm{MeV}\, \mathrm{fm}^3$; $t_{3}^{[n]}$ ($n=1,3,5,7$), $C^{[n]}$ and $D^{[n]}$ ($n=2,4,6,8$): $\mathrm{MeV}\, \mathrm{fm}^{n+3}$; $x_0$ and $x_{3}^{[n]}$ ($n=1,3,5,7$) are dimensionless.
}
\begin{ruledtabular}
\begin{tabular}{ccccccccc}
X&Dm07&Dm05&Dm03&Dm01&D01&D03&D05&D07 \\ \hline
$t_0$ &
$-1838.65$ & $-1838.65$ & $-1838.65$ & $-1838.65$ &
$-1838.65$ & $-1838.65$ & $-1838.65$ & $-1838.65$ \\
$t_{3}^{[1]}$ &
$12926.3$  & $12926.3$  & $12926.3$  & $12926.3$  &
$12926.3$  & $12926.3$  & $12926.3$  & $12926.3$  \\
$t_{3}^{[3]}$ &
$-4248.13$ & $-4248.13$ & $-4248.13$ & $-4248.13$ &
$-4248.13$ & $-4248.13$ & $-4248.13$ & $-4248.13$ \\
$t_{3}^{[5]}$ &
$2649.97$  & $2649.97$  & $2649.97$  & $2649.97$  &
$2649.97$  & $2649.97$  & $2649.97$  & $2649.97$  \\
$t_{3}^{[7]}$ &
$-127.291$ & $-127.291$ & $-127.291$ & $-127.291$ &
$-127.291$ & $-127.291$ & $-127.291$ & $-127.291$ \\
$x_0$ &
$0.331623$ & $0.312314$ & $0.293006$ & $0.273698$ &
$0.254389$ & $0.235081$ & $0.215773$ & $0.196464$ \\
$x_{3}^{[1]}$ &
$0.746536$ & $0.633826$ & $0.521118$ & $0.408409$ &
$0.295701$ & $0.182992$ & $0.070283$ & $-0.042425$ \\
$x_{3}^{[3]}$ &
$-4.62263$ & $-3.91835$ & $-3.21408$ & $-2.50981$ &
$-1.80555$ & $-1.10128$ & $-0.397013$& $0.307262$ \\
$x_{3}^{[5]}$ &
$-7.61305$ & $-7.09574$ & $-6.57845$ & $-6.06115$ &
$-5.54386$ & $-5.02656$ & $-4.50927$ & $-3.99196$ \\
$x_{3}^{[7]}$ &
$-3.15560$ & $-2.80526$ & $-2.45504$ & $-2.10484$ &
$-1.75469$ & $-1.40438$ & $-1.05416$ & $-0.703829$ \\
$C^{[2]}$ &
$-315.391$ & $-121.836$ & $71.7175$  & $265.271$ &
$458.825$  & $652.380$  & $845.934$  & $1039.49$ \\
$D^{[2]}$ &
$1392.53$  & $1005.42$  & $618.314$  & $231.206$ &
$-155.901$ & $-543.011$ & $-930.119$ & $-1317.23$ \\
$C^{[4]}$ &
$-5.17687$ & $-8.33604$ & $-11.4951$ & $-14.6543$ &
$-17.8134$ & $-20.9726$ & $-24.1317$ & $-27.2908$ \\
$D^{[4]}$ &
$-22.7285$ & $-16.4102$ & $-10.0919$ & $-3.77368$ &
$2.54458$  & $8.86288$  & $15.1811$  & $21.4994$  \\
$C^{[6]}$ &
$0.055782$ & $0.060957$ & $0.066131$ & $0.071306$ &
$0.076480$ & $0.081655$ & $0.086829$ & $0.092004$ \\
$D^{[6]}$ &
$0.037228$ & $0.026879$ & $0.016530$ & $0.006181$ &
$-0.004167$ & $-0.014517$ & $-0.024866$ & $-0.035215$ \\
$C^{[8]}$ &
$-0.001628$ & $-0.001698$ & $-0.001768$ & $-0.001839$ &
$-0.001909$ & $-0.001979$ & $-0.002050$ & $-0.002120$ \\
$D^{[8]}$ &
$-5.058\times10^{-4}$ & $-3.652\times10^{-4}$ & $-2.246\times10^{-4}$ &
$-8.399\times10^{-5}$ & $5.663\times10^{-5}$ & $1.972\times10^{-4}$ &
$3.378\times10^{-4}$ & $4.785\times10^{-4}$ \\
\end{tabular}
\end{ruledtabular}
\end{table*}
\begin{table*}
\caption{\label{tab:SP10}
The 22 parameters of the interactions SP10L45X.
Here the recombination of Skyrme parameters defined in Eq.~(\ref{eq:C2n})
and Eq.~(\ref{eq:D2n}) are used.
The units of parameters:
$t_0$: $\mathrm{MeV}\, \mathrm{fm}^3$; $t_{3}^{[n]}$ ($n=1,3,5,7,9$), $C^{[n]}$ and $D^{[n]}$ ($n=2,4,6,8,10$): $\mathrm{MeV}\, \mathrm{fm}^{n+3}$; $x_0$ and $x_{3}^{[n]}$ ($n=1,3,5,7,9$) are dimensionless.
}
\begin{ruledtabular}
\begin{tabular}{ccccccccc}
X&Dm07&Dm05&Dm03&Dm01&D01&D03&D05&D07 \\ \hline
$t_0$ &
$-1838.88$ & $-1838.88$ & $-1838.88$ & $-1838.88$ &
$-1838.88$ & $-1838.88$ & $-1838.88$ & $-1838.88$ \\
$t_{3}^{[1]}$ &
$12927.0$  & $12927.0$  & $12927.0$  & $12927.0$  &
$12927.0$  & $12927.0$  & $12927.0$  & $12927.0$  \\
$t_{3}^{[3]}$ &
$-4334.99$ & $-4334.99$ & $-4334.99$ & $-4334.99$ &
$-4334.99$ & $-4334.99$ & $-4334.99$ & $-4334.99$ \\
$t_{3}^{[5]}$ &
$2898.99$  & $2898.99$  & $2898.99$  & $2898.99$  &
$2898.99$  & $2898.99$  & $2898.99$  & $2898.99$  \\
$t_{3}^{[7]}$ &
$-477.033$ & $-477.033$ & $-477.033$ & $-477.033$ &
$-477.033$ & $-477.033$ & $-477.033$ & $-477.033$ \\
$t_{3}^{[9]}$ &
$212.964$  & $212.964$  & $212.964$  & $212.964$  &
$212.964$  & $212.964$  & $212.964$  & $212.964$  \\
$x_0$ &
$0.331118$ & $0.311884$ & $0.292650$ & $0.273417$ &
$0.254183$ & $0.234949$ & $0.215715$ & $0.196481$ \\
$x_{3}^{[1]}$ &
$0.745822$ & $0.633297$ & $0.520773$ & $0.408248$ &
$0.295723$ & $0.183199$ & $0.070674$ & $-0.041850$ \\
$x_{3}^{[3]}$ &
$-4.59069$ & $-3.89277$ & $-3.19486$ & $-2.49695$ &
$-1.79903$ & $-1.10112$ & $-0.403207$& $0.294711$ \\
$x_{3}^{[5]}$ &
$-7.21694$ & $-6.71607$ & $-6.21522$ & $-5.71435$ &
$-5.21350$ & $-4.71263$ & $-4.21177$ & $-3.71090$ \\
$x_{3}^{[7]}$ &
$-3.14162$ & $-2.79184$ & $-2.44216$ & $-2.09242$ &
$-1.74276$ & $-1.39297$ & $-1.04324$ & $-0.693462$ \\
$x_{3}^{[9]}$ &
$-3.17760$ & $-2.80998$ & $-2.44247$ & $-2.07489$ &
$-1.70740$ & $-1.33976$ & $-0.972199$& $-0.604576$ \\
$C^{[2]}$ &
$-313.147$ & $-119.109$ & $74.9264$  & $268.964$ &
$463.002$  & $657.040$  & $851.077$  & $1045.11$ \\
$D^{[2]}$ &
$1397.65$  & $1009.57$  & $621.505$  & $233.428$ &
$-154.645$ & $-542.721$ & $-930.795$ & $-1318.87$ \\
$C^{[4]}$ &
$-6.21986$ & $-9.38697$ & $-12.5540$ & $-15.7211$ &
$-18.8882$ & $-22.0553$ & $-25.2224$ & $-28.3895$ \\
$D^{[4]}$ &
$-22.8126$ & $-16.4784$ & $-10.1442$ & $-3.81005$ &
$2.52413$  & $8.85835$  & $15.1925$  & $21.5267$ \\
$C^{[6]}$ &
$0.073508$ & $0.078700$ & $0.083892$ & $0.089084$ &
$0.094276$ & $0.099469$ & $0.104661$ & $0.109853$ \\
$D^{[6]}$ &
$0.037399$ & $0.027014$ & $0.016630$ & $0.006246$ &
$-0.004138$ & $-0.014522$ & $-0.024906$ & $-0.038206$ \\
$C^{[8]}$ &
$-0.003309$ & $-0.003382$ & $-0.003455$ & $-0.003528$ &
$-0.003601$ & $-0.003674$ & $-0.003747$ & $-0.003820$ \\
$D^{[8]}$ &
$-5.258\times10^{-4}$ & $-3.798\times10^{-4}$ & $-2.338\times10^{-4}$ &
$-8.783\times10^{-5}$ & $5.818\times10^{-5}$ & $2.042\times10^{-4}$ &
$3.502\times10^{-4}$ & $4.962\times10^{-4}$ \\
$C^{[10]}$ &
$1.241\times10^{-5}$ & $1.256\times10^{-5}$ & $1.272\times10^{-5}$ &
$1.287\times10^{-5}$ & $1.303\times10^{-5}$ & $1.318\times10^{-5}$ &
$1.334\times10^{-5}$ & $1.349\times10^{-5}$ \\
$D^{[10]}$ &
$1.113\times10^{-6}$ & $8.044\times10^{-7}$ & $4.952\times10^{-7}$ &
$1.859\times10^{-7}$ & $-1.232\times10^{-7}$ & $-4.324\times10^{-7}$ &
$-7.416\times10^{-7}$ & $-1.050\times10^{-6}$ \\
\end{tabular}
\end{ruledtabular}
\end{table*}

\end{widetext}

\bibliography{SkyN5LORef}

\end{document}